\documentclass[fleqn,usenatbib]{mnras}
\usepackage{newtxtext,newtxmath}
\usepackage[T1]{fontenc}
\usepackage{xcolor}

\DeclareRobustCommand{\VAN}[3]{#2}
\let\VANthebibliography\thebibliography
\def\thebibliography{\DeclareRobustCommand{\VAN}[3]{##3}\VANthebibliography}

\usepackage{graphicx}	% Including figure files
\usepackage{amsmath}	% Advanced maths commands
\newcommand{\HI}{\mbox{H\hspace{0.15em}{\sc i}}}

\newcommand{\OI}{\mbox{O\hspace{0.15em}{\sc i}}}
\newcommand{\OII}{\mbox{O\hspace{0.15em}{\sc ii}}}
\newcommand{\OIII}{\mbox{O\hspace{0.15em}{\sc iii}}}

\newcommand{\CII}{\mbox{C\hspace{0.15em}{\sc ii}}}

\newcommand{\CIV}{\mbox{C\hspace{0.15em}{\sc iv}}}
\newcommand{\NaI}{\mbox{Na\hspace{0.15em}{\sc i}}}
\newcommand{\MgII}{\mbox{Mg\hspace{0.15em}{\sc ii}}}

\newcommand{\SII}{\mbox{S\hspace{0.15em}{\sc ii}}}
\newcommand{\NII}{\mbox{N\hspace{0.15em}{\sc ii}}}
\newcommand{\NV}{\mbox{N\hspace{0.15em}{\sc v}}}
\newcommand{\SiII}{\mbox{Si\hspace{0.15em}{\sc ii}}}
\newcommand{\SiIV}{\mbox{Si\hspace{0.15em}{\sc iv}}}
\newcommand{\FeII}{\mbox{Fe\hspace{0.15em}{\sc ii}}}

\newcommand{\ljmu}{1}
\newcommand{\cornell}{2}
\newcommand{\zhejiang}{3}
\newcommand{\weizmann}{4}
\newcommand{\taui}{5}
\newcommand{\caltech}{6}
\newcommand{\oxford}{7}
\newcommand{\syracuse}{8}
\newcommand{\esochile}{9}
\newcommand{\unc}{10}
\newcommand{\ncu}{11}
\newcommand{\dirac}{12}
\newcommand{\berkeley}{13}
\newcommand{\lbnl}{14}
\newcommand{\ifa}{15}
\newcommand{\ista}{16}
\newcommand{\finca}{17}
\newcommand{\ipac}{18}
\newcommand{\umn}{19}
\newcommand{\sorbonne}{20}
\newcommand{\cnrs}{21}
\newcommand{\lancaster}{22}
\newcommand{\coo}{23}
\newcommand{\cfa}{24}
\newcommand{\ice}{25}
\newcommand{\ieec}{26}
\newcommand{\okc}{27}
\newcommand{\warsawobs}{28}
\newcommand{\cardiff}{29}
\newcommand{\turku}{30}
\newcommand{\northwestern}{31}
\newcommand{\ciera}{32}
\newcommand{\skai}{33}
\newcommand{\cddd}{34}
\newcommand{\carnegie}{35}
\newcommand{\mitkavli}{36}
\newcommand{\tcd}{37}
\newcommand{\icen}{38}
\newcommand{\queens}{39}
\newcommand{\ncnr}{40}
\newcommand{\tarapaca}{41}
\newcommand{\ucla}{42}
\newcommand{\warwick}{43}
\newcommand{\maryland}{44}
\newcommand{\jssi}{45}
\newcommand{\gsfc}{46}

\newcommand{\spc}{\hspace{0.28cm}}

\title[AT2024wpp]{AT\,2024wpp: An Extremely Luminous Fast Ultraviolet Transient Powered by Accretion onto a Black Hole}

\author[D. A. Perley et al.]{
{Daniel A. Perley}$^{\ljmu}$\thanks{E-mail: d.a.perley@ljmu.ac.uk},
{Anna Y. Q. Ho}$^{\cornell}$,
{Zo\"e McGrath}$^{\ljmu}$,
{Michael Camilo}$^{\cornell}$,
{Cassie Sevilla}$^{\cornell}$,
\newauthor
{Ping Chen}$^{\zhejiang,\weizmann}$,
{Genevieve Schroeder}$^{\cornell}$,
{Taya Govreen-Segal}$^{\taui}$,
{Aleksandra Bochenek}$^{\ljmu}$,
{Yu-Jing Qin}$^{\caltech}$,
\newauthor
{James H. Gillanders}$^{\oxford}$,
{Benjamin Amend}$^{\syracuse}$,
{Joseph P. Anderson}$^{\esochile}$,
{Igor Andreoni}$^{\unc}$,
{Amar Aryan}$^{\ncu}$,
\newauthor
{Eric C. Bellm}$^{\dirac}$,
{Joshua S. Bloom}$^{\berkeley,\lbnl}$,
{Thomas de Boer}$^{\ifa}$,
{Jonathan Carney}$^{\unc}$,
{Ilaria Caiazzo}$^{\ista}$,
\newauthor
{Ken C. Chambers}$^{\ifa}$,
{Panos Charalampopoulos}$^{\finca}$,
{Ting-Wan Chen}$^{\ncu}$,
{Tracy X. Chen}$^{\ipac}$,
{Eric R. Coughlin}$^{\syracuse}$,
\newauthor
{Michael Coughlin}$^{\umn}$,
{Michel Dennefeld}$^{\sorbonne,\cnrs}$,
{Georgios Dimitriadis}$^{\lancaster}$,
{Christoffer Fremling}$^{\coo,\caltech}$,
\newauthor
{Danielle Frostig}$^{\cfa}$,
{Avishay Gal-Yam}$^{\weizmann}$,
{Llu\'is Galbany}$^{\ice,\ieec}$, %csic
{Anjashay Gangopadhyay}$^{\okc}$,
\newauthor
{Melzie Ghendrih}$^{\lancaster}$,
{Matthew J. Graham}$^{\caltech}$,
{Mariusz Gromadzki}$^{\warsawobs}$,
{Steven L. Groom}$^{\ipac}$,
\newauthor
{Claudia P. Guti\'errez}$^{\ieec,\ice}$,
{K.-Ryan Hinds}$^{\ljmu,\cornell,\caltech}$,
{Mark E. Huber}$^{\ifa}$,
{Cosimo Inserra}$^{\cardiff}$,
{Benjamin C. Kaiser}$^{\unc}$,
\newauthor
{Mansi M. Kasliwal}$^{\caltech}$,
{Niilo E. Koivisto}$^{\turku}$,
{Chien-Cheng Lin}$^{\ifa}$,
{Chang Liu}$^{\northwestern,\ciera,\skai}$,
{Thomas B. Lowe}$^{\ifa}$,
\newauthor
{Eugene Magnier}$^{\ifa}$,
{Ashish A. Mahabal}$^{\caltech,\cddd}$,
{Andrew Milligan}$^{\lancaster}$,
{Paloma Minguez}$^{\ifa}$,
{Geoffrey Mo}$^{\caltech,\carnegie,\mitkavli}$,
\newauthor
{Tom\'as E. M\"uller-Bravo}$^{\tcd,\icen}$,
{Matt Nicholl}$^{\queens}$,
{Priscila J. Pessi}$^{\okc,\ncnr}$,
{Giuliano Pignata}$^{\tarapaca}$,
{Josiah Purdum}$^{\coo}$,
\newauthor
{Nabeel Rehemtulla}$^{\northwestern,\ciera,\skai}$,
{R. Michael Rich}$^{\ucla}$,
{Anwesha Sahu}$^{\warwick}$,
{Avinash Singh}$^{\okc}$,
{Stephen J. Smartt}$^{\oxford,\queens}$,
\newauthor
{Ian A. Smith}$^{\ifa}$,
{Jesper Sollerman}$^{\okc}$,
{Gokul Srinivasaragavan}$^{\maryland,\jssi,\gsfc}$,
{Shubham Srivastav}$^{\oxford}$,
\newauthor
{Robert D. Stein}$^{\maryland,\jssi,\gsfc}$,
{Steve Schulze}$^{\northwestern}$,
{Jack W. Tweddle}$^{\oxford}$,
{Richard Wainscoat}$^{\ifa}$,
{Jacob L. Wise}$^{\ljmu}$,
\newauthor
{Lin Yan}$^{\coo}$,
{David R. Young}$^{\queens}$
\\
\\
Affiliations are listed at the end of the manuscript.
}

\date{Accepted 2026 April 7. Received 2026 April 5; in original form 2026 January 7}

\pubyear{2026}

\begin{document}
\label{firstpage}
\pagerange{\pageref{firstpage}--\pageref{lastpage}}
\maketitle

\begin{abstract}
We present the discovery of AT\,2024wpp (``Whippet''), a fast and luminous 18cow-like transient.   At a redshift of $z=0.0868$, revealed by Keck Cosmic Web Imager spectroscopy of its faint star-forming host, it is the fourth-nearest example of its class to date.  Rapid identification of the source in the Zwicky Transient Facility data stream permitted ultraviolet-through-optical observations to be obtained prior to peak, allowing the first determination of the peak bolometric luminosity ($2\times10^{45}$ erg s$^{-1}$), maximum photospheric radius ($10^{15}$ cm), and total radiated energy ($10^{51}$ erg) of an 18cow-like object.  We present results from a comprehensive multiwavelength observing campaign, including a far-UV spectrum from the Cosmic Origins Spectrograph on the Hubble Space Telescope and deep imaging extending $>$100 days post-explosion from the Very Large Telescope, Hubble Space Telescope, Very Large Array, and Atacama Large Millimetre Array.
We interpret the observations under a model in which a rapidly-accreting central engine blows a fast ($\sim$\,0.2\,$c$) wind into the surrounding medium and irradiates it with X-rays.  The high Doppler velocities and intense ionization within this wind prevent identifiable spectroscopic features from appearing in the ejecta or in the surrounding circumstellar material.  Weak H and He signatures do emerge in the spectra after 35 days in the form of double-peaked narrow lines.  Each peak is individually narrow (full width $\delta v \sim 3000$\,km\,s$^{-1}$) but the two components are separated by $\Delta v \sim 6600$ km~s$^{-1}$, indicating stable structures of denser material, possibly representing streams of tidal ejecta or an ablated companion star. 
\end{abstract}


%%%%%%%%%%%%%%%%%%%%%%%%%%%%%%%%%%%%%%%%%%%%%%%%%%

%%%%%%%%%%%%%%%%% BODY OF PAPER %%%%%%%%%%%%%%%%%%

\section{Introduction}

The introduction of wide-field synoptic sky surveys in the past two decades has greatly expanded our view of the processes involved in the deaths of stars (for a review, see \citealt{Inserra+2019}).  Among the new classes of transients being identified are supernovae that are unexpectedly over- or under-luminous \citep{Quimby+2011,GalYam+2019,Kasliwal+2012}, those which are unusually fast or unusually long-lasting \citep{Arcavi+2017,Gillanders+2020,Schulze+2024}, or with spectra differing dramatically from previous populations of events \citep{GalYam+2022,Schulze+2025}.

Among this menagerie of unusual transients, one particular type of event stands out as particularly notable. 
While fast-rising, fast-decaying optical transients with peak luminosities characteristic of supernovae are known from a variety of sources \citep{Drout+2014,Tanaka+2016,Pursiainen+2018,Rest+2018}, the real-time discovery of nearby events suitable for late-time and multi-wavelength follow-up \citep{Prentice+2018,Perley+2019,Margutti+2019,Coppejans+2020,Perley+2021,Ho+2023sample} demonstrated that a subset of them --- referred to either as luminous fast blue optical transients (LFBOTs) or 18cow-like objects (after the prototypical event AT\,2018cow) --- consistently demonstrate properties across the EM spectrum that put them at the extremes of known transient parameter space.
Hallmarks include: (1) particularly fast optical evolution, with a rise time of no more than a few days; (2) a high optical luminosity ($M_g \sim -20$ mag); (3) a hot ($>$\,15000 K) and featureless thermal spectrum that persists well after peak with little or no cooling; (4) luminous and variable X-ray emission; (5) luminous radio and millimetre emission.  Individual objects have shown even more exotic properties:  the prototype event AT\,2018cow left behind a hot, UV-luminous, slowly-fading ``remnant'' \citep{Sun+2022,Sun+2023,Chen+2023,Inkenhaag+2023,Inkenhaag+2025}; while the recent event AT\,2022tsd emitted a series of very fast ($\sim$ minutes-duration, with $<$30\,s variability) optical flares, some of which outshone the initial transient itself \citep{Ho+2023tsd}.   All 18cow-like objects to date have been found in star-forming galaxies, typically with subsolar metallicity \citep{Ho+2020koala,Coppejans+2020,Wiseman+2020,Perley+2021}, although they are generally not found in the youngest or brightest regions \citep{Sun+2023,Chrimes+2024b}.

The properties above impose strong constraints on the origin of these objects.   The fast optical evolution and absence of late-time SN features imply little ejected matter, while the high luminosity implies substantial release of energy.  Fast, stochastic X-ray (and in the case of AT\,2022tsd, optical) variability implicates an unstable central engine (an energetic compact object).  The radio properties imply a fast (nearly relativistic) shock propagating in dense material.   

Despite the abundance of clues, there is no consensus on the nature of these objects.  The host-galaxy properties, and inference of substantial pre-existing circumstellar material from radio/submillimetre observations, point towards an association with massive stars.  In a core-collapse scenario, the very low ejecta masses would require either an ultra-stripped progenitor or a fallback scenario involving a collapse to a black hole \citep{Perley+2019,Margutti+2019}.  On the other hand, the sustained high photospheric temperatures, featureless spectra, and optical light curves that fade as power-laws before levelling out on a plateau (in at least one case, i.e. the ``remnant'' phase of AT2018cow) are suggestive of tidal disruption events (TDEs), and 18cow-like objects have also been interpreted as disruptions of solar-mass stars around intermediate-mass black holes (IMBHs;  \citealt{Perley+2019,Kuin+2019,Mummery+2024}).  More recently, \cite{Metzger+2022} has proposed a scenario involving a binary merger of a massive star and a black hole (see also \citealt{Tsuna+2025}), which could simultaneously explain both the massive-star-like and TDE-like properties, but specific predictions of this model have not been established.

\begin{figure*}
    \includegraphics[width=\textwidth]{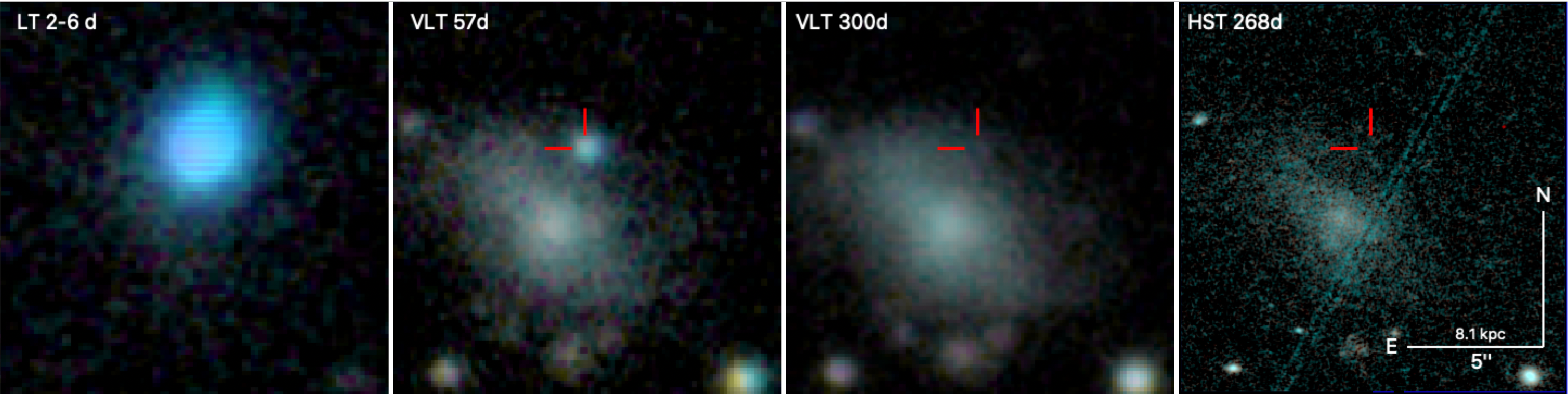}
    \vspace{-0.5cm}
\caption{Imaging sequence showing AT\,2024wpp and its host galaxy.  The leftmost panel is a stack of LT $g$/$r$/$i$ data from the first five nights of follow-up.  The left-centre panel shows VLT/FORS2 $g$/$r$/$i$ observations at 57\,observer frame days, when the source had faded substantially.  The right two panels show the host galaxy after the transient had faded below detectability: in VLT $g$/$r$/$i$ (right-centre) and HST F555W/F814W (right).  The diagonal feature in the HST image is an artefact from a moving object in the F555W image.}
    \label{fig:image}
\end{figure*}

Observational study of this population remains challenging, due to the combination of the intrinsic rarity ($<$0.1\% of the core-collapse supernova rate; \citealt{Ho+2023sample}) and rapid evolution (which requires high-cadence surveys and immediate follow-up) of these events.  However, both discovery and follow-up capabilities continue to improve, as does our awareness of the types of observations that may be necessary to obtain a breakthrough in understanding their origins.  

The recent LFBOT AT\,2024wpp is among the closest and brightest members of this population.   Since our initial announcement of the discovery \citep{TNSAN2024.272} and confirmation \citep{TNSAN2024.276,TNSAN2024.280} of this event in September 2024, it has been the subject of several studies in the literature presenting data across the electromagnetic spectrum \citep{Pursiainen+2025,Ofek+2025,Nayana+2025,LeBaron+2026}.   In this paper, we document its discovery in the ZTF data stream and present an extensive suite of early-through-late time multiwavelength follow-up observations.  
The observations presented here extend the optical temporal coverage both earlier and later in time than previous papers, and include other types of data that have not yet been reported for this event (and have rarely or never been reported for any previous LFBOT):  FUV spectroscopy, multi-epoch late-time optical spectroscopy, and high-frequency submillimetre observations.  These observations differentiate LFBOTs even further from known populations of transients yet point towards a picture in which most of these differences can largely be attributed to a single feature:  powerful X-ray radiation from the central object.  At the same time, they also raise new questions about the fundamental nature of the progenitor system.

This paper is organized as follows.  Our discovery of the object and multi-wavelength follow-up observations are described in \S \ref{sec:observations}.  The observational characteristics of the event are reviewed in \S \ref{sec:characteristics} and implications discussed in \S \ref{sec:discussion}.  Conclusions are summarized in \S \ref{sec:conclusions}.   Throughout the paper we assume a cosmology with $\Omega_M=0.3$, $\Omega_\Lambda=0.7$, $H_0$ = 70 km s$^{-1}$ Mpc$^{-1}$ and estimate Galactic reddening via the dust maps of \cite{Schlafly+2011} and extinction curve of \cite{Fitzpatrick1999}; for AT\,2024wpp we use a value of $E_{B-V} = 0.026$ mag.   Unless otherwise stated, uncertainties are 1$\sigma$ and magnitudes are expressed in the AB system.

\section{Observations}
\label{sec:observations}

\subsection{P48}
\label{sec:p48}

AT\,2024wpp (``Whippet'') was discovered by the Zwicky Transient Facility (ZTF; \citealt{Bellm+2019,Graham+2019}), a combined private/public time-domain optical sky survey that employs a 47 deg$^{2}$ field-of-view camera \citep{Dekany+2020} on the refurbished Samuel Oschin 48-inch Schmidt telescope (P48) at Palomar Observatory.  (For additional details about the ZTF observing and data processing system, see \citealt{Masci+2019,Patterson+2019,Mahabal+2019,Duev+2019}.)   The source is first detected in two exposures taken in immediate succession beginning at MJD 60578.43651 (2024-09-25 10:28:34 UTC), although the S/N in each was too low to generate a real-time alert.   Combining the two exposures in flux space gives $g=21.14\pm0.27$ (4$\sigma$ detection). 

By the following night the transient had brightened to $g$~=~17.52~$\pm$~0.02 mag and was clearly detected in observations in all three ZTF filters, and it was assigned the internal name ZTF24abjjpbo.   It was registered to the Transient Name Server and given the identifier AT\,2024wpp after this detection caused it to pass the Bright Transient Survey \citep{Fremling+2020,Perley+2020} software filter.\footnote{All of the detections were in observations taken as part of either the ZTF partnership survey or Caltech-allocated observing time, so the source is not registered in public ZTF brokers.}  Additionally, its blue colour, rapid rise, and proximity to a faint and potentially intermediate-redshift galaxy (based on a photometric redshift estimate of $z=0.11\pm0.05$ from the Legacy Survey; \citealt{Zhou+2023}) led to it being identified as an LFBOT candidate under the filtering and scanning system described in \cite{Ho+2023sample}, and it was prioritized for immediate follow-up, described in subsequent sections.

We used the IPAC forced-photometry pipeline \citep{Masci+2019}, fixed at the average position of the transient from the alert data ($\alpha$\,=\,02$^{\rm h}$42$^{\rm m}$05$^{\rm s}$.48, $\delta$\,=\,$-$16$^{\circ}$57${'}$23${''}.06$; J2000), to obtain final P48 photometry and pre-explosion upper limits.  These are reported in Table~\ref{tab:photometry}.

\begin{table}
	\centering
	\caption{UV/Optical/NIR Photometry of AT\,2024wpp}
	\label{tab:photometry}
	\begin{tabular}{lccc}
		\hline
Facility & MJD & filter & mag$^{a}$ \\
		\hline
P48   & 60578.4365 & {\it g   } &  21.14 $\pm$ 0.27 \\
P48   & 60579.3709 & {\it g   } &  17.57 $\pm$ 0.02 \\
P48   & 60579.4308 & {\it r   } &  17.97 $\pm$ 0.04 \\
P60   & 60579.4667 & {\it r   } &  18.10 $\pm$ 0.04 \\
P60   & 60579.4926 & {\it r   } &  18.02 $\pm$ 0.04 \\
LT    & 60580.0492 & {\it u   } &  16.37 $\pm$ 0.02 \\
LT    & 60580.0502 & {\it g   } &  17.04 $\pm$ 0.03 \\
LT    & 60580.0512 & {\it r   } &  17.44 $\pm$ 0.04 \\
LT    & 60580.0522 & {\it i   } &  17.80 $\pm$ 0.03 \\
LT    & 60580.0531 & {\it z   } &  18.10 $\pm$ 0.06 \\
...   & ...        & {\it ... } &  ...     \\
		\hline
	\end{tabular}
{\par \begin{flushleft}
Note --- Only the first few lines are shown here.  A complete table containing all measurements is available in the supplementary material. \\
$^{a}$\, AB magnitudes, not corrected for Galactic extinction.  The host-galaxy contribution to the flux has been subtracted.
\end{flushleft}}
\end{table}

\subsection{Ground-based follow-up Imaging}
\label{sec:moreimaging}

We acquired optical imaging observations at several ground-based facilities to track the light curve and colour evolution of AT\,2024wpp.   Observations were acquired using IO:O on the Liverpool Telescope (LT) in the $ugriz$ filters, the Liverpool Infra-Red Imaging Camera (LIRIC)  on the Liverpool Telescope in the FELH1500 (approximately $H$-band) filter, the Rainbow Camera on the Palomar 60-inch telescope in $ugri$ filters, the Panoramic Survey Telescope and Rapid Response System (Pan-STARRS or PS) in $grizy$ filters, the Wide-Field Infrared Survey Telescope (WINTER) in $J$ and $H_s$ filters, the Goodman High-Throughput Spectrograph (GHTS) on the Southern Astrophysical Research Telescope (SOAR) in $gr$ filters, the ESO Faint Object Spectrograph and Camera v2 (EFOSC2) on the New Technology Telescope (NTT) in $ugriz$ filters, and the Focal Reducer/Low Dispersion Spectrograph 2 (FORS2) on the Very Large Telescope (VLT) in $ugri$ filters.  These were reduced using standard techniques.

While the host galaxy of AT\,2024wpp is relatively faint and its nucleus is offset significantly (by 3\arcsec) from the transient position, flux from the host can contribute significantly to the flux from the transient in images at later times (Figure \ref{fig:image}).
Image subtraction of early-time $griz$ observations (from LT and P60) was performed using pre-explosion imaging from Pan-STARRS as a template.  No template was used for the early $u$-band observations, since the host galaxy is not a significant contributor to the flux at this time.  
For deeper observations at later times, we use either Legacy Survey \citep{Dey+2019} images taken in the same filter, or deep imaging at late times from our own campaign, as the template.    Additional details of these facilities and the reduction and image subtraction techniques can be found in Appendix \ref{sec:observationdetails}.

Due to the very blue nature of the transient ($g-i \approx -0.8$ mag at early times close to maximum, and $g-i \approx -0.3$ mag even at 40 days), differences in the transmission profiles of different filter sets, detectors, and atmospheric properties will manifest as systematic offsets between light curves from different facilities.  The effect is particularly large for facilities using non-standard filters, such as the Gunn $g$-band used by EFOSC2.  We performed a simple empirical correction for this effect by applying a constant offset to align the light curves and remove any offsets.  A table of offsets applied is given in Table \ref{tab:filteroffsets}.  It should be cautioned that this technique does not account for the time-evolving colour of the object, and because the choice of which instrument to treat as the baseline is arbitrary a systematic uncertainty in the flux calibration comparable to the characteristic offsets will remain in any resulting SED analysis.  However, the method is adequate for the SED and light curve analysis performed here.  Photometry is summarized in Table \ref{tab:photometry} and Figure \ref{fig:lightcurve}.

\begin{table}
	\centering
	\caption{Instrument-specific corrections applied}
    \label{tab:filteroffsets}
	\begin{tabular}{llr}
		\hline
Facility & filter  & offset \\
		\hline
P60  & $u$ & +0.04 \\
     & $g$ & $-$0.07 \\
     & $r$ & $-$0.03 \\
     & $i$ & $-$0.02 \\
PS   & $g$ & +0.06 \\
     & $r$ & +0.02   \\
     & $i$ & +0.06   \\
     & $z$ & +0.08   \\
NTT  & $g$ & $-$0.17 \\
     & $r$ & $-$0.07 \\
     & $i$ & +0.04 \\
      \hline
	\end{tabular}
{\par \begin{flushleft}
Note --- Values indicate the magnitude offsets added to the original photometry to obtain the values given in Table \ref{tab:photometry}. \\
\end{flushleft}}
\end{table}

\subsection{Swift}
\label{sec:swift}

Sixty-three observations of AT\,2024wpp were obtained by the Neil Gehrels \emph{Swift} Observatory \citep{Gehrels2004}, using both the X-Ray Telescope (XRT; \citealt{swift_xrt}) and the Ultraviolet-Optical Telescope (UVOT; \citealt{swift_uvot}). 

\subsubsection{XRT}
\label{sec:xrt}

To obtain XRT count rates we used the tools\footnote{\url{https://www.swift.ac.uk/API}} developed by the \emph{Swift} team \citep{Evans2007,Evans2009}. In the settings, centroiding was turned on, and the position error was set to the default of 1\,arcmin. Visual inspection revealed that during the third snapshot (orbit) of the first observation (OBS ID 00016843001), the source position was impacted by a cosmic ray hit in the CCD baseline region. Therefore, for the first observation we obtained count rates for each of the non-impacted orbits. For the remaining observations we obtained count rates for the total observation.

We fit a single spectrum for all XRT observations within the first ten days (OBS ID 00016843002--00016848008), except for the first epoch (due to the cosmic ray hit). The fit assumed a Milky Way $N_H = 2.7 \times 10^{20}\,$cm$^{-2}$ \citep{Willingale2013}. We found that the data are well described by a power law with index $\Gamma=1.7^{+0.3}_{-0.3}$ (90\% confidence). The reduced chi squared of the fit is $\chi_r^2=1.2$. From WebPIMMS\footnote{https://heasarc.gsfc.nasa.gov/cgi-bin/Tools/w3pimms/w3pimms.pl}, applying an intrinsic host $N_H = 6.3\times10^{20}\,$cm$^{-2}$ from our own analysis (\S \ref{sec:uvspecanalysis}), the conversion factor from count rate to unabsorbed flux is $4.865\times10^{-11}$ erg\,cm$^{-2}$. 

The X-ray counterpart of AT\,2024wpp is not detected in most observations after approximately 15 days (although it briefly becomes detectable again after 40 days), and for the generally short observations over this period we stack observations taken close in time (within 2--3 days) to obtain deeper limits.  The X-ray light curve is given in Table \ref{tab:xrtlightcurve} and plotted in Figure \ref{fig:xrtlightcurve}.

\begin{table}
	\centering
	\caption{\emph{Swift} XRT light curve of AT\,2024wpp}
	\label{tab:xrtlightcurve}
\begingroup
\renewcommand{\arraystretch}{1.1} 
	\begin{tabular}{lll}
		\hline
MJD start & MJD end   & \multicolumn{1}{c}{Flux$^{a}$} \\
          &           & ($10^{-14}$ erg cm$^{-2}$ s$^{-1}$)\\
		\hline
60580.406 & 60580.412 & \spc   129.1 $^{  +47.0 }_{  -38.0}$\\
60580.479 & 60580.488 & \spc    85.6 $^{  +30.9 }_{  -24.9}$\\
60580.606 & 60580.611 & $<$ 220.2                \\
60580.673 & 60580.690 & \spc    58.1 $^{  +18.2 }_{  -15.1}$\\
60580.738 & 60580.745 & \spc   109.6 $^{  +38.5 }_{  -31.5}$\\
60581.785 & 60582.515 & \spc   135.1 $^{  +18.7 }_{  -18.7}$\\
60583.679 & 60583.748 & \spc    48.3 $^{  +23.7 }_{  -18.0}$\\
60584.021 & 60584.879 & \spc    77.5 $^{   +9.6 }_{   -9.6}$\\
60585.587 & 60585.606 & \spc    71.2 $^{  +17.2 }_{  -17.2}$\\
60587.810 & 60587.829 & \spc    47.8 $^{  +15.8 }_{  -13.0}$\\
60588.406 & 60588.479 & $<$ 49.1                 \\
60589.651 & 60589.670 & \spc    20.9 $^{  +13.2 }_{   -9.4}$\\
60590.433 & 60590.450 & \spc    30.6 $^{  +14.0 }_{  -10.9}$\\
60590.454 & 60590.583 & \spc    17.3 $^{   +7.5 }_{   -5.9}$\\
60590.642 & 60590.846 & \spc     9.4 $^{   +5.1 }_{   -3.9}$\\
60591.816 & 60591.898 & \spc    17.5 $^{  +11.4 }_{   -8.1}$\\
60592.466 & 60592.485 & \spc    14.1 $^{   +9.8 }_{   -7.0}$\\
60593.323 & 60594.639 & $<$ 19.9                 \\
60595.284 & 60595.563 & \spc    16.9 $^{   +9.9 }_{   -7.3}$\\
60596.270 & 60597.665 & $<$  8.3                 \\
60599.606 & 60600.351 & $<$ 24.5                 \\
60601.183 & 60602.704 & $<$ 16.3                 \\
60603.603 & 60604.282 & $<$ 23.7                 \\
60605.377 & 60606.570 & $<$ 23.9                 \\
60607.140 & 60609.065 & $<$ 15.7                 \\
60610.680 & 60612.800 & $<$ 22.3                 \\
60617.301 & 60619.610 & $<$ 16.4                 \\
60621.689 & 60623.865 & $<$ 22.8                 \\
60625.672 & 60625.815 & \spc    14.0 $^{   +8.0 }_{   -5.9}$\\
60630.721 & 60630.793 & $<$ 36.4                 \\
60631.711 & 60632.221 & \spc    12.9 $^{   +6.1 }_{   -4.8}$\\
60632.613 & 60636.611 & $<$ 18.8                 \\
60637.384 & 60637.461 & \spc    10.1 $^{   +6.8 }_{   -4.9}$\\
60642.231 & 60647.741 & $<$ 26.9                 \\
60656.092 & 60656.504 & $<$ 31.2                 \\
60663.022 & 60663.618 & \spc    11.1 $^{   +6.8 }_{   -5.0}$\\
60677.589 & 60677.667 & $<$ 15.6                 \\
60687.257 & 60697.657 & $<$ 13.8                 \\
60697.051 & 60697.657 & $<$ 22.9                 \\
60919.095 & 60947.341 & $<$ 10.4                 \\
      \hline
	\end{tabular}
\endgroup    
{\par \begin{flushleft}
$^{a}$\,Flux in the XRT bandpass (0.3--10 keV). %, in units of $10^{-14}$ erg cm$^{-2}$ s$^{-1}$.  
Limits are 3$\sigma$.  Uncertainties correspond to a 68\% confidence interval on each measurement.\\
\end{flushleft}}
\end{table}

\subsubsection{UVOT}
\label{sec:uvot}

We used HEAsoft\footnote{HEASoft version 6.33.2 and \emph{Swift} version {\tt Swift\_Rel5.8(Bld47)\_30Jun2023}} set for \emph{Swift} UVOT photometry. We extracted source counts from a 5.0\arcsec\ radius aperture centered on AT\,2024wpp using the task {\tt uvotsource} on all individual exposures. During this process, exposures that suffered from small-scale sensitivity issues were flagged and discarded from further analysis\footnote{Those refer to images with our target detected in areas of low sensitivity on the detector. We adopted the default level of {\tt Low} to remove the smallest number of sources, which is designed to cover all the most strongly affected locations. See \url{https://www.swift.ac.uk/analysis/uvot/sss.php}. }. We also summed the exposures of each epoch using the task {\tt uvotimsum}, and then extracted source counts the same way as we did for the individual exposures. The individual exposures suffering from small-scale sensitivity issues were discarded before summing.   The source counts were converted into flux densities and magnitudes based on the most recent UVOT calibrations\footnote{Update to the 20240201 release.   \\ \url{https://swift.gsfc.nasa.gov/caldb/docs/uvot/uvot_release_history.html}} \citep{Breeveld2011}.

All late-time UVOT images contain significant flux from the host galaxy. 
We estimate the host contribution by taking synthetic photometry of the host SED model over each UVOT band, and apply an aperture correction factor to determine the predicted flux within our 5$\arcsec$ transient aperture (see \S \ref{sec:host} for additional details).
These in-aperture fluxes ($F_{UVW2}$ = 2.42, $F_{UVM2}$ = 2.71, $F_{UVW1}$ = 3.04, $F_U$ = 4.67, $F_B$ = 11.71, $F_V$ = 18.12; all units $\mu$Jy)
are then subtracted from the measurements to obtain the host-corrected fluxes and corresponding magnitudes given in Table \ref{tab:photometry}.

\subsection{Hubble Space Telescope Imaging}
\label{sec:hstobs}

We obtained two epochs of \emph{Hubble Space Telescope} (HST) photometry of AT\,2024wpp under HST Cycle 32 Director's Discretionary Time proposal 17889 (PI Ho), using the ultraviolet/visible (UVIS) channel of the Wide Field Camera 3 (WFC3).
The two epochs were on 2024 December 19 and 2025 June 20, each in four filters: F225W, F336W, F555W, and F814W.  Reduced observations were obtained from the \emph{HST} archive.  The source was well detected in all filters in the first epoch, and not detected in any filters during the second epoch (rightmost panel of Figure \ref{fig:image}).

Photometry was performed using a 0.1925$\arcsec$ radius aperture using standard techniques as given in the WFC3 handbook; no host correction was applied as no significant flux above the background was detected in the second epoch in any filter.  Additional details are provided in the Appendix (\S \ref{sec:hstphot}).

\subsection{Ground-Based Spectroscopy}
\label{sec:morespectra}

Observations were acquired using low-resolution spectrographs at many ground-based facilities: the Spectral Energy Distribution Machine (SEDM) on P60, GHTS on SOAR, the Spectrograph for the Rapid Acquisition of Transients (SPRAT) on LT, the Double Beam Spectrograph (DBSP) on P200, the Alhambra Faint Object Spectrograph and Camera (ALFOSC) on the Nordic Optical Telescope (NOT), the Keck Cosmic Web Imager (KCWI) on Keck II, EFOSC2 on the NTT, the Low Resolution Imaging Spectrometer (LRIS) on Keck, the Kast Double Spectrograph on the Lick Shane 3m, Binospec on the Multiple Mirror Telescope (MMT), the Inamori Magellan Areal Camera and Spectrograph (IMACS) on Magellan, and FORS2 on VLT.  The list of observations is provided in Table \ref{tab:spectroscopy_log}.  Descriptions of the observations and data reduction for each facility are given in the Appendix (\S \ref{sec:observationdetails}).  Following the initial reductions for each instrument, we apply an additional correction to each spectrum to match the flux scale from our interpolated photometry.   For early-time spectra we multiply by a further power-law correction term to match the $g-i$ colour inferred from the interpolated photometry.

Most spectra are featureless and show no lines from either the transient or the host galaxy, in emission or in absorption.  However, the KCWI IFU (Appendix \S \ref{sec:kcwi}) spectrum revealed narrow emission lines characteristic of a star-forming galaxy at a common redshift of $z=0.0868$.  
Together with the absorption lines seen in the HST ultraviolet spectrum (\S \ref{sec:uvspectra}) and some weak intermediate-width H and He features that emerged in the transient spectrum at later times (\S \ref{sec:specevolution}), this establishes the distance scale to this system and confirms its highly luminous nature.  The host galaxy spectrum is analysed in more detail in section \S \ref{sec:host}.

\begin{table*}
	\centering
	\caption{Table of optical spectroscopic observations}
	\label{tab:spectroscopy_log}
	\begin{tabular}{llrclll}
		\hline
		Observation date & MJD & $\Delta t_{\rm rest}$ & Facility & Exp. time & Grism/grating & Remark \\
         &  & (d) &  & (s) &  &  \\
		\hline
2024-09-26 11:12:53 & 60579.467 &  0.98 & P60/SEDM   & 2160    & --- & a \\
2024-09-27 03:59:30 & 60580.166 &  1.53 & SOAR/GHTS  & 3$\times$300  & 400M1 & \\
2024-09-27 04:22:02 & 60580.182 &  1.64 & LT/SPRAT   & 2$\times$600  & blue & \\
2024-09-27 09:03:53 & 60580.378 &  1.82 & P200/DBSP  & 900        & B600+R316 & \\
2024-09-28 09:01:12 & 60581.376 &  2.74 & SOAR/GHTS  & 3$\times$300      & 400M1 & \\
2024-09-29 04:07:34 & 60582.172 &  3.47 & NOT/ALFOSC & 1200     & Grism 4 & \\
2024-09-29 05:14:53 & 60582.219 &  3.51 & LT/SPRAT   & 1000      & blue & \\
2024-10-01 07:18:39 & 60584.305 &  5.43 & SOAR/GHTS  & 3$\times$300  & 400M1 & \\
2024-10-02 11:00:48 & 60585.459 &  6.49 & Keck/KCWI  & 2$\times$1320 / 8$\times$300  & BL+RL & b \\ 
2024-10-04 03:50:29 & 60587.160 &  8.06 & NTT/EFOSC2 & 1500    & Gr11+Gr16 & \\
2024-10-06 02:51:14 & 60589.119 &  9.86 & NOT/ALFOSC & 900     & Grism 4 & \\
2024-10-06 13:18:46 & 60589.555 & 10.26 & Keck/LRIS  & 240     & B400+R400 & \\
2024-10-07 08:12:15 & 60590.341 & 10.99 & Lick/Kast  & 1$\times$1560 / 3$\times$600 & 600/4310+300/7500 & \\
2024-10-08 08:35:48 & 60591.358 & 11.92 & Lick/Kast  & 2$\times$1560 / 5$\times$600        & 600/4310+300/7500 & \\
2024-10-09 13:12:45 & 60592.551 & 13.02 & Keck/LRIS  & 300     & B400+R400 & \\
2024-10-10 08:20:09 & 60593.347 & 13.76 & MMT/Binospec & 3$\times$600 & 270  & c \\ 
2024-10-13 08:13:03 & 60596.342 & 16.51 & P200/DBSP  & 1200    & B600+R316 & \\
2024-10-14 06:22:17 & 60597.265 & 17.36 & NTT/EFOSC2 & 1800    & Gr11+Gr16 & \\
2024-10-14 09:03:27 & 60597.377 & 17.46 & P200/DBSP  & 3000    & B600+R316 & \\
2024-10-16 13:51:24 & 60599.577 & 19.48 & HST/COS    &  8409       & G140L & \\
2024-10-17 19:47:27 & 60600.825 & 20.63 & HST/STIS   &  4474   & G230L & \\ 
2024-10-20 01:37:39 & 60603.068 & 22.70 & NOT/ALFOSC & 1500    & Grism 4 & \\
2024-10-23 09:44:38 & 60606.406 & 25.77 & P200/DBSP  & 3600    & B600+R316 & \\
2024-11-03 09:44:05 & 60617.406 & 35.89 & Keck/LRIS  & 900     & B400+R400 & \\
2024-11-06 04:19:12 & 60620.180 & 38.44 & Magellan/IMACS & 2$\times$1500    & 300  & \\
2024-11-08 10:10:45 & 60622.424 & 40.51 & Keck/LRIS  & 5$\times$900 & B600+R400 & d \\
2024-11-21 02:39:08 & 60635.111 & 52.18 & VLT/FORS2  & 2$\times$1320 & 300V & \\
2024-11-24 01:07:09 & 60638.047 & 54.88 & VLT/FORS2  & 2$\times$1380 & 300V & \\
2024-12-01 02:24:41 & 60645.100 & 61.37 & VLT/FORS2  & 2$\times$1380 & 300V & \\
2025-01-01 04:40:10 & 60676.195 & 89.98 & Keck/KCWI  & 8$\times$275 / 2$\times$1230 & BL+RL & e \\
		\hline
	\end{tabular}
{\par \begin{flushleft}
Notes: (a) Affected by camera problem. 
(b) Transient is saturated; used to obtain initial host redshift.  (c) Affected by unstable sensitivity function.  (d) This spectrum is independently reduced and reported in \cite{LeBaron+2026} as part of a data sharing agreement. (e) Transient only marginally detected; used for host galaxy analysis.
\end{flushleft}}
\end{table*}

\subsection{HST Ultraviolet Spectroscopy}
\label{sec:uvspectra}

We obtained observations of AT\,2024wpp with the \emph{Hubble Space Telescope} (\emph{HST}), using both the Cosmic Origins Spectrograph (COS; \citealt{Green+2012}) and the Space Telescope Imaging Spectrograph (STIS; \citealt{Woodgate+1998}) under program ID GO\#17477 (PI Perley).  The COS observations (4 orbits) employed the G140L grating and were carried out on 2024 October 16 between 13:51:24 and 19:04:33.  The STIS observations (2 orbits) employed the G230L grating and were carried out on 2024 October 17 between 19:47:27 and 21:52:57.  We use the pipeline reductions (1D extractions) from the \emph{HST} archive.  

As a comparison, we also present previously unpublished COS and STIS spectroscopy acquired using the same setup as above for the Type Ibn supernova SN\,2023iuc on 2023 May 27, observed under program ID GO\#16714 (PI Perley).  Pipeline reductions were acquired from the archive in the same way as for AT\,2024wpp.

\subsection{Very Large Array}
\label{sec:vla}

We initiated observations of AT\,2024wpp using the National Science Foundation's Karl G. Jansky Very Large Array (VLA) on 2024-10-08 at a mid-time of 09:39:35 UT 
and a mid-frequency of 10~GHz (4~GHz bandwidth, X-band).  Additional epochs at mid-frequencies of 3~GHz (S-band, 2~GHz bandwidth), 6~GHz (C-band, 4~GHz), 10~GHz, 15~GHz (Ku-band, 6~GHz bandwidth), 22~GHz (K-band, 8~GHz bandwidth), and 33~GHz (Ka-band, 8~GHz bandwidth) were acquired from 2024-10-25 to 2025-07-21 ($\sim 27$--$275~$ rest-frame days), until the source associated with AT\,2024wpp faded beyond detection. 

We reduced and analysed the data using the Common Astronomy Software Applications \citep[\texttt{CASA}]{McMullin+2007,CASA2022} VLA pipeline and produced calibrated continuum images using the VLA Imaging Pipeline\footnote{\url{https://science.nrao.edu/facilities/vla/data-processing/pipeline/vipl}}. To determine the flux density of the transient and RMS for each image, we use \texttt{pwkit/imtool} \citep[]{2017ascl.soft04001W}.  Images and flux-density measurements were produced both for the full receiver bandwidth and also in sub-bands of 0.5 GHz, 1 GHz, or 2 GHz bandwidth (depending on the frequency and S/N).
The full-bandwidth flux densities are presented in Table~\ref{tab:radiomm_Observations}; measurements divided by sub-band are provided in the supplementary material.

\begin{figure}
    \includegraphics[width=\columnwidth]{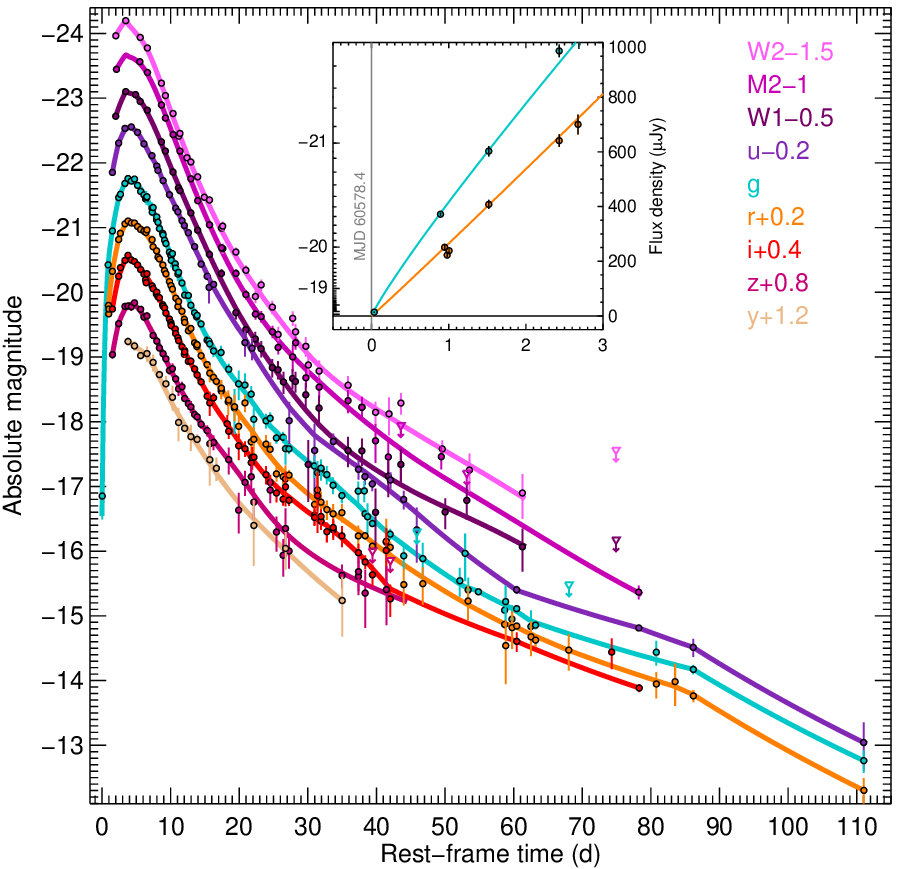}
    \vspace{-0.25cm}
\caption{Multi-band UVOIR light curve of AT\,2024wpp.  Magnitudes have been corrected for Galactic extinction, and a standard 2.5\,log$_{10}$(1+$z$) $k$-correction has been applied to convert to equivalent rest-frame absolute magnitudes.  Additional small offsets (indicated in the legend) have been applied to each band for clarity. The inset shows a zoom-in (with the $y$-axis linear in flux) around the time of explosion, including the P48 observation almost immediately after first light.}
    \label{fig:lightcurve}
\end{figure}

\begin{figure}
    \includegraphics[width=\columnwidth]{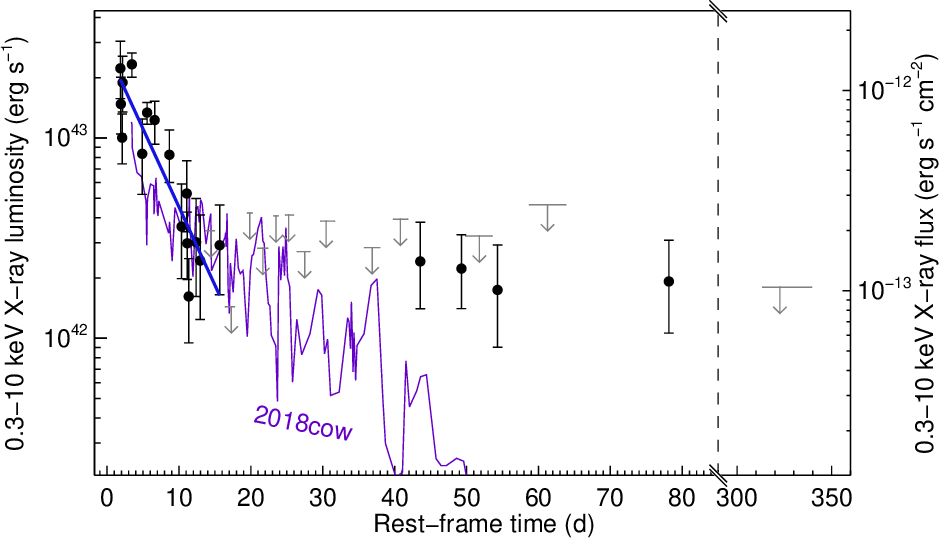}
    \vspace{-0.25cm}
\caption{\emph{Swift}-XRT light curve of AT\,2024wpp (black dots).  The slanted line shows an exponential model fit to the early ($t < 20$\,d) data.  The purple line is the light curve of AT\,2018cow (as seen at the distance of AT2024wpp) for comparison.  Arrows indicate 3-$\sigma$ limits.}
    \label{fig:xrtlightcurve}
\end{figure}

\begin{figure}
    \includegraphics[width=\columnwidth]{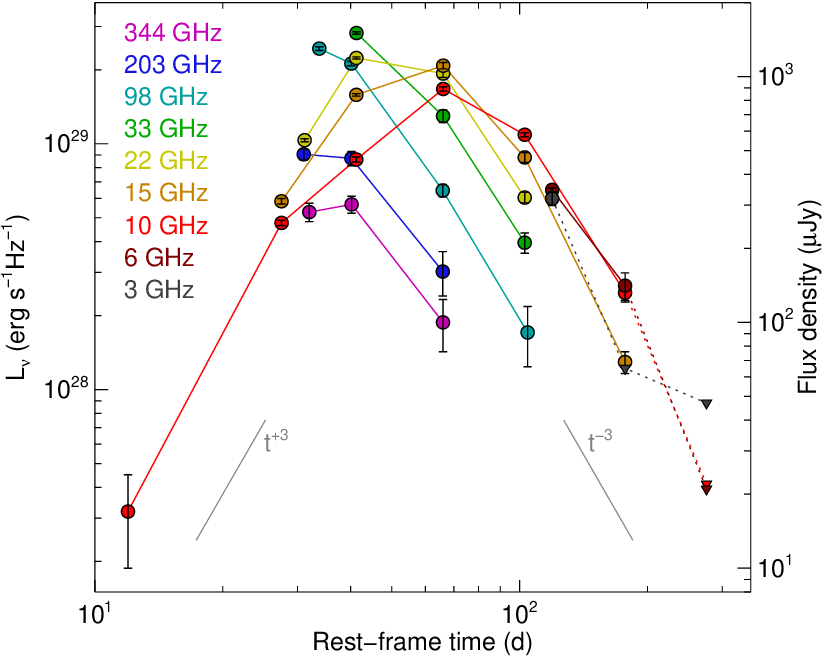}
    \vspace{-0.25cm}
\caption{Radio/millimetre/submillimetre light curve of AT\,2024wpp from VLA and ALMA observations.  Dotted lines connect measurements to upper limits.  The slanted grey lines at the bottom left and right indicate slopes of $t^{3}$ and $t^{-3}$, respectively, which describe the power-law evolution of the radio transient before peak and after peak, respectively.}
    \label{fig:radiolightcurve}
\end{figure}

\begin{table}
	\centering
\caption{VLA and ALMA radio observations of AT\,2024wpp.}
	\label{tab:radiomm_Observations}
	\begin{tabular}{ccccc}
		\hline
		 \multicolumn{2}{c}{Observation mid-time} & band & $\nu_{\rm obs}$ & $F_{\nu}$ \\
            (UTC) & (MJD) & & (GHz) & ($\mu Jy$) \\
		\hline
            \multicolumn{5}{c}{{\it VLA Observations}}\\
            \hline
2024-10-08 09:42:21 & 60591.4044 & X  & 10     & 17 $\pm$ 7 \\
2024-10-25 07:17:18 & 60608.3037 & X  & 10     & 254 $\pm$ 6 \\
2024-10-25 06:52:37 & 60608.2865 & Ku & 15     & 311 $\pm$ 8 \\
2024-10-29 06:45:33 & 60612.2816 & K  & 22     & 551 $\pm$ 8 \\
2024-11-09 05:18:29 & 60623.2212 & X  & 10     & 460 $\pm$ 11 \\
2024-11-09 05:36:16 & 60623.2335 & Ku & 15     & 843 $\pm$ 10 \\
2024-11-09 06:01:33 & 60623.2511 & K  & 22     & 1191 $\pm$ 12 \\
2024-11-09 06:34:16 & 60623.2738 & Ka & 33     & 1504 $\pm$ 18 \\
2024-12-06 05:21:56 & 60650.2236 & X  & 10     & 891 $\pm$ 18 \\
2024-12-06 05:03:42 & 60650.2109 & Ku & 15     & 1108 $\pm$ 33 \\
2024-12-06 04:38:31 & 60650.1934 & K  & 22     & 1029 $\pm$ 38 \\
2024-12-06 04:07:09 & 60650.1716 & Ka & 33     & 691 $\pm$ 40 \\
2025-01-15 03:21:28 & 60690.1399 & X  & 10     & 580 $\pm$ 11 \\
2025-01-15 03:03:14 & 60690.1272 & Ku & 15     & 469 $\pm$ 22 \\
2025-01-15 02:38:02 & 60690.1097 & K  & 22     & 322 $\pm$ 16 \\
2025-01-15 02:06:40 & 60690.0880 & Ka & 33     & 211 $\pm$ 20 \\
2025-02-02 00:43:16 & 60708.0300 & C  & 6      & 346 $\pm$ 6 \\
2025-02-02 00:23:05 & 60708.0160 & S  & 3      & 319 $\pm$ 20 \\
2025-04-05 21:21:46 & 60770.8901 & S  & 3      & $<$65 \\
2025-04-05 21:05:57 & 60770.8791 & C  & 6      & 141 $\pm$ 18 \\
2025-04-05 20:50:36 & 60770.8685 & X  & 10     & 132 $\pm$ 11 \\
2025-04-05 20:34:43 & 60770.8574 & Ku & 15     & 69 $\pm$ 7 \\
2025-07-21 15:04:45 & 60877.6283 & S  & 3      & $<$47 \\
2025-07-21 14:48:47 & 60877.6172 & C  & 6      & $<$21 \\
2025-07-21 14:33:11 & 60877.6064 & X  & 10     & $<$22 \\
            		\hline
            \multicolumn{5}{c}{{\it ALMA Observations}}\\
            \hline
2024-11-01 02:08:29 & 60615.0892 &  3 &  97.5  & 1300 $\pm$ 22 \\
2024-10-29 02:53:14 & 60612.1203 &  5 & 202.95 &  483 $\pm$ 25 \\
2024-10-30 03:37:00 & 60613.1507 &  7 & 343.5  &  281 $\pm$ 24 \\
2024-11-08 01:07:44 & 60622.0470 &  3 &  97.5  & 1130 $\pm$ 22 \\
2024-11-08 01:33:23 & 60622.0648 &  5 & 202.95 &  465 $\pm$ 31 \\
2024-11-08 02:33:19 & 60622.1065 &  7 & 343.5  &  302 $\pm$ 24 \\
2024-12-06 00:47:31 & 60650.0330 &  3 &  97.5  &  344 $\pm$ 20 \\
2024-12-06 01:12:28 & 60650.0503 &  5 & 202.95 &  161 $\pm$ 33 \\
2024-12-06 03:19:08 & 60650.1383 &  7 & 343.5  &  100 $\pm$ 24 \\
2025-01-16 21:09:14 & 60691.8814 &  3 &  97.5  &   91 $\pm$ 25 \\
            \hline
            \end{tabular}
{\par \begin{flushleft}
Note --- Uncertainties on $F_\nu$ are $1\sigma$, upper limits are 3 times the rms of the image.  VLA observations further divided by sub-band are available in the supplementary material.
\end{flushleft}}
\end{table}

\subsection{Atacama Large Millimetre Array}
\label{sec:alma}

We observed AT\,2024wpp using the Atacama Large Millimetre/submillimetre Array (ALMA) with three frequencies (Band 3, Band 5, and Band 7) under project code 2023.1.01730.T (joint with the VLA, project ID VLA/24B-338, PI Ho). We obtained three epochs of multi-frequency (Bands 3, 5, and 7) observations, and one final epoch of Band 3 only. The data were automatically calibrated with the ALMA pipeline\footnote{\url{https://almascience.nrao.edu/processing/science-pipeline}} through the ALMA Quality Assurance (QA) initiative. We downloaded the pipeline calibrated images from the ALMA archive and report the peak flux density and rms reported in the QA2 results in Table \ref{tab:radiomm_Observations}\footnote{To check the QA2 results, we also used \texttt{imtool} to measure the flux density and rms and found the values to be consistent with the QA2 values.}. 
The source was clearly detected in all epochs, and faded over the course of the observations.  The ALMA and VLA measurements are plotted together in Figure \ref{fig:radiolightcurve}.

\section{Observational Characteristics}
\label{sec:characteristics}

The basic characteristics of AT\,2024wpp establish it as an example of the empirical class of LFBOTs:  most notably, a high luminosity, fast rise and fade, persistent blue colour, and luminous X-ray and millimetre/radio emission.  We perform further analysis of the observational data and discuss the intrinsic properties of the event in more detail below.

\subsection{UV/Optical Light Curve}
\label{sec:lc}

The multi-band light curve of AT\,2024wpp for the first 120 days is shown in Figure \ref{fig:lightcurve}.  The time of first light is taken to be MJD 60578.40 based on a power-law fit to the early $g$-band data, shown in the figure inset.\footnote{This time is tightly constrained by our initial P48 detection: despite its low S/N, the flux is only 1\% of the peak value, implying that the observation must have occurred very close to the true explosion time (within about 1 hour) \emph{if} the early rise takes the form of a standard power-law.  It could potentially be argued that such a fortunate occurrence is not probable and the early rise of the light curve therefore probably exhibited more complicated behaviour, with an initial (easier-to-catch) slow rise followed by a faster one towards the peak.  In any case, our conclusions are not strongly sensitive to the precise choice of explosion time.}  This time will be used as $t_0$ (the reference point for all elapsed times quoted) throughout this paper.

The evolution is well-sampled close to peak in the optical bands, and shows a dramatic rise over the first day following detection of more than 3 magnitudes, then more gradually rises to peak two days later.  The time of peak is nearly the same in all bands (3--4 days after first detection).

After peak a rapid decay phase sets in.   From 10 days onward, 
the evolution in all bands can be reasonably well-approximated by a steep power-law (with $F \sim t^{-3}$ in the UV and $F \sim t^{-2.7}$ in the optical).  The colour reddens only moderately during this entire period, remaining quite blue ($g-r \sim 0$) as late as 111 rest-frame days, the time of our last detection.

\begin{figure}
    \includegraphics[width=\columnwidth]{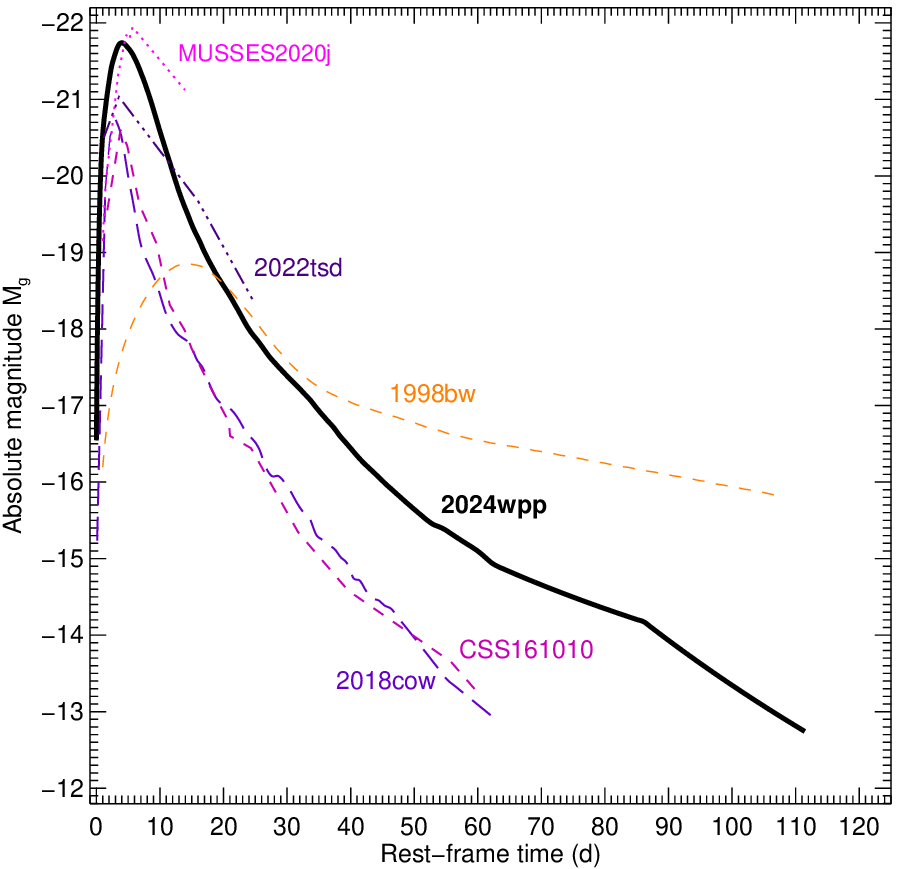}
    \vspace{-0.25cm}
\caption{Comparison of the $g$-band light curve of AT\,2024wpp to two other nearby LFBOTs with late-time observations (CSS161010 and AT\,2018cow) and two other high-luminosity LFBOTs (AT\,2024tsd and MUSSES2020j; the MUSSES2020j curve has been approximately $k$-corrected from a rest-frame wavelength of 3020\,\AA).  The Ic-BL SN\,1998bw is also shown for comparison.}
    \label{fig:comparelc}
\end{figure}

The $g$-band light curve of AT\,2024wpp is contrasted with several other LFBOTs (as well as the gamma-ray burst associated SN\,1998bw) in Figure \ref{fig:comparelc}, roughly matched to the same rest-frame central wavelength ($\lambda_c \approx 4500\,$\AA).   Data for the comparison objects are obtained from \citealt{Perley+2019} (AT\,2018cow), \citealt{Gutierrez+2024} (CSS161010), \citealt{Ho+2023tsd} (AT\,2022tsd), and \citealt{Jiang+2022} (MUSSES2020j; we apply an approximate extra $k$-correction of 1 mag to convert their observed $r$-band light curve to rest-frame $g$-band).  
AT\,2024wpp follows a similar light-curve evolution as other known nearby LFBOTs but it rises for longer than other events (reaching maximum about two days later), is more luminous at peak (by about 1 mag), and remains overluminous by about 2 mag throughout the decline.   Its rise time and peak luminosity are similar to those of MUSSES2020j, although the decay of AT\,2024wpp is somewhat faster.

\begin{figure}
    \includegraphics[width=\columnwidth]{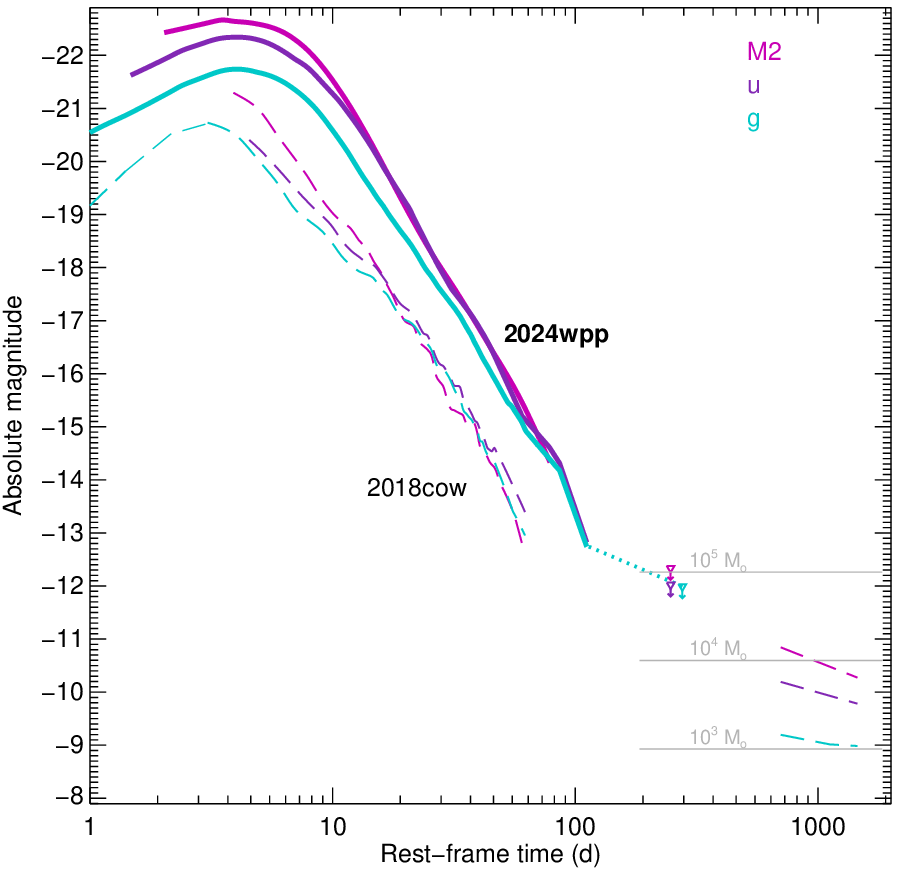}
    \vspace{-0.25cm}
\caption{Light curves of AT\,2024wpp in $g$(/F555W), $u$(/F336W), and \textit{UVM2}(/F225W) band compared to AT\,2018cow out to late times.  AT\,2024wpp is shown as solid lines, with the HST and VLT upper limits as downward arrows, with a dotted line connecting the last $g$-band detection to the first F555W-band limit.  Measurements for AT\,2018cow are shown as dashed lines.  The horizontal grey lines indicate predicted late-time optical plateau luminosities in the TDE model of Mummery et al.\,(2024).}
    \label{fig:latelc}
\end{figure}

AT\,2018cow showed a UV/optical plateau at several years post-explosion.  The late-time HST limits on AT\,2024wpp at 240 days (Figure \ref{fig:latelc}) place constraining limits on any similar feature in this event: the limits are about 1 mag higher than of the luminosity of the first AT\,2018cow plateau-phase detection but are significantly earlier.  (Potential implications will be discussed in \S \ref{sec:plateau}.)

\subsection{Constraints on Host Extinction}
\label{sec:hostextinction}

The blue nature of the transient and its location in the outlying regions of a small, face-on galaxy suggest that the extinction towards the line of sight to the transient is low.  We attempt to quantify this further by placing a limit on the equivalent width (EW) of the Na\,D $\lambda$5890 line using the Keck/LRIS spectrum at 6.5 days.  Assuming the line is unresolved, the 3$\sigma$ limit on the equivalent width is $<$0.18 \AA.  This would correspond to a limit of $E_{B-V} < 0.034$ mag given the relation of \cite{Poznanski+2012}, although this relation is calibrated on higher-extinction sightlines and the more recent relation of \cite{Maxted2025} would imply a less constraining limit of $E_{B-V} < 0.12$ mag.  We will generally assume host extinction to be negligible throughout this work (but c.f. \S \ref{sec:host}).

\subsection{Blackbody Evolution}
\label{sec:bb}

\begin{figure*}
    \includegraphics[width=\textwidth]{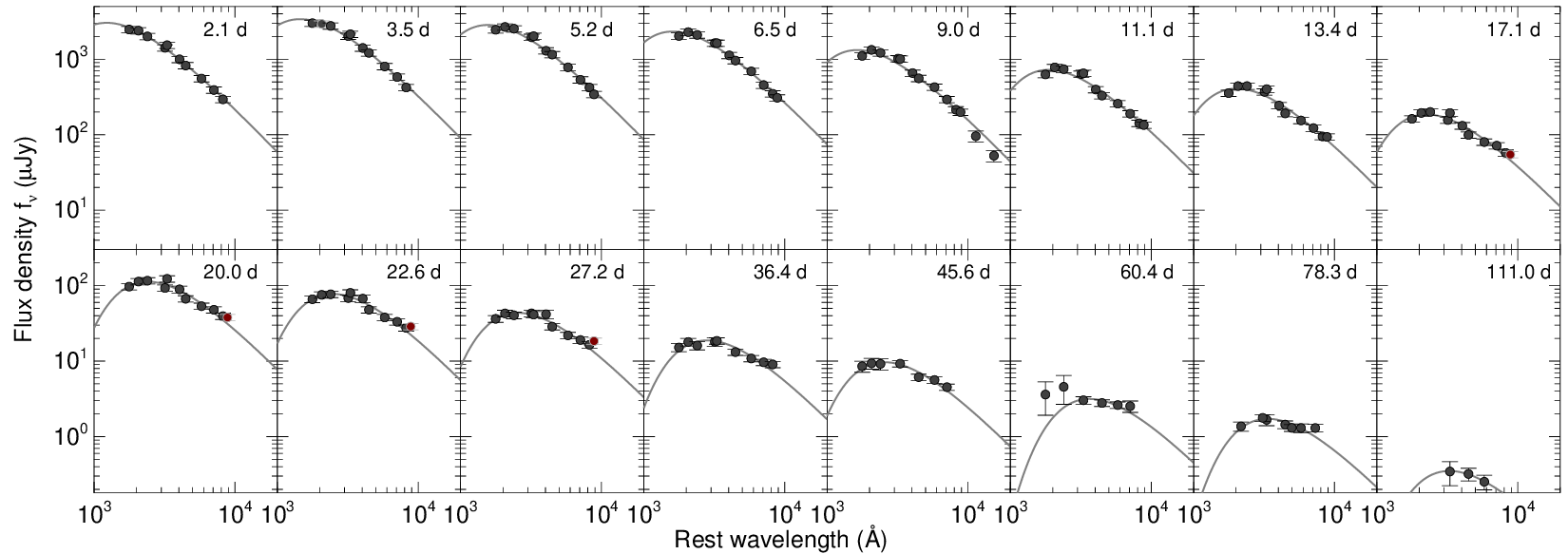}
\vspace{-0.5cm}    
\caption{Blackbody fits to the UVOIR SEDs of AT\,2024wpp at select epochs. Times are given in rest-frame days.  Red points indicate PS1 $y$-band measurements likely affected by near-IR excess.}
    \label{fig:bbfits}
\end{figure*}

\begin{figure}
    \includegraphics[width=\columnwidth]{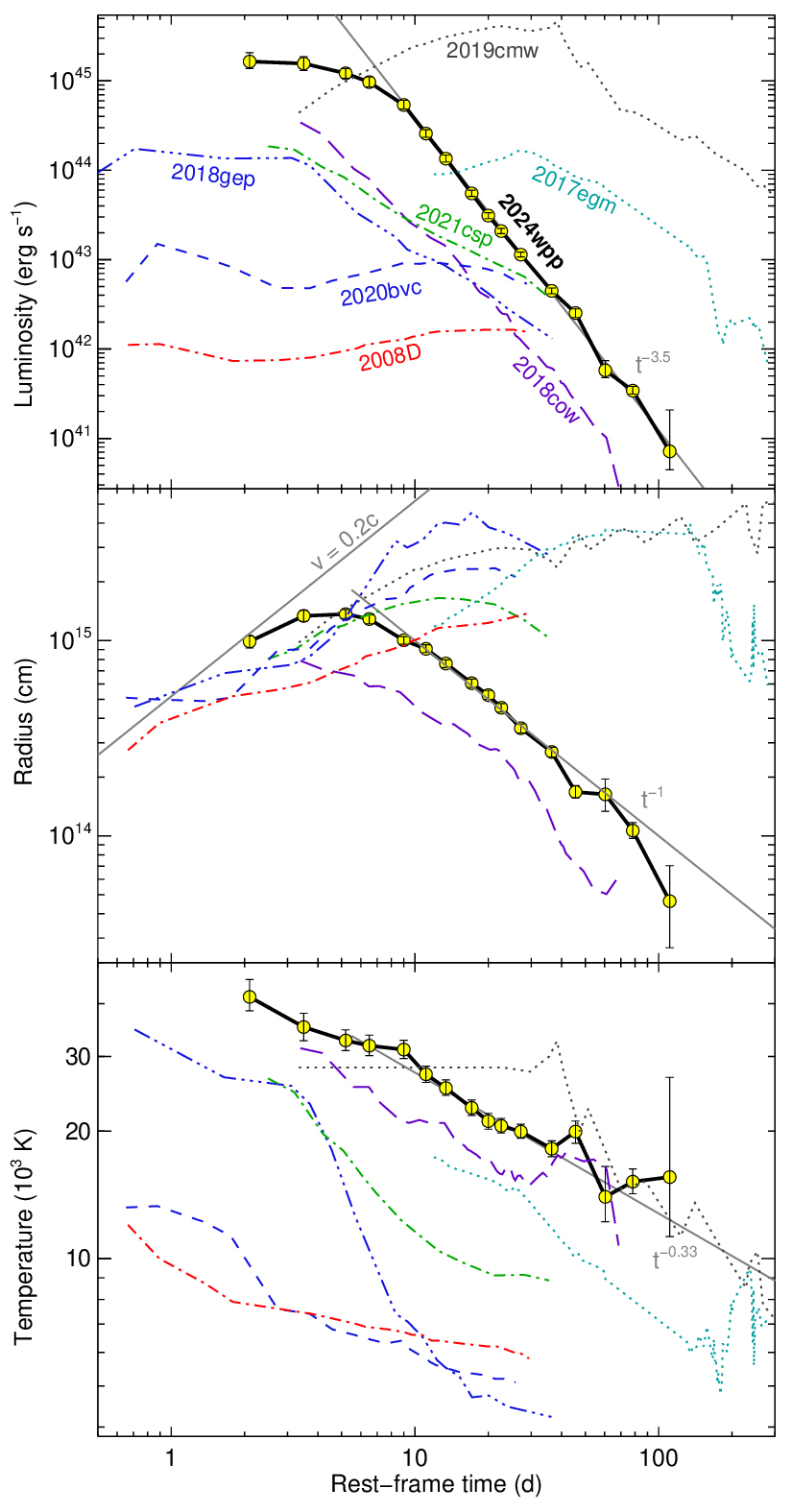}
\vspace{-0.5cm}    
\caption{Evolution of the blackbody parameters of AT\,2024wpp and other fast or luminous optical transients:  SN\,2008D (Ib), SN\,2020bvc (Ic-BL), SN\,2018gep (Ic-BL), SN\,2021csp (Icn), AT\,2018cow (LFBOT), SN\,2017egm (SLSN-I), and AT\,2019cmw (TDE candidate). }
    \label{fig:physevol}
\end{figure}

\begin{table}
\centering
\caption{Parameters from blackbody fits to UVOIR observations}
\label{tab:bbfits}
\begingroup
\renewcommand{\arraystretch}{1.3} 
\begin{tabular}{c c | c c c }
\hline
MJD & $\Delta t_{\rm rest}$ & log$_{10}$($L$) & log$_{10}$($R$)  & log$_{10}$($T$) \\
 & (days) & (erg s$^{-1}$)& (cm)  & (K)  \\
\hline
60580.68 &   2.10 & $45.21^{+0.10}_{-0.08}$ & $15.00^{+0.03}_{-0.03}$ & $ 4.62^{+0.04}_{-0.03}$ \\ 
60582.20 &   3.51 & $45.19^{+0.08}_{-0.08}$ & $15.13^{+0.03}_{-0.03}$ & $ 4.55^{+0.03}_{-0.03}$ \\ 
60584.05 &   5.21 & $45.08^{+0.05}_{-0.06}$ & $15.14^{+0.02}_{-0.02}$ & $ 4.52^{+0.03}_{-0.02}$ \\ 
60585.46 &   6.51 & $44.98^{+0.06}_{-0.05}$ & $15.11^{+0.02}_{-0.02}$ & $ 4.50^{+0.02}_{-0.02}$ \\ 
60588.18 &   9.02 & $44.73^{+0.05}_{-0.05}$ & $15.00^{+0.02}_{-0.02}$ & $ 4.49^{+0.02}_{-0.02}$ \\ 
60590.46 &  11.13 & $44.41^{+0.04}_{-0.04}$ & $14.96^{+0.02}_{-0.02}$ & $ 4.44^{+0.02}_{-0.02}$ \\ 
60592.96 &  13.43 & $44.12^{+0.04}_{-0.04}$ & $14.88^{+0.02}_{-0.02}$ & $ 4.40^{+0.02}_{-0.02}$ \\ 
60596.98 &  17.14 & $43.74^{+0.04}_{-0.03}$ & $14.78^{+0.02}_{-0.02}$ & $ 4.36^{+0.02}_{-0.02}$ \\ 
60600.14 &  20.06 & $43.49^{+0.03}_{-0.03}$ & $14.72^{+0.02}_{-0.02}$ & $ 4.32^{+0.02}_{-0.02}$ \\ 
60602.96 &  22.66 & $43.31^{+0.03}_{-0.03}$ & $14.66^{+0.02}_{-0.02}$ & $ 4.31^{+0.02}_{-0.02}$ \\ 
60607.96 &  27.27 & $43.05^{+0.03}_{-0.03}$ & $14.55^{+0.02}_{-0.02}$ & $ 4.30^{+0.02}_{-0.02}$ \\ 
60617.96 &  36.49 & $42.64^{+0.03}_{-0.03}$ & $14.43^{+0.02}_{-0.03}$ & $ 4.26^{+0.02}_{-0.02}$ \\ 
60627.96 &  45.72 & $42.40^{+0.05}_{-0.05}$ & $14.22^{+0.03}_{-0.03}$ & $ 4.30^{+0.03}_{-0.03}$ \\ 
60644.04 &  60.55 & $41.76^{+0.11}_{-0.08}$ & $14.21^{+0.08}_{-0.09}$ & $ 4.15^{+0.07}_{-0.06}$ \\ 
60663.50 &  78.51 & $41.53^{+0.05}_{-0.04}$ & $14.03^{+0.04}_{-0.04}$ & $ 4.18^{+0.03}_{-0.03}$ \\ 
60699.03 & 111.28 & $40.85^{+0.46}_{-0.21}$ & $13.67^{+0.18}_{-0.24}$ & $ 4.19^{+0.24}_{-0.14}$ \\ 
\hline
\end{tabular}
\endgroup
%{\par \begin{flushleft}
%Note --- Times above are in the rest frame.
%\end{flushleft}}
\end{table}

We interpolate the photometry shown in Figure \ref{fig:lightcurve} using a combination of direct polynomial fits and local regression, and sample the resulting model light curves at a list of selected times that roughly match the mean times of the \emph{Swift}/UVOT epochs (at early times) or the HST and VLT multi-band observations (at late times) to form a series of co-eval SEDs between approximately 2 and 111 rest-frame days after the assumed explosion time.  
The uncertainties on the fluxes used in the fit are a quadrature combination of (up to) three terms.  For late-time photometry ($\Delta t >50$\,d), we include a photometric error term equal to the photometric uncertainty on the most precise nearby measurement.\footnote{Photometric errors are negligible relative to other sources of error at earlier times when the afterglow is brighter.}   For the Swift/UVOT filters, we include a host subtraction uncertainty equal to the in-aperture host flux uncertainty.  For all bands, we further include a 10\% systematic error term to approximate the effect of calibration uncertainties, variation in the effective central wavelength, deviations from a perfect blackbody, and uncertainty in the interpolation model.\footnote{The value of 10\% was was chosen to achieve $\hat{\chi}^2\approx1$ on average across all fits.}
These are then fit to a Planck function at each epoch.  (We correct for Galactic extinction but do not correct for any host extinction, for the reasons described in the previous subsection.)  Pan-STARRS~1 $y$-band measurements after 16 rest-frame days were excluded from the fit due to the possibility of NIR excess affecting these measurements.
The resulting SED fits are shown in Figure \ref{fig:bbfits}.  In general, the photometry shows good consistency with the simple blackbody model except for a slight excess in $y$-band emerging at later times: likely the same type of NIR excess seen in AT\,2018cow \citep{Perley+2019}.

The evolution of the blackbody parameters (luminosity, temperature, and effective radius) is provided in Table \ref{tab:bbfits} and displayed in Figure \ref{fig:physevol}, with some additional objects shown for comparison: the prototypical LFBOT AT\,2018cow (from \citealt{Perley+2019}), the fast-rising Ic-BL SN\,2018gep \citep{Ho+2019}, the Ic-BL SN\,2020bvc \citep{Ho+2020bvc}\footnote{The derived luminosities for AT\,2020bvc shown in the plot have been corrected for a minor error in their originally published values.}, the X-ray discovered Ib SN\,2008D (\citealt{Soderberg+2008}; data from \citealt{Modjaz+2009}), the nearby Type I superluminous supernova (SLSN-I) SN\,2017egm \citep{Zhu+2023}, and the ambiguous nuclear transient (and possible high-mass TDE) AT\,2019cmw \citep{Wise+2026}. (We chose one comparison object of each type based on the availability and quality of high-quality multi-band data, and in the case of AT\,2019cmw due to its particularly extreme properties.)  We set $t_0$ for SN\,2017egm and AT\,2019cmw based on a linear extrapolation of the early radius evolution to $R=0$.
The behaviour of AT\,2024wpp resembles that of AT\,2018cow but is even more extreme: in particular, its luminosity exceeds that of AT\,2018cow by about an order of magnitude at almost all epochs.   Indeed, the peak observed bolometric luminosity at the time of the first \emph{Swift} UVOT epoch exceeds that of \emph{any} known supernova or other transient at the equivalent phase of within a few days of explosion (excepting gamma-ray burst afterglows, which are nonthermal), and it remains among the highest out to at least 20\,d.  As the temperature evolution is quite similar to AT\,2018cow itself, the higher luminosity is primarily a consequence of the larger photospheric radius at most epochs.

Thanks to our immediate identification of this source as a candidate of interest and our resulting early observations with \emph{Swift}, P60, and LT,
AT\,2024wpp is unique among LFBOTs for having sufficient multiband photometry to observe the photospheric radius expand early in its evolution.  
While these measurements are somewhat uncertain due to the very high temperatures ($>30$kK, such that the true blackbody peak is in the extreme UV and thus poorly constrained), they are consistent with the notion that the photosphere is expanding at very high speeds of 0.1--0.2$c$ for a few days before rapidly transitioning into a phase of photospheric contraction.  Specifically, between the inferred explosion time (shortly before the first P48 detection) and 2.1\,d, we infer an expansion rate of $0.17c$.  From 2.1\,d to 3.5\,d (the first two days with blackbody fits, rest frame) we infer a photospheric expansion rate of $v=0.1c$.  

From integrating the bolometric light curve, we infer a total radiated energy of $1.0\times10^{51}\,$erg.

\subsection{High-Energy Evolution}

The X-ray light curve of AT\,2024wpp is shown in Figure \ref{fig:xrtlightcurve}.  The light curve of AT\,2018cow (the only other LFBOT with a well-sampled light curve spanning early to late times) is shown for comparison:  the luminosities of the two events and the decay rate of the early light curves are quite similar.  AT\,2018cow showed significant variability on $\sim$2 day timescales starting approximately 15\,d after peak; unfortunately, the S/N of the observations does not permit a detailed search for short-timescale variability in AT\,2024wpp.  However, the source does rebrighten in X-rays after 40 days to become marginally detectable again, which is distinctive behaviour not previously seen in AT\,2018cow.

\subsection{Spectroscopic Evolution}
\label{sec:specevolution}

\begin{figure}
    \includegraphics[width=\columnwidth]{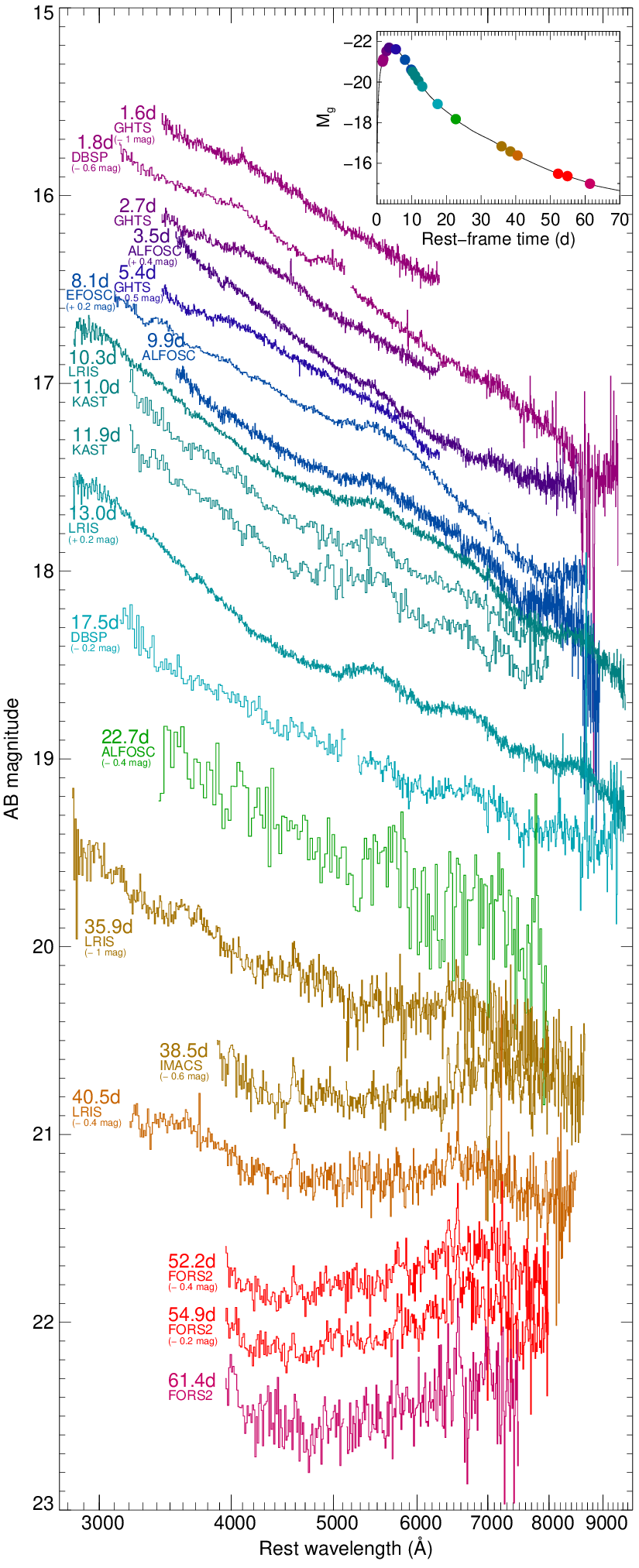}
    \vspace{-0.5cm}
\caption{Sequence of selected optical spectra showing the spectral evolution of AT\,2024wpp.  The $y$-axis is magnitude (i.e., logarithmic $F_\nu$); most curves are shown at their actual values (corrected for Galactic extinction) but small offsets (as indicated) have been applied to maintain a non-overlapping sequence.   The inset shows the $g$-band light curve with the times (and contemporaneous $g$-band luminosities) of the indicated spectral epochs shown as coloured circles.
}
    \label{fig:allspectra}
\end{figure}

A series showing the optical spectroscopic evolution of AT\,2024wpp is given in Figure \ref{fig:allspectra}.  We plot as AB magnitude, i.e. $-$2.5$\times$log$_{10}$($F_\nu$\,/\,3630\,Jy), instead of the traditional $F_\lambda$, to better illustrate variations on top of the steep blue continuum.  All spectra are featureless and blue, with any deviation from a blackbody at the level of no more than about $\approx$0.1 mag ($\approx$10\%) at any epoch.  Flux-calibrating longslit spectra to this level of accuracy across a wide wavelength range is challenging, and some of the residuals (as well as small differences in slopes) that remain are likely instrumental in origin.   However, there are several features seen consistently in spectra from multiple instruments (and reduced independently by different individuals using different standard-stars and independent techniques) that we are confident are real.

During the first 6 days there are no reliable spectral features in any of our spectra.  However, starting with the EFOSC2 spectrum at 8.4\,d we recognize a prominent blue dip at approximately 5000\,\AA\ in the rest frame, resembling a similar broad feature seen in AT\,2018cow and AT\,2020mrf; its profile remains roughly stable until at least day 13.4.   Beginning at 11 days, a second broad feature appears alongside it, visible as a narrower dip centred at about 6100~\AA.

\begin{figure}
    \includegraphics[width=\columnwidth]{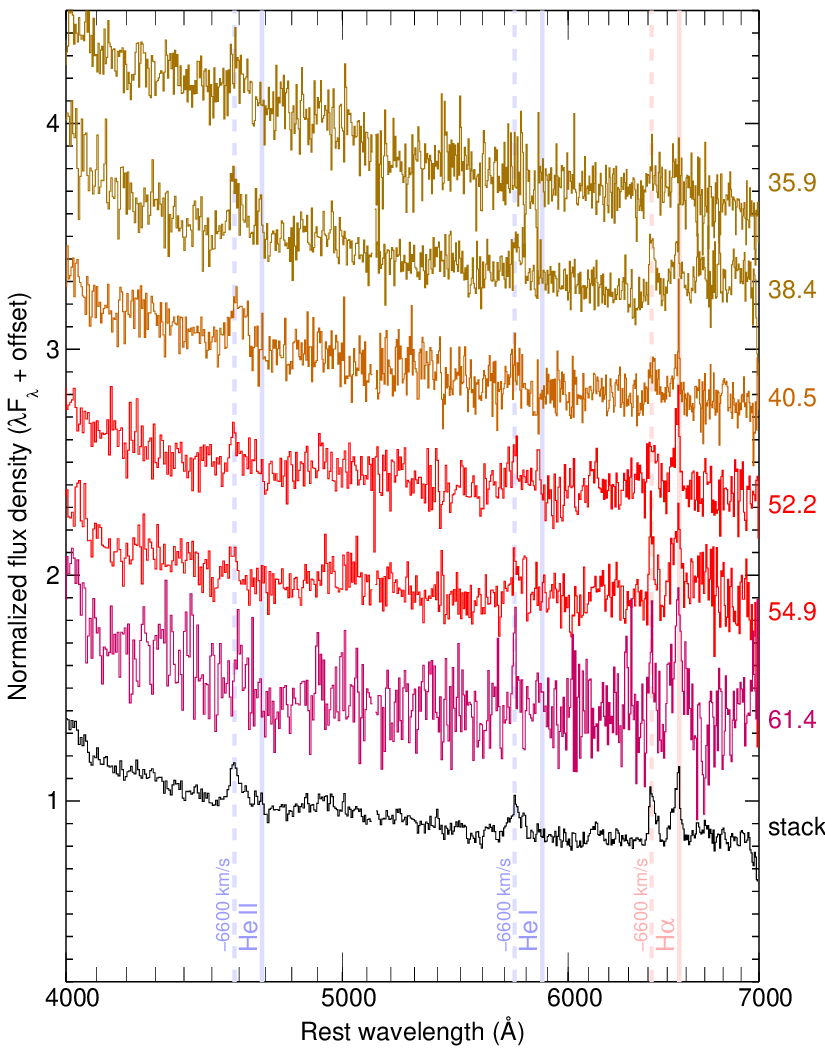}
    \vspace{-0.25cm}
\caption{Late-time spectral sequence showing the emergence of weak emission lines of H and He.   The numbers at right are rest-frame times in days after $t_0$; the bottom spectrum (in black) is a normalized stack of the middle four spectra.  The H$\alpha$ line has two distinct components, one centred at close to the systemic redshift and one blueshifted by $-6600$ km~s$^{-1}$; the latter are marked with dashed lines.  The He lines show only the blueshifted component clearly.}
    \label{fig:latespectra}
\end{figure}

\begin{figure}
    \includegraphics[width=\columnwidth]{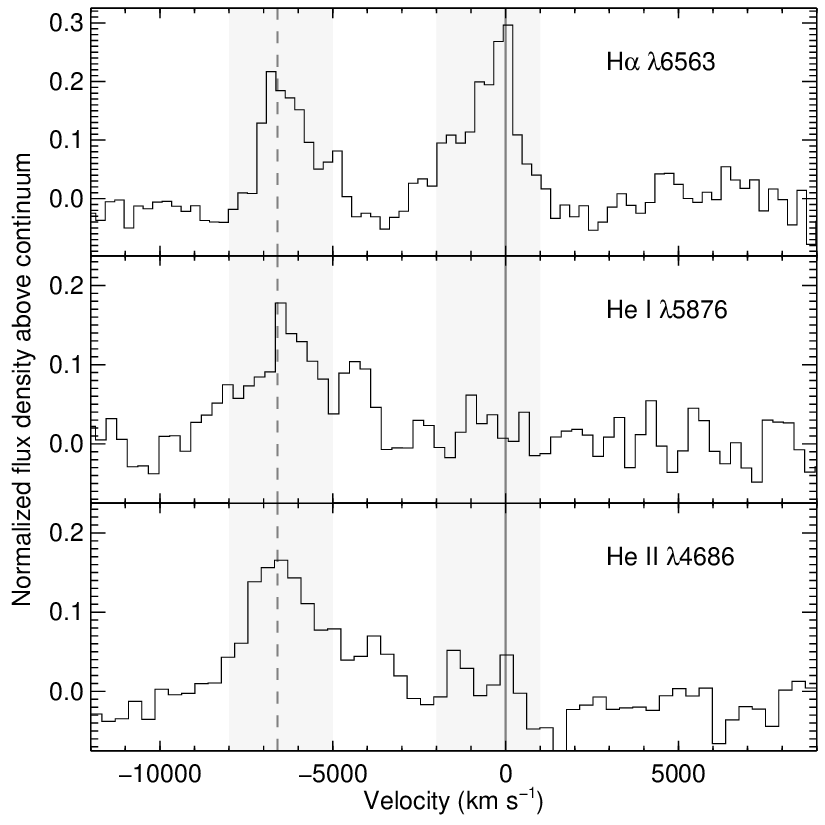}
    \vspace{-0.25cm}
\caption{Velocity profiles for the late-time emission features, from a stacked and normalized combination of four deep spectra obtained between 38--55\, rest-frame days.  The shaded regions indicate the inferred maximum velocity extents of each component.}
    \label{fig:linevelocities}
\end{figure}

Due to the fading of the source and the full moon, there is then a gap in observations of about 20 days (the NOT/ALFOSC spectrum on day 23 was taken under poor conditions and is not constraining).  When observations resume at 36 days, the broad features above are no longer recognizable, although some even broader deviations in the continuum are apparent.  

More notably, in the 36\,d spectrum and in all subsequent observations, intermediate-width emission features are visible at or near the rest-frame wavelengths of H$\alpha$, He\,I\,$\lambda$5876, and He\,II\,$\lambda$4686 (Figure \ref{fig:latespectra}; stacked and zoomed in on each line feature in Figure \ref{fig:linevelocities}).  This makes AT\,2024wpp only the third LFBOT to show identifiable spectral features.  The velocity profiles of these lines are extremely unusual:  the H$\alpha$ line in particular shows a clear double-peaked profile\footnote{The line at the position of H$\alpha$ in the host frame is close to, but inconsistent with, the wavelength of He\,I\,$\lambda$6678 blueshifted by the amount seen in the other lines.}, with one component peaking at or near the systemic redshift and another peaking at a relative \emph{blueshift} of $\Delta v = -$6600 km\,s$^{-1}$.  The two lines have consistent velocity widths (full width at zero intensity) of $\delta v \approx 2000$ km\,s$^{-1}$, much narrower than the separation between them.  The flux returns essentially to the continuum level in between the two components, suggesting they are physically separate systems.  There is no obvious evolution during the 26-day span between the first and the last spectrum in which the lines are detected.

\begin{figure}
    \includegraphics[width=0.48\textwidth]{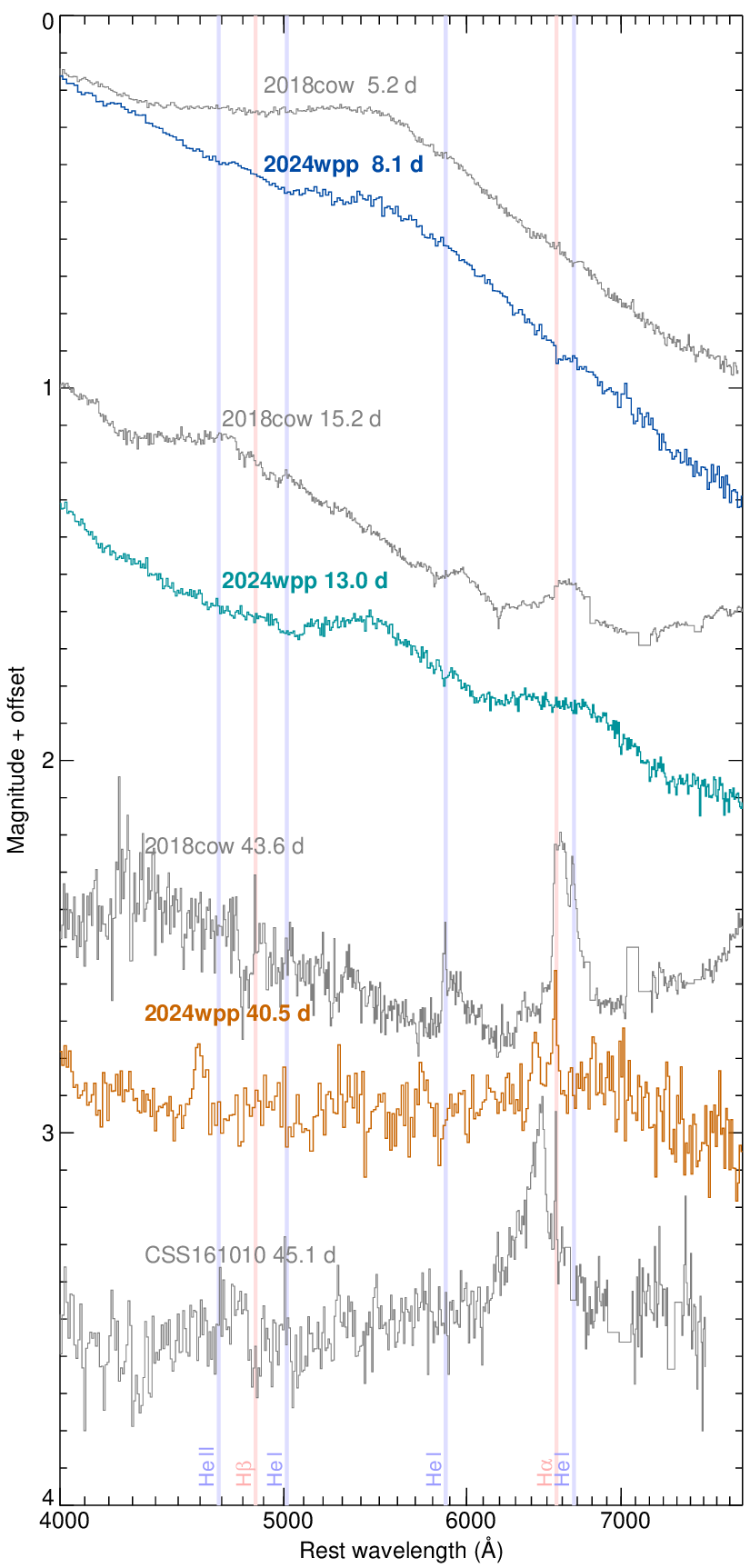}
    \vspace{-0.25cm}
\caption{Comparison between spectra of AT\,2024wpp at early, intermediate, and late times to those of AT\,2018cow at similar phases.   CSS161010 is also shown for comparison to the late time spectrum. Vertical lines mark the rest wavelengths of transitions, although note that the He transitions in AT\,2024wpp are offset by $-$6600 km s$^{-1}$.}
    \label{fig:comparespectra}
\end{figure}

The late-time behaviour is qualitatively similar to what has been seen in the other two events with late-time spectroscopy (AT\,2018cow and CSS161010), indicating that some of these properties are likely to be generic to the LFBOT class.  Specifically: both AT\,2018cow and CSS161010 showed emission only from light elements, the lines in both events developed starting at about 1 month, both had intermediate velocity widths, and both showed large asymmetries in the line profiles.  However in detail the three events are distinct: AT\,2018cow showed primarily redshifted profiles while the other two events show blueshifts;  CSS161010 showed only H in its late-phase spectra (though He may have been detected at earlier times); and AT\,2024wpp is the only event to show a clear double-peaked component.   The three events are compared (at different phases) in Figure \ref{fig:comparespectra}, using spectra from \cite{Perley+2019} and \cite{Gutierrez+2024}.

\subsection{Ultraviolet Spectroscopic Analysis}
\label{sec:uvspecanalysis}

\begin{figure}
    \includegraphics[width=\columnwidth]{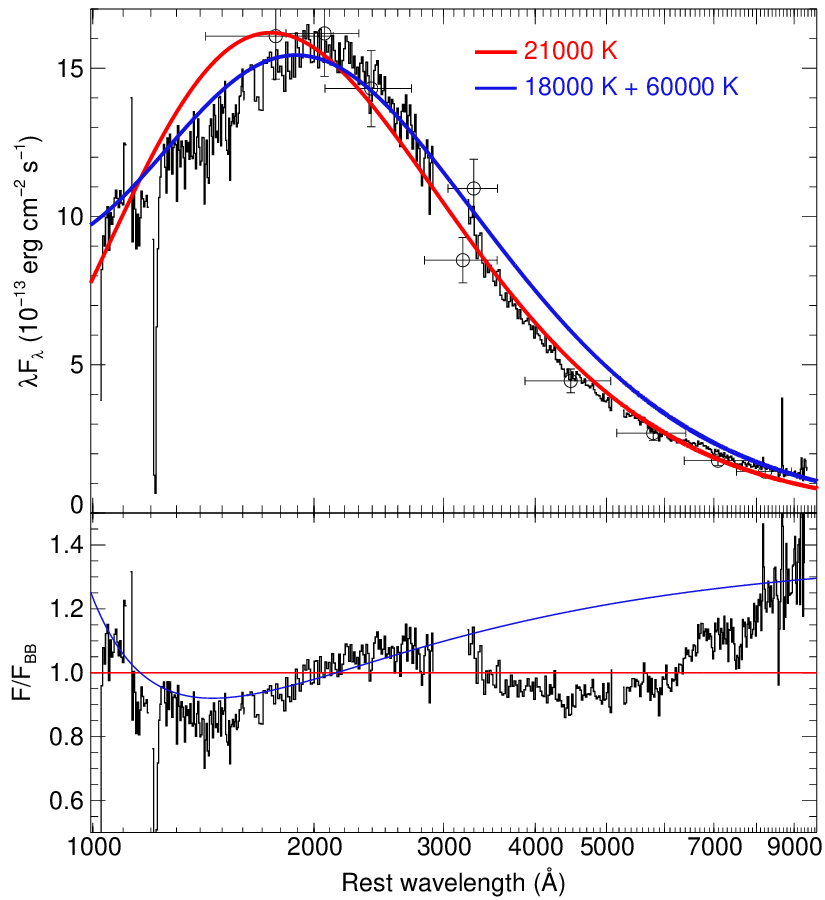}
    \vspace{-0.25cm}
\caption{Combined HST/COS, HST/STIS, and ground-based optical spectra of AT\,2024wpp at approximately 20 rest-frame days after explosion.  The upper panel shows the flux-calibrated spectrum, corrected for Galactic extinction but with no further adjustments.  Interpolated photometry from our empirical light curve model is overlaid.  The red curve is that of a 21000\,K perfect blackbody; the blue curve shows a two-component model.  The inset shows the spectrum divided by the simple blackbody model.}
    \label{fig:uvoptspec}
\end{figure}

A combined ultraviolet+optical spectrum of AT\,2024wpp at 20 rest-frame days (merging the COS spectrum, STIS spectrum, and a ground-based spectrum from P200) is plotted in Figure \ref{fig:uvoptspec}.  Individual spectra have been binned to $R\approx200$ (1400~km\,s$^{-1}$ resolution) to reduce noise and emphasize broad features, and regions of the spectrum affected by strong geocoronal absorption or emission have been removed.  The spectrum is remarkably simple, showing a smooth continuum covering the entire spectral range.   The strongest visible feature is \HI\,$\lambda$1216 (Lyman $\alpha$) at the host redshift, which can be attributed to the host galaxy (as described later in this section).  There are no other strong features in emission or absorption at any of the positions of strong resonance lines present in the UV and evident in previous UV spectra of hot transients: for example, \CII\,$\lambda$1334, \NV\,$\lambda$1240, \OI\,$\lambda$1302, \SiIV\,$\lambda$1400, and \CIV\,$\lambda$1550 are all absent as broad lines.

Also plotted in Figure \ref{fig:uvoptspec} is a simple single-temperature blackbody with $T$\,=\,21000\,K, which is generally within 20\% of the measured flux at all wavelengths (except for narrow absorption lines and noise).   The figure inset shows the spectrum normalized to this model.  We experimented fitting a multi-temperature blackbody to the UV alone, but the resulting fit (shown as the blue curve) substantially overpredicts the optical region of the spectrum.  It is possible that unidentified broad features are responsible for the deviation: this is likely true in the NIR region of the spectrum at minimum, since the $y$-band photometry shows a probable excess during this period (\S \ref{sec:bb}).  However, these deviations are relatively minor and in general the simple blackbody model describes all phases of the transient (between 1000$-$10000 \AA) remarkably well.

\begin{figure}
    \includegraphics[width=\columnwidth]{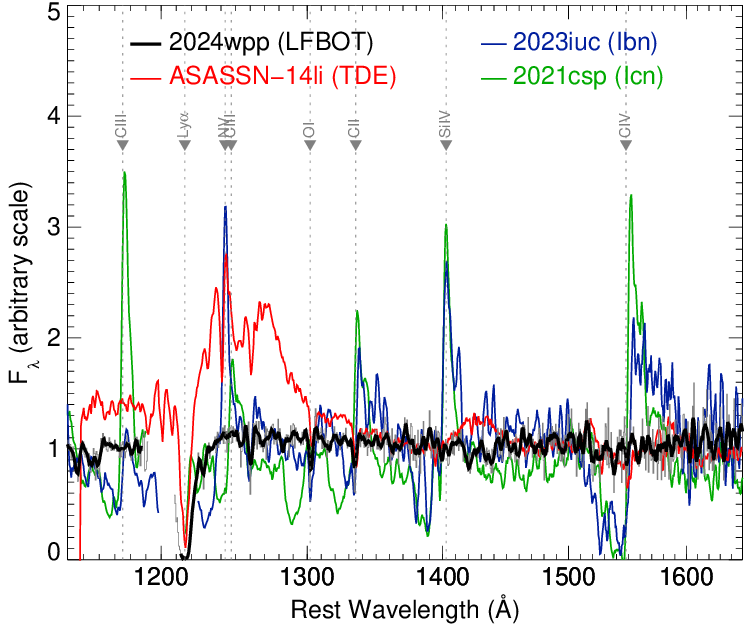}
    \vspace{-0.25cm}
\caption{Comparison of the FUV spectrum of AT\,2024wpp to other UV-luminous transients (Ibn/Icn) observed at similar phases.  The strong P-Cygni lines seen in these other spectra are absent in AT\,2024wpp.  The spectrum of ASASSN-14li (a TDE) lacks the P-cygni metal features, but contains other broad features absent in AT\,2024wpp.}
    \label{fig:uvcompare}
\end{figure}

The remarkable contrast between AT\,2024wpp's featureless spectrum and known classes of optical/UV transients in this regard is further emphasized in Figure \ref{fig:uvcompare}, which compares the COS spectrum of AT\,2024wpp 
to three other fast and luminous blue transients observed with HST at FUV wavelengths: SN\,2021csp (Icn, from \citealt{Perley+2022}), 
SN\,2023iuc (Ibn), 
and ASASSN-14li (TDE, from \citealt{Cenko+2016}).  
The UV spectra of the two comparison interacting SNe are dominated by strong, broad P-cygni features of light elements, notably including strong \CIV\ in all three cases, and thus do not resemble AT\,2024wpp.   P-cygni metal features are much weaker in ASASSN-14li, and a broad, strong intrinsic Lyman-$\alpha$ feature is present; none of these features are seen in AT\,2024wpp.

\begin{figure*}
    \includegraphics[width=\textwidth]{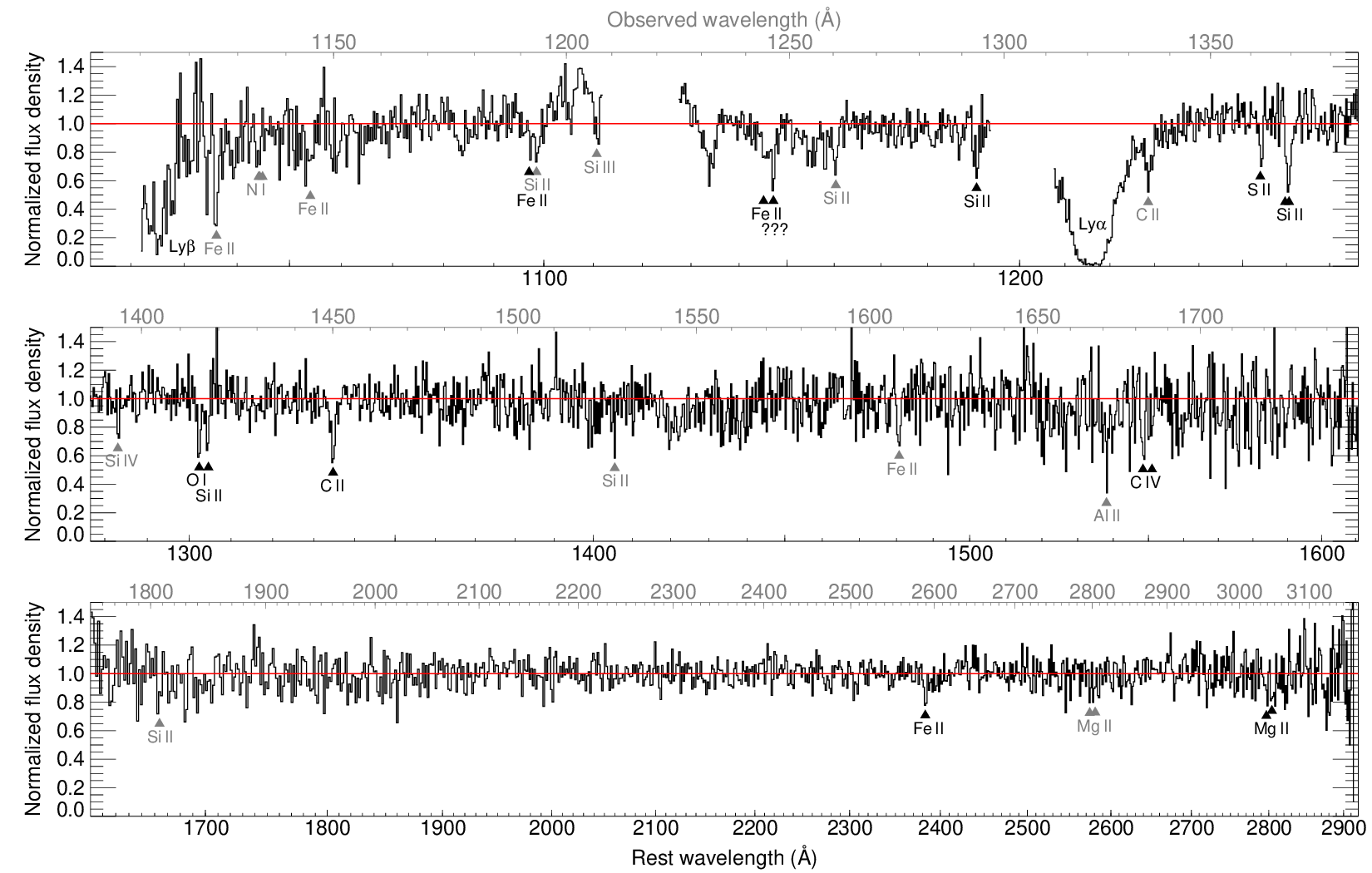}
    \vspace{-0.5cm}
\caption{UV spectrum of AT\,2024wpp, normalized by a polynomial model.  Absorption line signatures are identified:  those associated with the host are marked in black, and those with the Milky Way Galaxy in grey.}
    \label{fig:uvlines}
\end{figure*}

The spectrum of AT\,2024wpp does contain a number of narrow, weak absorption features, which are highlighted in Figure \ref{fig:uvlines}.  All of these lines are unresolved and can be associated with common ISM features, either in the host galaxy or the Milky Way galaxy.  We fit the high-S/N host lines in the COS spectrum (\SiII, \OI, \CII, \SII) with a Gaussian absorption model convolved with the COS line-spread function and find that the line is consistent with being saturated in all cases, so it is not possible to derive abundances from these observations.  We also measure the equivalent widths in the STIS and optical (early-time LRIS) observations by direct integration.   We report the resulting values in Table \ref{tab:abslineEW}.

\begin{table}
	\centering
	\caption{Rest-frame equivalent widths of narrow UV/optical absorption lines detected in the host system.}
	\label{tab:abslineEW}
	\begin{tabular}{lcc}
		\hline
Species & $\lambda_{\rm rest}$ & EW$_{\rm rest}$ \\
 & (\AA)  & (\AA) \\
		\hline
\SiII & 1190.42  & 0.28 $\pm$ 0.08 \\
\SII &  1253.81  & 0.20 $\pm$ 0.08 \\
\SII  & 1259.52  & 0.33 $\pm$ 0.07 \\ 
\SiII$^{c}$ & 1260.42  & 0.60 $\pm$ 0.09 \\
\OI   & 1302.17  & 0.30 $\pm$ 0.08 \\
\SiII & 1304.37  & 0.30 $\pm$ 0.08 \\
\CII  & 1334.53  & 0.46 $\pm$ 0.07 \\
\SiII & 1526.71  & 0.23 $\pm$ 0.14 \\ % low significance
\CIV  & 1548.20  & 0.34 $\pm$ 0.12 \\
\FeII & 2382.77  & 0.59 $\pm$ 0.19 \\
\MgII$^{a}$ & 2800     & 1.00 $\pm$ 0.45 \\
\NaI  & 5889.95  & $<$0.18 \\
      \hline
	\end{tabular}
{\par \begin{flushleft}
$^{a}$\,Given EW is that of both members of the doublet combined.
\end{flushleft}}
\end{table}

The broad Lyman-$\alpha$ feature (the blue wing of which is cut off by geocoronal contamination) is well-fit by a damped Lyman-$\alpha$ profile and indicates a hydrogen column through the host of (6.5 $\pm$ 0.4) $\times$ $10^{20}$ cm$^{-2}$.

\subsection{Radio evolution}
\label{sec:radioevolution}

To infer the properties of the forward shock and CSM, we model the radio emission from AT\,2024wpp using the traditional synchrotron self-absorption (SSA) framework \citep{Chevalier+1998,Soderberg+2005}. This framework assumes a nonthermal power-law electron energy distribution.\footnote{We leave modelling via a \textit{thermal} relativistic-Maxwellian prescription \citep{Ho+2022,Margalit+2021,Margalit+2024} for future work.} The SSA framework has been assumed for previously discovered LFBOTs to infer the shock radius $R_{\rm p}$, shock speed $\beta_{\rm sh}$, magnetic field strength $B_{\rm p}$, total energy $U_{\rm tot}$, and density of the ambient medium $n_e$ \citep{Ho+2019,Coppejans+2020,Ho+2020koala,Ho+2022,Yao+2022,Chrimes+2024b}.

\begin{figure}
    \includegraphics[width=\columnwidth]{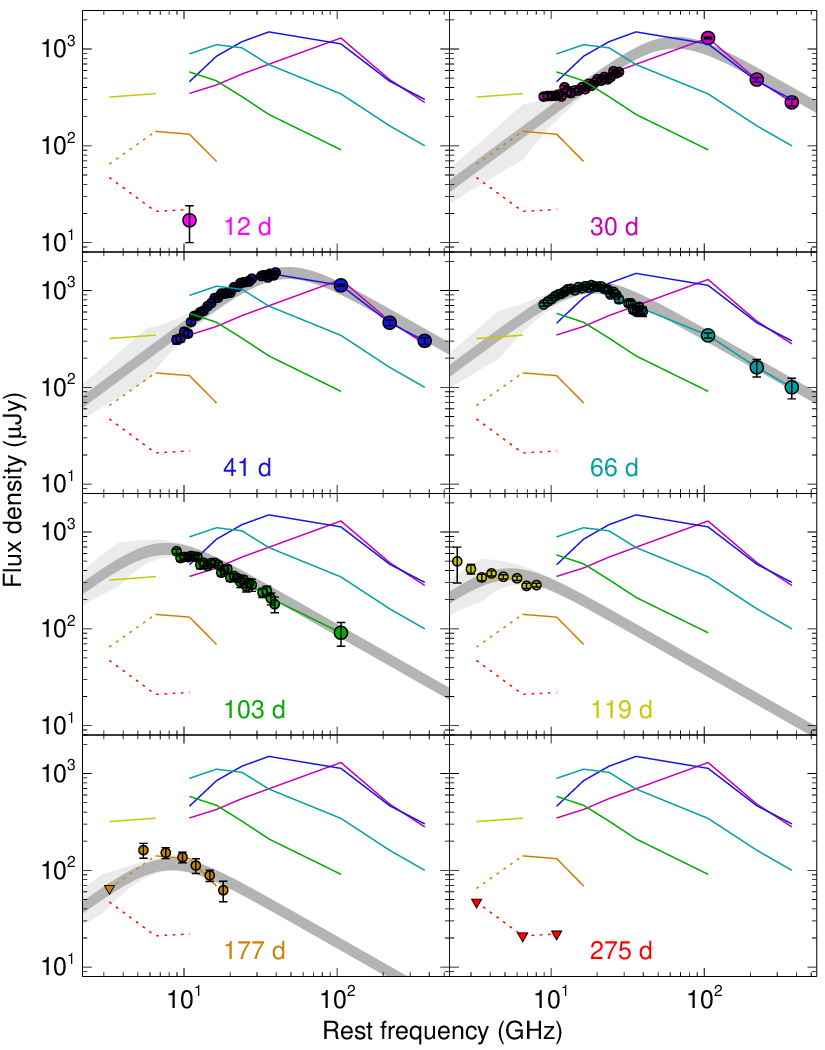}
    \vspace{-0.5cm}
\caption{Evolution of the radio SED with time from combined VLA+ALMA observations.   Each panel shows the SED taken from measurements at similar times (rest-frame days post-explosion are indicated in each panel) as circles with error-bars; the broken power-law SED used to infer the shock properties at that epoch is added as a thick grey line.  The light grey envelope around the line shows a model for the impact of scintillation modulation of the afterglow flux.  Thin, coloured lines show SEDs formed from (band-averaged) measurements at similar times and are plotted in every panel for ease of comparison.}
    \label{fig:radioseds}
\end{figure}

Using multiple epochs of nearly co-eval radio observations ($\sim$30 days\footnote{For the first epoch, the time-evolution of the transient at low frequencies between 27 days (when the X and Ku observations were taken) and 30--34 days (when the higher-frequency observations were taken) is significant, as the low-frequency light curve is rapidly brightening at this time (Figure \ref{fig:radiolightcurve}), leading to a discontinuity in the SED if adjustments are not made.  We apply a correction factor of 1.37 (consistent with a source rising as $F \sim t^{+2.5}$ between 27 and 31 days) to the X and Ku measurements, which removes the discontinuity; corrected SEDs are used throughout the analysis described below and in Figure \ref{fig:radioseds}.}, $\sim$41 days, $\sim$66 days, $\sim$103 days, $\sim$119 days, and $\sim$177 days rest frame), we track the evolution of the peak flux $F_{\nu,{\rm peak}}$ and frequency $\nu_{\rm peak}$ for AT\,2024wpp (Figure \ref{fig:radioseds}), which can be used to determine the aforementioned shock properties.  We first perform initial fits of a soft broken power-law ($F_\nu = F_{\nu,{\rm peak}}(({\nu/\nu_{\rm peak})^{-s\beta_1}+(\nu/\nu_{\rm peak})^{-s\beta_2}})^{-1/s}$) to the sub-band gridded radio data at 41\, and 66\,d (the only two epochs where the peak is continuously sampled) to estimate values of the spectral index below the peak, the spectral index above the peak, and the sharpness parameter; these are found to be $\beta_1 = +$1.17, $\beta_2 = -$0.90, and $s=$ 1.5, respectively.  The low-frequency index $\beta_1$ is significantly shallower than predicted under the normal shock model (as has been previously seen in LFBOTs; e.g. \citealt{Nayana+2021,Bright+2022,Ho+2022}).  The exact choice of spectral indices is not extremely important, as the primary motivation for performing these fits is to identify the frequency and flux of the peak of the SED.  Deviation in the low-frequency index could indicate additional populations of lower-energy electrons contributing to the flux or potentially other components such as a reverse shock.  

We then fix $\beta_1$, $\beta_2$, and $s$ to the values above, and fit each SED using the band-integrated flux density measurements to determine the peak frequency ($\nu_{\rm p}$) and flux ($F_{\rm p}$) for all six epochs; the basic shock parameters are inferred from these values via equations 2--6 from \cite{Chrimes+2024b}.  We assume an electron energy power-law index of $p=3$, a volume filling factor of $f=0.5$, and equipartition such that the magnetic energy density ($\epsilon_B$) and energy density of electrons ($\epsilon_e$) contribute equally to the post shock energy ($\epsilon_B$=$\epsilon_e$=1/3).  Uncertainties (95\% confidence) on each parameter are determined from a Monte Carlo procedure.  Flux modulations induced by interstellar scintillation (see following text) were included in the uncertainty analysis, as was an overall 10\% flux calibration uncertainty component applied uniformly to all bands.  The resulting values of the shock parameters are provided in Table \ref{tab:radioTab}.\footnote{The fourth epoch (at 103\,d) converges to a broken power-law with $\nu_{p}$ = 7 GHz but is consistent with a $\nu^{-0.90}$ power-law with no break within uncertainties, so only limiting values of the parameters are quoted for this epoch.}

\begin{figure}
    \includegraphics[width=\columnwidth]{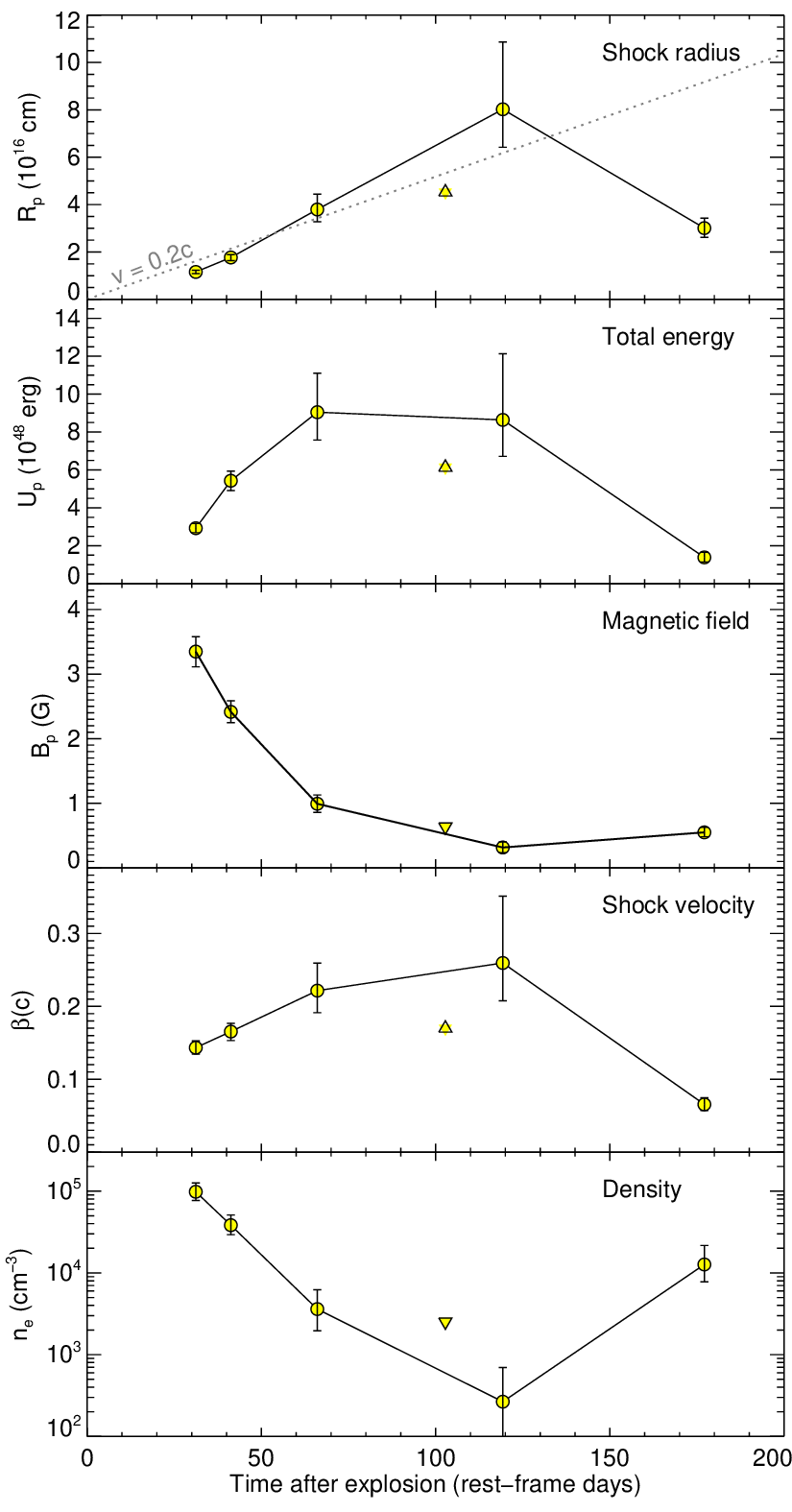}
    \vspace{-0.5cm}
\caption{Evolution of radio shock parameters from fitting a Chevailer self-absorption model to the radio SEDs.}
    \label{fig:radioevol}
\end{figure}

Figure \ref{fig:radioevol} shows the evolution of the shock properties of AT\,2024wpp as inferred by this prescription.
We see an initial increase in shock radius from $\sim$30--120 days. The velocity inferred from the change in radius over this time span is $\sim$0.2\,$c$, 
placing the fastest ejecta in the mildly relativistic shock speed regime, similar to other LFBOTs like CSS161010 and AT\,2023fhn \citep{Gutierrez+2024,Chrimes+2024b}. 

After $\sim$119 days the shock properties undergo a significant transition:  the peak flux of the SED begins to decline dramatically, even as the peak frequency remains essentially constant. 
Similar rapid late-phase declines have been seen in other LFBOTs \citep{Margutti+2019,Coppejans+2020,Chrimes+2024b}, and have been sometimes interpreted as the result of the shock reaching the limit of a confined ``bubble'' of circumstellar material \citep{Ho+2019}.  Our modelling of AT\,2024wpp after the transition point does not provide direct evidence that the shock has transitioned into a regime of lower density: because the peak frequency changes little even as the flux drops sharply, the electron density we infer at 177\,d density is higher than what is seen at 119\,d, and the radial extent of the shock is smaller.  This could indicate that the shock encountered a denser structure (such as a torus) in some directions, and a region of lower density in others.  

As our simplified model assumes spherical symmetry, it is not possible to test this possibility in detail at this stage.  The non-standard evolution could also indicate larger deviations from the basic model.  The above analysis assumes that the forward shock dominates the emission at all times, but it is also possible that a reverse shock contributes or even dominates at some times, and in this scenario a rapid drop in the light curve could occur when the shock traverses the ejecta.  More detailed modelling of this transition will be left for future work.

\begin{table*}
\centering
\caption{Shock parameters at different epochs, derived from radio measurements assuming equipartition.}
\label{tab:radioTab}
\begingroup
\renewcommand{\arraystretch}{1.3} 
\begin{tabular}{c c c | c c c c c}
\hline
$\Delta t_{\rm rest}$ & $\nu_{\rm p}$ & $F_{\rm p}$  & $R_{\rm p}$ & $B_{\rm p}$  & $U_{\rm tot}$ & $\beta_{\rm sh}$ & $n_e$ \\
(days) & (GHz)& ($\mu$Jy)  & ($10^{15}$ cm) & (G) & ($10^{48}$ erg) & ($c$) & (cm$^{-3}$) \\
\hline
% t  & nu_p  & F_p  &   R_p  &   B  &  U   & beta &   n_e
  30.5$^{+0.8}_{-3.5}$ &  59.4$^{+4.2}_{-4.3}$ & 1837$^{+130}_{-130}$ & 11.58$^{ +0.75}_{ -0.71}$ &  3.35$^{+0.23}_{-0.24}$ &  2.9$^{+0.3}_{-0.2}$ & 0.14$^{+0.01}_{-0.01}$ & 98300$^{+27800}_{-21300}$\\
  40.8$^{+0.4}_{-0.6}$ &  44.1$^{+3.2}_{-3.2}$ & 2393$^{+144}_{-150}$ & 17.68$^{ +1.21}_{ -1.30}$ &  2.42$^{+0.17}_{-0.17}$ &  5.4$^{+0.5}_{-0.5}$ & 0.17$^{+0.01}_{-0.01}$ & 38400$^{+12700}_{ -9020}$\\
  66.0$^{+0.1}_{-0.1}$ &  17.5$^{+2.4}_{-2.3}$ & 1699$^{+159}_{-129}$ & 37.94$^{ +6.45}_{ -5.17}$ &  0.99$^{+0.14}_{-0.13}$ &  9.0$^{+2.1}_{-1.5}$ & 0.22$^{+0.04}_{-0.03}$ &  3610$^{ +2620}_{ -1650}$\\
 103.1$^{+1.3}_{-0.4}$ & $<$ 10.4           & $>$ 803           & $>$45.24           & $<$ 0.64           & $>$ 6.1           & $>$0.17           & $<$ 2525          \\
 119.3$^{+0.1}_{-0.1}$ &   5.0$^{+1.4}_{-1.2}$ &  577$^{ +84}_{ -59}$ & 80.24$^{+28.38}_{-16.02}$ &  0.32$^{+0.09}_{-0.08}$ &  8.6$^{+3.5}_{-1.9}$ & 0.26$^{+0.09}_{-0.05}$ &   266$^{  +429}_{  -186}$\\
 177.1$^{+0.1}_{-0.1}$ &   7.7$^{+1.1}_{-0.9}$ &  182$^{ +22}_{ -21}$ & 30.08$^{ +4.18}_{ -3.88}$ &  0.55$^{+0.08}_{-0.07}$ &  1.4$^{+0.3}_{-0.2}$ & 0.07$^{+0.01}_{-0.01}$ & 12700$^{ +8970}_{ -4910}$\\
\hline
\end{tabular}
\endgroup
{\par \begin{flushleft}
Note --- Frequencies and times above are in the rest frame.   The time values in the leftmost column give the mean time of all observations used to construct that epoch, with the associated uncertainties indicating the span (earliest and latest).  Power-law indices for the fit were $\alpha_1=$\ 1.17 and $\alpha_2=$\ $-$0.9, respectively.
\end{flushleft}}
\end{table*}

The measurements plotted in Figure \ref{fig:radioseds} show some mild deviation from the model at the lowest frequencies in the 
30\,d, 41\,d, and 119\,d SEDs.   We investigated whether this could be a consequence of interstellar scintillation.  The maps of \cite{Walker+2001} indicate a transition frequency of $\nu_0 \approx 7$\,GHz and limiting Fresnel angle of $\theta_F \approx 4.5$\,mas at the sky location of AT\,2024wpp.  From the equations in \cite{Walker+1998}, and assuming a source expanding radially at 0.2$c$,
large ($m>0.3$) scintillation effects are expected below 13\,GHz at 30\,d, below 11\,GHz at 41d, and below 6\,GHz at 119\,d (the expected modulation range is shown as a light grey envelope in Figure \ref{fig:radioseds}), in agreement with the deviations observed. 

\subsection{Host galaxy properties}
\label{sec:host}

Photometry of the host is taken from Legacy Survey Tractor forced photometry on pre-explosion images from DECam in the $griz$ filters, from WISE \citep{Wright+2010} in the (mid-IR) W1 and W2 filters, and from the GALEX archive \citep{Bianchi+2017}.
We additionally perform photometry of the host using our late-time VLT and NTT reference images and a 7\arcsec\ radius aperture.  We also take stacks of UVOT imaging from January 2025 ($\Delta t_{\rm rest}$ $\approx$ 100--110\,d), September 2025 ($\Delta t_{\rm rest}$ $\approx$ 313--339\,d) and January 2026 ($\Delta t$ $\approx$ 434--450\,d) in each band, calculate photometry using a 7\arcsec\ radius aperture\footnote{Additionally, we calculate photometry within a 5\arcsec\ radius at the transient location for the purpose of estimating the host contamination to the UVOT images, as described in \S \ref{sec:uvot}.}, and average the three epochs together in flux space, weighted by exposure time.  The transient contribution is negligible in the later two epochs, and minimal (less than the 1$\sigma$ uncertainty) in the January 2025 epoch, but we subtract estimates of the transient flux at 105\,d (0.75$\mu$Jy in $UVW1$ and $UVW2$) from the January 2025 measurements prior to averaging. 
Host-galaxy magnitudes are provided in Table \ref{tab:hostphotometry}.

We use the python package \texttt{prospector} \citep{Leja+2017,Johnson+2021} to derive the host galaxy properties from photometry.  This package models the galaxy's SED using the Flexible Stellar Population Synthesis (FSPS; \citealt{Conroy+2010FSPS}) package, and fits the model to the observations with \texttt{dynesty}. 
For each \texttt{prospector} model run, we use the basic parametric star-formation history template.  We fit without nebular emission, but remove the contribution of the nebular lines to the photometry ($<$4\% in all filters) prior to fitting using the values in Table \ref{tab:hostlineflux}. We choose the Chabrier initial mass function \citep{Chabrier2003}, the Calzetti dust attenuation model \citep{Calzetti+2000}, and a star-formation history of the form $t\,{\rm e}^{-t/\tau}$.

After an initial fit, we estimate the stellar mass and fix the stellar metallicity to log($Z/Z_\odot$) = $-$0.6 according to the mass-metallicity relation \citep{Gallazzi+2005}.  We then re-run the model fit with metallicity fixed to this new value to obtain our final set of galaxy parameters.   

\begin{figure}
    \includegraphics[width=\columnwidth]{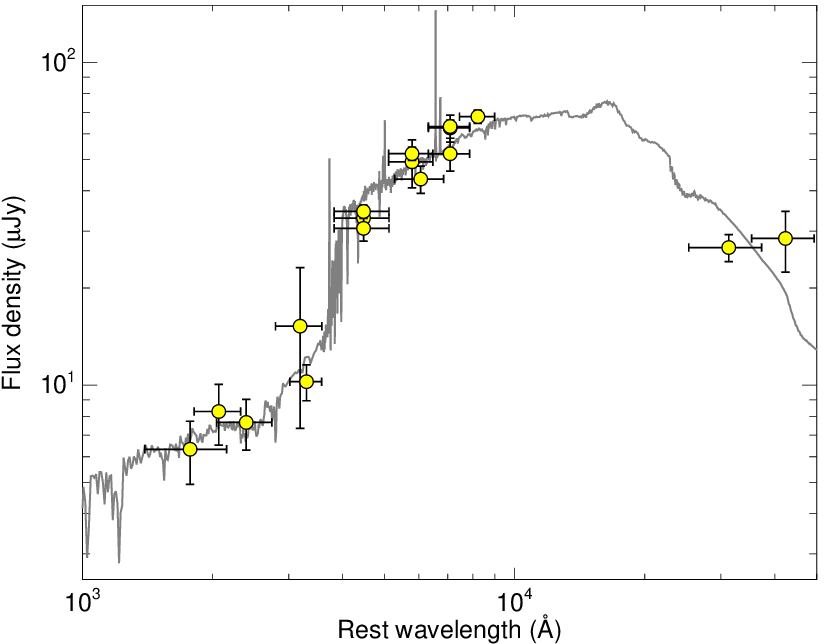}
    \vspace{-0.5cm}
\caption{Spectral energy distribution of the host galaxy of AT\,2024wpp.  Measurements are shown as circles with error bars; the result of the Prospector SED fit is shown as grey line.  (Observed fluxes are shown here, with the observed nebular line fluxes re-added to the Prospector continuum-only model for consistency.  No extinction correction is applied.)}
    \label{fig:hostsed}
\end{figure}

The SED model implies a galaxy dominated by intermediate-age stars ($\approx$ 1.8 Gyr) but which is currently forming stars at a relatively low rate ($\approx$ 0.07 $M_\odot$ yr$^{-1}$).
The results are summarized in Table \ref{tab:hostmodelproperties}.  A plot of the model, compared to the observed photometry, is provided in Figure \ref{fig:hostsed}.

\begin{figure}
    \includegraphics[width=\columnwidth]{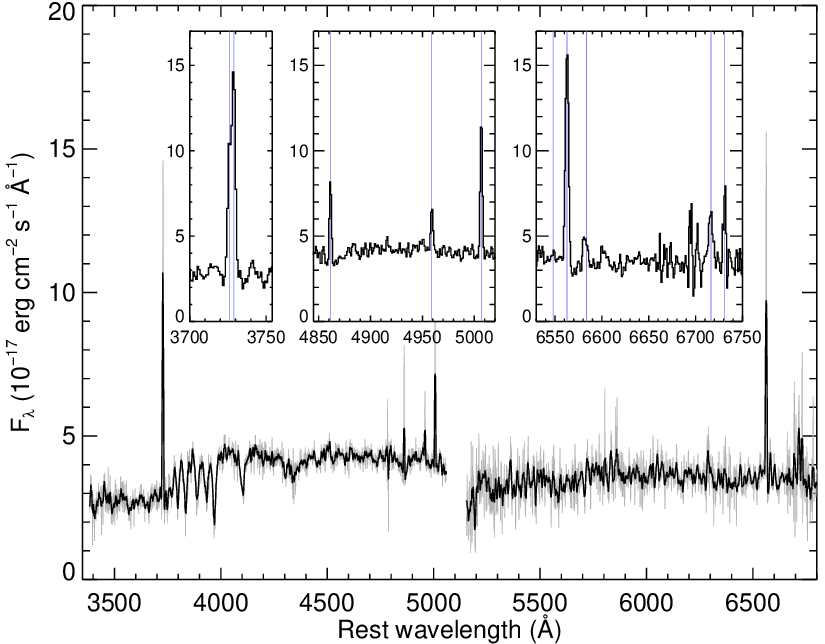}
    \vspace{-0.5cm}
\caption{KCWI integral field unit spectrum of the host galaxy of AT\,2024wpp.  The grey line shows the original spectrum, the black line is lightly smoothed using Savitzky-Golay convolution.  The three insets show zoom-ins on regions of the spectrum around strong emission lines (rest-frame wavelengths marked in blue): $[$\OII$]$ (left), H$\beta$+$[$\OIII$]$ (centre), H$\alpha$+$[$\NII$]$+$[$\SII$]$ (right).
}
    \label{fig:hostspectrum}
\end{figure}

We use the 2025 January 1st KCWI IFU observation ($\Delta t_{\rm rest}$ = 90\,d) to extract the spectrum of the host over an elliptical aperture of radius set to 2$\times$ the FWHM of the galaxy along each axis.  The extracted spectrum is shown in Figure \ref{fig:hostspectrum}.  To measure fluxes of major strong emission lines, we first subtract the stellar continuum using an SED model from the photometric fit.  We then fit a Gaussian profile at the rest wavelength of each corresponding line of interest.  The redshifts and widths of weaker lines are fixed to those of nearby stronger lines.  The extracted fluxes are provided in Table \ref{tab:hostlineflux}.

The H$\alpha$/H$\beta$ ratio of 3.4 (after correction for Milky Way extinction) indicates modest extinction towards the star-forming regions in this galaxy ($A_V = 0.6$ mag)\footnote{The significantly higher extinction from what is inferred from the SED is in some tension with the SED fitting results, but it is common for the star-forming regions responsible for the line fluxes to be extinguished more than the somewhat older stars that dominate the continuum flux; \citealt{Calzetti+1997}.}, and after correcting for this internal extinction the H$\alpha$ flux implies a modest star-formation rate of 0.1 $M_\odot$ yr$^{-1}$.    We use a variety of line diagnostics to estimate the metallicity (using the recent calibrations of \citealt{Curti+2017}) and find a (gas-phase) oxygen abundance of 12+log[O/H] in the range of 8.4--8.6 or a metallicity of about 0.5--0.8 Solar \citep{Asplund+2021}, consistent with expectations given the stellar mass and the low-redshift mass-metallicity relation (e.g., \citealt{Sanchez+2017}).  This suggests that the metallicity of the progenitor is not extreme, although it should be noted that the transient location is well outside the central region of the galaxy that dominates the star-formation.  

There is insufficient signal to estimate the properties at the specific site of AT\,2024wpp from our data.  There is no clear detection of narrow lines at this location in our spectroscopy, nor any evidence of a distinct star-forming region in the HST data (Figure \ref{fig:image}), although the observations are relatively shallow.

\begin{table}
	\centering
	\caption{UVOIR photometry of the host galaxy}
	\label{tab:hostphotometry}
	\begin{tabular}{lcc}
		\hline
Facility & filter & mag \\
		\hline
GALEX      & NUV  & 22.00  $\pm$ 0.49 \\
Swift/UVOT & UVW2 & 22.11 $\pm$ 0.22 \\
Swift/UVOT & UVM2 & 21.85 $\pm$ 0.21 \\
Swift/UVOT & UVW1 & 21.86 $\pm$ 0.17 \\
Swift/UVOT & UVU  & 21.07  $\pm$ 0.45 \\
Swift/UVOT & UVB  & 20.62  $\pm$ 0.68 \\
VLT/FORS2  & u   & 21.50  $\pm$ 0.13  \\
VLT/FORS2  & g   & 20.20  $\pm$ 0.07  \\
NTT/EFOSC  & g   & 20.28  $\pm$ 0.09  \\
LS/DECam   & g   & 20.15  $\pm$ 0.05  \\
LS/DECam   & r   & 19.68  $\pm$ 0.05  \\
NTT/EFOSC  & r   & 19.74  $\pm$ 0.17  \\  
VLT/FORS2  & R   & 19.87  $\pm$ 0.10  \\
NTT/EFOSC  & i   & 19.66  $\pm$ 0.12  \\
LS/DECam   & i   & 19.45  $\pm$ 0.05  \\
VLT/FORS2  & i   & 19.46  $\pm$ 0.10  \\
LS/DECam   & z   & 19.36  $\pm$ 0.05  \\
WISE   & W1 & 20.34  $\pm$ 0.10  \\
WISE   & W2 & 20.26  $\pm$ 0.21  \\
      \hline
	\end{tabular}
{\par \begin{flushleft}
Note --- All magnitudes are AB, and have not been corrected for Galactic extinction.
\end{flushleft}}
\end{table}

\begin{table}
	\centering
	\caption{Host galaxy properties}
	\label{tab:hostmodelproperties}
	\begin{tabular}{lll}
		\hline
Parameter & Unit & Value  \\
		\hline
\multicolumn{3}{c}{{\it SED fitting}}\\
        \hline
log$_{10}$$M_*$ & $M_\odot$ & 8.96  $\pm$ 0.06  \\
Age   &  Gyr                & 1.8   $\pm$ 0.5   \\
SFR   & $M_\odot$ yr$^{-1}$ & 0.07 $\pm$ 0.01 \\
sSFR  & Gyr$^{-1}$            & 0.07  $\pm$ 0.02 \\
$A_V$ & mag                   & 0.03  $\pm$ 0.03 \\
      \hline
\multicolumn{3}{c}{{\it IFU Spectroscopy}}\\
        \hline
SFR   & $M_\odot$ yr$^{-1}$ & 0.1      \\ 
$A_V$ & mag                 & 0.6      \\
Metallicity & 12+log$_{10}$[O/H]  & 8.5      \\ 
        \hline
	\end{tabular}
\end{table}

\subsection{Constraints on the rate of 18cow-like and 24wpp-like transients}

In seven years of operation, ZTF has detected three LFBOTs at $z<0.1$:  AT\,2018cow, AT\,2023vth, and AT\,2024wpp.  Within this redshift range, all LFBOTs will peak at magnitudes brighter than the  $<18.5$ mag threshold to meet criteria for follow-up under the ZTF Bright Transient Survey, and most will remain above this limit for at least 3 days, ensuring detection even if the peak itself is missed due to a cadence gap.  We can thus provide a provisional updated rate constraint using the current sample. 

The comoving volume within $z<0.1$ is 0.30 Gpc$^{3}$ (all-sky); accounting for the active survey footprint (35\% of the sky) approximately 0.1 Gpc$^{3}$ is being surveyed (on the timescale of the survey cadence) at any given time.  Assuming that all LFBOTs in this volume were successfully classified, this corresponds to a rate (95\% confidence interval) of 0.9--12.5 yr$^{-1}$ Gpc$^{-3}$, or about 0.001--0.01\% of the core-collapse supernova rate ($10^5$ yr$^{-1}$ Gpc$^{-3}$; \citealt{Perley+2020}).  The rate may be somewhat higher if some LFBOTs were missed due to host or Galactic extinction, confusion with CV's, or gaps in coverage longer than a few days due to weather.  However, we note that of the three $z<0.1$ events, only one actually passed the standard BTS survey coverage cuts (AT\,2024wpp itself was affected by a two-week ZTF coverage gap around the time of peak; AT\,2023vth has no coverage $\gtrsim$20\,d post-peak), suggesting that within these distances recovery of LFBOTs is not dependent on maintaining ideal survey cadence.  Additionally, we find that all ZTF superluminous supernovae (SLSNe) within $z<0.1$ have a detectable host galaxy in PS1, so unless LFBOTs are even more biased towards low-luminosity hosts than SLSNe they are unlikely to be confused with CVs during ZTF scanning efforts.  This rate is consistent with other recent estimates from ZTF \citep{Ho+2023sample}, but pushes the upper limit of the interval further downward.   It implies that LFBOTs are among the rarest transients known, perhaps even rarer than gamma-ray bursts ($\approx$50--100 yr$^{-1}$ Gpc$^{-3}$ after beaming correction; \citealt{Ghirlanda+2022}) and superluminous supernovae (2--10 yr$^{-1}$ Gpc$^{-3}$; Figure 8 of \citealt{Perley+2020}).  The rate may be similar to that of high-luminosity ambiguous nuclear transients \citep{Hinkle+2025,Wiseman+2025}.

For events similar to AT\,2024wpp specifically, the rate must be substantially lower: we define 24wpp-like events to have equal or greater luminosity and rise/decline times than AT 2024wpp itself ($M_{g,{\rm\,peak}}=-$21.8, $M_{u,{\rm\,peak}}=-$22.4, 
$t_{\rm rise}=4$\,d, $t_{\rm fade,1/2} = 4$\,d), factors which would only increase the recovery rate.  Accounting for $k$-correction effects, any such event would be detectable to $z=0.24$: the lack of any other such objects in ZTF to date implies an upper limit on the rate\footnote{We do not set a lower limit due to the a-posteriori nature of the fact that this calculation is defined by the properties of AT\,2024wpp in the first place: its detection is a precondition to producing an estimate.} of $<0.7$ Gpc$^{-3}$ yr$^{-1}$.

More precise constraints on the rate and luminosity function of 18cow/24wpp-like events will require a more in-depth recovery analysis of the ZTF transient data stream and systematic reclassification of other fast transients, and will be presented in future work.

\subsection{Additional constraints from recent publications}

Following our discovery and public announcement of this object in September 2024, several additional papers on this object have appeared in the literature or as preprints.  Measurements in these works are generally consistent with ours, although our observations extend the temporal coverage to both earlier and later times as well as to higher radio frequencies.  Our late-time measurements in particular should be more robust than those currently reported as we use host-galaxy template subtraction for all of our observations.   Some additional information derived from these sources that extends what can be inferred from our own data is summarized below; any differences are also highlighted.

\vspace{0.2cm}
\textbf{Polarimetry:} \cite{Pursiainen+2025} report on early imaging polarimetry spanning from 6--14\,d, and do not detect any polarization.  This indicates a relatively symmetrical explosion geometry over this timespan, and contrasts with early polarimetry of AT\,2018cow, which showed a detection of 7\% polarization at 5.7\,d (but only nondetections afterwards).

\textbf{No detected flares:} We have inspected all of our late-time optical imaging and no clear intra-exposure variability is visible, suggesting that AT\,2024wpp is not undergoing optical flares with the same frequency as AT\,2022tsd.   A more comprehensive investigation (although to shallower limits) using the Large Array Survey Telescope has been performed by \cite{Ofek+2025}, who also find no flares.  They place a limit of $<$0.11 hr$^{-1}$ (or $<$2.6\ d$^{-1}$).

\textbf{Near-infrared excess:} AT\,2018cow showed a prominent excess in flux after 20\,d that can be approximately modelled as a power-law \citep{Perley+2019} or as thermal radiation from dust grains heated by the transient \citep{Metzger+2023}.  We were able to obtain only limited follow-up of AT\,2024wpp at wavelengths beyond 1$\mu$m (a single set of observations at 9\,d, which rules out a NIR excess at that time).  We do see a slight excess in the PS1 $y$ band beginning at $\approx$17 days, but it is difficult to securely associate it with a broader NIR excess.  However, \cite{LeBaron+2026} confirm that a NIR excess is present in this source redward of 1$\mu$m, similar to that seen in AT\,2018cow, after 24\,d.\footnote{\cite{Pursiainen+2025} also report a NIR excess at 20\,d, but the $y-J$ colour implied by their data ($y-J \approx +1$ AB mag) is unexpectedly red for such a short wavelength interval and conflicts with the more modest excess reported by \cite{LeBaron+2026}, so it may be due to a calibration issue.}

\textbf{X-ray rebrightening and hardening:}  \cite{Nayana+2025} report on Chandra, XMM, and NuSTAR observations of AT\,2024wpp which clearly confirm the rebrightening that is apparent in \emph{Swift} XRT observations (\S \ref{sec:xrt}), and note that the spectrum becomes much harder during this period, consistent with a Compton hump feature.  The X-ray and optical luminosities are comparable during this period.  They do not report any extremely rapid or large-amplitude variations in the X-ray data.

\textbf{Radio inversion:}  \cite{Nayana+2025} also present late-time radio observations, primarily from the Australian Compact Telescope Array.  They report rapid evolution of the peak frequency of the radio SED to lower frequencies during the first $\sim$120 days, followed by an ``inversion'' in which the low-frequency radio spectral index increases again at late times (133--161 days). 
While our observations provide good agreement with most of those reported by \cite{Nayana+2025}, our late-time ($>$30\,d), high-frequency ($>$15\,GHz) radio observations are inconsistent with (in all cases, substantially brighter than) those reported in their study.  As a result, we infer much slower evolution of the peak frequency between 50--120 days, closer to standard expectation.  We do observe an increase (``inversion'') in the spectral index below 10 GHz between 119\,d and 177\,d, however, consistent with the idea that the radio behaviour has departed strongly from the expectations from the standard model after 120 days.

\vspace{0.2cm}
Additionally, we note that our own UVOT photometry implies somewhat higher peak temperatures and luminosities than those provided by \cite{Pursiainen+2025}, although somewhat less than \cite{LeBaron+2026}.  These differences do not affect our derived conclusions.  For simplicity and independence, in our analysis below we generally refer only to results derived from our own observations.  The additional information provided above (lack of polarimetry, lack of flaring, and X-ray rebrightening) are all fully consistent with our results.  Our conclusions are also broadly insensitive to the radio discrepancy, as they are derived primarily from the earlier observations.

\section{Interpretation and Discussion}
\label{sec:discussion}

\subsection{Ejecta mass constraints and inner pre-explosion material}

The evolution of the UVOIR SED during the first few days (a blackbody expanding as $R \sim vt$) is consistent with a sphere of fast-expanding, optically thick ejecta.  
An initial estimate of the ejecta mass in the event can be derived from basic physical considerations. For energy deposited at the centre of the ejecta, photons will diffuse out of the ejecta on a timescale \citep{Kasen2017}

\vspace{-0.25cm}
\begin{equation}
    t_\mathrm{diff} = \left[ \frac{3}{4\pi} \frac{\kappa M_\mathrm{ej}}{vc} \right]^{1/2}. 
\end{equation}

Solving for $M_{\rm ej}$, this can be expressed in fiducial form as:

\vspace{-0.25cm}
\begin{equation}
     \frac{M_\mathrm{ej}}{M_\odot} 
    = \left(\frac{t_\mathrm{diff}}{8.4\,\mathrm{d}}\right)^{2}
    \left( \frac{\kappa}{0.1\,\mathrm{cm}^{2}\,\mathrm{g}^{-1}} \right)^{-1}
    \left( \frac{v}{0.1c} \right)^{}.
\end{equation}

\noindent The rise time of the light curve in AT\,2024wpp is $\approx1.9$\,d (from inferred time of first light to the peak of the bolometric light curve). Adopting $\kappa_{0.1}=1$ for hydrogen depleted CSM\footnote{We choose this as a fiducial value on account of the relatively weak H emission at late times (relative to He emission) in this event and in AT\,2018cow \citep{Fox+2019}.} and $v=0.2c$ as an estimate of the photospheric velocity we find $M_\mathrm{ej}\approx0.1\,M_\odot$. The corresponding kinetic energy is $4\times10^{51}\,$erg. 

The inferred bolometric radiated energy from integrating the light curve is close to $10^{51}$\,erg, suggesting that the radiative efficiency of the transient is very high.  An explosion within a stellar progenitor with a typical size of a few AU would not be able to radiate efficiently after expanding to the observed maximum photospheric radius of 50--100\,AU without an additional energy source.  Radioactive decay is clearly insufficient even without considering the low ejecta mass (at least 300 $M_\odot$ of $^{56}$Ni would be required to explain the peak luminosity).  This leaves shock interaction between the ejecta and pre-existing material, heating from a central engine, or some combination.

Shock interaction with dense circumstellar matter is well known to be an effective mechanism for producing luminous transients \citep{Khatami+2024}.  In this context, the early light curve would be powered by shock-heating of the CSM material with the breakout occurring at the bolometric peak, and the derivation for the diffusion time can be interpreted as the shock breakout time, from a CSM with $M(R>R_{bo})\simeq 0.1 M_\odot$ where the breakout radius, $R_{bo}$ is approximately the blackbody radius at the time of the peak $R_{bo} \sim 10^{15}$ cm. This automatically gives a Thomson optical depth of $\tau\simeq c/v$ at breakout. In this scenario, the $\sim$1--4$\times10^{51}$ erg carried by the fast ejecta imply that the fast ejecta has a mass of $M(v\gtrsim 0.2c)\sim 0.05M_\odot-0.2M_\odot$.   It would also have to be sufficiently optically thin for photons to escape ($\tau < c/v$), although this criterion is satisfied automatically (if $M_{\rm sh} \approx M_{\rm ej}$) as this condition is already established by Equation 1.

While ejecta-CSM interaction can thus provide a satisfactory explanation for the initial rise to peak, it is unclear whether interaction alone can explain the post-peak behaviour. 
In the basic scenario seen in other types of interacting SNe, interaction results in the formation of a dense shell of swept-up material at the leading edge of the ejecta; the energy is deposited in forward and reverse shocks very close to this shell and the photosphere is also located in this region during the regime where interaction continues to power the light curve \citep{Smith+2014}.  This shell is a physical structure and must be expanding with time.  In contrast, in AT\,2024wpp the blackbody radius contracts without much change in temperature; inconsistent with the optically thick, spherical shock that would otherwise be suggested by the blackbody-like spectrum that dominates throughout the entire observed evolution of this transient.   Additionally, the highly luminous (and in some other events, highly variable: \citealt{Ho+2019,Yao+2022}) X-rays are also not characteristic of optically-thick shock interaction.\footnote{It is possible that this is due to the X-rays being absorbed in the region upstream of the shock, which, as discussed in subsequent sections, is not expected for this event.}

We note that there exist scenarios in which the model of the shock as an optically thick, geometrically thin, spherical structure is not appropriate.  For example, the CSM may be aspherical, or it may even be optically thin despite the thermal shape of the SED (which can be produced, for example, by X-rays reprocessed by ionized matter), in which case arguments connected to the size of the photosphere should be treated with caution.    However, the extensive differences between this event and other interaction-powered transients will motivate us to consider other scenarios to explain the power source out to late times.

\subsection{Power from a central X-ray source}

The other potential power source is input from a central engine, probably in the form of X-ray heating.    This imposes its own requirements on the density profile, although because the power source is now in the centre the requirement applies to the ejecta below the photosphere, rather than ISM above it.   The early optical luminosity exceeds the X-ray luminosity by two orders of magnitude, so we require the X-ray optical depth $\tau \approx \kappa M_{\rm ej} / (4\pi R_{\rm ej}^2)$ to be at least a few but not $\gg$10, or the X-rays would not escape at all.   

The optical depth through a shell of material at radius $R$ at bolometric peak is:

\vspace{-0.25cm}
\begin{equation}
\tau = 1.6 \left(\frac{\kappa}{0.1\ {\rm cm}^2 {\rm g}^{-1}}\right)\left(\frac{M_{\rm ej}}{0.1\ M_\odot}\right)  \left(\frac{R}{10^{15}\ {\rm cm}}\right)^{-2}
\end{equation}

In the soft X-rays, for material that is only partly ionized, $\kappa_{0.1} \approx 10^4$ and thus $\tau \approx 16000$ for the ejecta mass inferred from the UV/optical analysis:  no X-rays would escape.  This could indicate that the ejecta is highly asymmetric, or that the observed X-rays do not originate from the engine (although the fast variability seen in AT\,2018cow suggests that latter interpretation is unlikely to hold for that system).   Alternatively, the X-rays themselves could be keeping the ejecta almost completely ionized, which is plausible for an X-ray source radiating at $10^{45}$ erg s$^{-1}$ (see following section).

\subsection{Featureless spectra from ionization and heating}

If shock interaction plays even a partial role in explaining the early light curve of this event, the requirement for a large amount of pre-existing material to decelerate the early ejecta seems to pose a contradiction with the lack of features in any of our spectra prior to 50\,d.  This includes optical spectra taken at peak when the interaction should have been ongoing, as well as the the 20\,d HST spectrum, which covers many very strong UV resonance lines normally seen in SN spectroscopy that are entirely absent.   Our limits on the fluxes of any emission lines, or equivalent widths of any broad absorption lines, are 1--2 orders of magnitude below what is seen in comparably well-observed Type Ibn supernovae (Figure \ref{fig:uvcompare}).

A lack of lines can be satisfied even in the presence of significant dense material if the absorbing material (including metals) is highly ionized.  Additionally, emission lines are suppressed when the gas temperature is high due to the strong temperature dependence of the recombination coefficient.
The luminous X-rays seen in AT\,2024wpp (and other LFBOTs, but generally \emph{not} seen in interacting SNe\footnote{Only two SNe Ibn have been detected in X-rays to date:  SN\,2006jc \citep{Immler+2008} and SN\,2022ablq \citep{Pellegrino+2024}.  The peak luminosities of these events are 3--4 orders of magnitude lower than that of SN\,2024wpp.}) provide a potential way to achieve this condition.
To determine when the metals are sufficiently ionized, we use the X-ray absorption criterion from \citet{GovreenSegal+2025}. 
The conditions for using their framework are that $L_\nu \propto \nu^{\alpha}$ where $-2 \leq \alpha \leq 0$, and that the density profile has the most matter near an inner radius. 
Our observed X-ray spectral index fulfils the first criterion, while the steep density profile inferred by the radio observations (\S \ref{sec:densityprofile}) fulfils the second criterion.

We can use the lack of X-ray photoabsorption to place a limit on the density in the matter between the source and the observer before absorption lines develop.  The criterion for lack of X-ray photoabsorption can be expressed in terms of the parameter $W$, which is related to the inverse optical depth for absorption at 1 keV (Equation 7 of \citealt{GovreenSegal+2025}); photoionization-induced transparency leading to a lack of spectral features is achieved when $W\gg1$.  This parameter can be expressed as: 
\begin{equation}
    W \approx \left(\frac{L_X}{10^{43}\ {\rm erg}}\right) \left(\frac{r}{10^{15}\,\mathrm{cm}}\right)^{-3} \left(\frac{n_e}{10^{10}\mathrm{cm}^{-3}}\right)^{-2} \left(\frac{Z}{Z_\odot}\right)^{-1},
\end{equation}
where $L_x$ is the X-ray luminosity in 0.3-10 keV, $r$ is the radius of the (potentially) absorbing matter, and $Z$ is the metallicity.  
The density requirement is thus:
\begin{equation}
    n_e \ll {10^{10}\mathrm{cm}^{-3}} \left(\frac{L_X}{10^{43}\ {\rm erg}}\right)^{1/2} \left(\frac{r}{10^{15}\,\mathrm{cm}}\right)^{-3/2}  \left(\frac{Z}{Z_\odot}\right)^{-1/2}.
\end{equation}
At 3\,d (approximately bolometric maximum and the time of the initial spectroscopy) we have $L_X = 2\times10^{43}$ erg s$^{-1}$, which implies a density of $n_e \ll 10^{10}$ cm$^{-3}$ at the photospheric radius of $r=10^{15}$\,cm. This is similar to the average density of pre-explosion material within this radius if the ejecta swept up 0.1\,$M_\odot$ during the first few days (as inferred in the interaction model), and thus is (marginally) consistent with the picture in which X-ray radiation can suppress any absorption lines even if interaction is still ongoing. 

To inhibit emission lines, we can derive the ionization parameter ($\xi$).  This quantity is directly related to the matter temperature (see \citealt{GovreenSegal+2025}, Figure 4): for power-law spectra with $-1 < \alpha < -1/2$, the matter is heated to the radiation Compton temperature if $\xi \gtrsim 1-2$. 
The ionization parameter is:
\begin{equation}
    \xi=\frac{\dot{N_\gamma}}{4\pi r^2 nc}\simeq 1.7 \left(\frac{L_X}{10^{43}\ {\rm erg}}\right) \left(\frac{r}{10^{15}\,\mathrm{cm}}\right)^{-2}\left(\frac{n_e}{10^{10} {\rm ~cm^{-3}}}\right)^{-1}
\end{equation}

For the same parameters implied by our constraints from absorption, $\xi\gtrsim 3$.  This implies that the matter is heated to the Compton temperature, which is above a few keV (depending on the hard X-ray emission at this time, which is unobserved), suppressing the line emission significantly. 

Similar arguments can be applied at later times to explain the absence of stellar wind features in the HST spectrum at 20\,d: for $L_X \sim 2 \times 10^{42}$ erg\,s$^{-1}$, we obtain $n_e\ll 10^8 {\rm cm^{-2}}\left(\frac{r}{10^{16} {\rm~cm}}\right)^{3/2}$, $n < 10^{8}$ cm$^{-3}$ at $10^{16}$\,cm, which is consistent with the constraints inferred from the radio shock properties.

\subsection{Blackbody radius contraction}

After a few days the inferred photospheric radius from blackbody fitting (referred to here as the blackbody radius) begins to recede back towards the centre of the explosion---not just in the moving frame of radially-expanding ejecta (as is always seen in homologous outflows as the density drops), but also physically in the frame of the progenitor.  Steady contraction continues throughout our observations, reaching only $\approx10^{14}$\,cm ($\approx$7\,AU) at 100\,d---a size comparable to a supergiant star.  Similar behaviour was observed in AT\,2018cow, and remains one of the distinctive hallmarks of this class of event (at least among non-nuclear transients: broadly similar behaviour is seen in some TDEs; \citealt{Gezari2021}). 

While it is tempting to interpret the recessing blackbody radius literally in terms of a physically contracting envelope or disk, the smooth transition from an extremely fast-moving outflow suggests that this is not a new structure being revealed but a recession of the photosphere deep into the ejecta as the optical depth drops---as is normally seen in SNe at late times, although in this case the recession sets in much earlier and in material that is travelling outward orders of magnitude faster. 
This is unlikely to occur in the standard picture of homologous expansion following an explosion, as the inner material is not expanding fast enough to become optically thin on the proper timescale (nor is there enough of it to maintain optical thickness on the scale of the remarkably small photospheric radius seen at late times).  However, it could occur in a picture where the later evolution of the transient is driven not by homologous expansion but a steady wind.

\cite{Piro+2020} and \cite{Uno+2020} independently proposed a model in which a central source (likely an accretion disk) produces a fast wind and, simultaneously, irradiates the wind from beneath with X-rays.  In this scenario, the apparent photospheric radius can shrink as the mass-loss rate (and therefore wind density) drops with time, causing the radius at which the photons thermalize to move inward closer to the source.

Modelling the full data set via this scenario is beyond the scope of this work, but qualitatively it provides a consistent picture with what can be inferred from our observations.  The initial explosion ejects relatively little matter and is followed by a fast, rarefied wind travelling at 0.1$c$.  The photosphere tracks the expanding ejecta out to $\delta t \approx 4$~d, at which point the outer material becomes optically thin and the ``photosphere'' (blackbody radius) moves behind the shock front into the wind, gradually migrating inward as the mass loss rate slows.  As X-rays keep the wind hot and ionized, the scattering of photons within the wind does not lead to strong line formation for the same reason as the X-rays suppress lines in the CSM.  Possible weaker features---originating from clumps/inhomogeneities or regions of less-than-complete ionization---are smeared out by the high Doppler velocities inherent in the wind (see \S \ref{sec:intermediatefeatures}).

\subsection{The density profile and outer wind}
\label{sec:densityprofile}

\begin{figure}
    \includegraphics[width=\columnwidth]{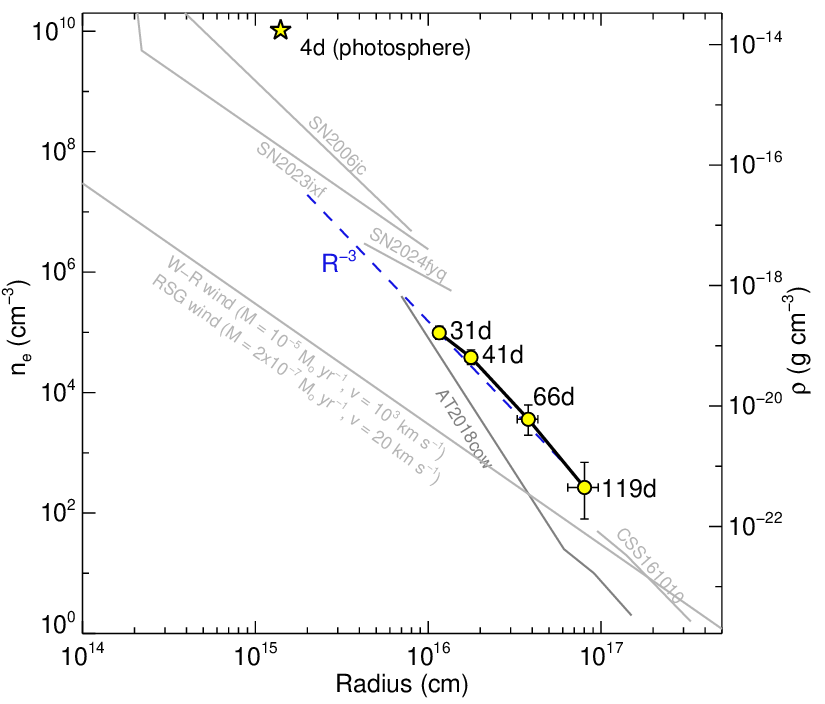}
\vspace{-0.3cm}
\caption{Constraints on the density profile of the circumstellar material surrounding the progenitor.  The measurements (shown as yellow circles) show abundant dense material close to the progenitor that drops off strongly with radius (approximately as $\rho \propto r^{-3}$), similar to previous LFBOTs, Type Ibn SNe, and ``CSM-enhanced'' Type II SNe (comparison objects are shown in grey).  The 4\,d measurement shown at the top as a star is approximate and assumes that the early peak is driven by CSM interaction.  (The final radio measurement at 176 days is omitted, as the shock prescription likely does not apply at this time, but would imply a density and radius similar to the 65\,d point.)}
    \label{fig:densProf}
\end{figure}

While the (apparent) photosphere contracts into the rarefying wind, the radio observations continue to probe the expansion of the initial ejecta in its optically thin phase, as it continues to expand into and shock the surrounding material, maintaining approximately the same velocity as the initial ejecta.  Prior to the transition in the radio behaviour at $\approx$100\,d, the shock properties are well described by the basic Chevalier model and the inferred values of $n_e$ and $R_p$ can be used to construct a density profile.

The inferred profile is shown in Figure \ref{fig:densProf}.  The density falls significantly faster than the $n_e \propto R^{-2}$  seen in a stellar wind:  the actual profile is closer to $R^{-3}$.  Interestingly, both this steep radial profile and the characteristic density itself ($\sim10^{5}$ cm$^{-3}$ at $10^{16}$ cm) are similar to what has been inferred from other interacting SNe.  For comparison we also add the profiles of two Type Ibn SNe:  SN\,2006jc \citep{Maeda+2022} and SN\,2024fyq \citep{BaerWay+2025}, the strongly-interacing Type IIP SN\,2023ixf \citep{Zimmerman+2024}, and two other 18cow-like events (18cow itself and CSS161010; \citealt{Ho+2019,Coppejans+2020}), which either follow the same profile or an interpolation/extrapolation of it.  While this could be a coincidence, it raises the possibility that enhanced mass loss in the final phases of stellar evolution, or due to the influence of an inspiraling companion, could play a common role in all of these transients.

\subsection{The origin of intermediate-time broad spectral features}
\label{sec:intermediatefeatures}

The optical spectra are not entirely devoid of features.  Two weak, very broad features are evident in spectra between 10--20 days, requiring at least some non-ionized, fast-moving material to be present.  Interestingly, aside from the blue underlying continuum, the spectra between 11--13 days superficially resemble Type Ic-BL supernovae: they are dominated by two strong features in the 5000-7000~\AA\ range with characteristic velocities of 0.1$c$ (crudely estimated from the peak-to-trough wavelength difference of the residuals after a blackbody fit); the optical luminosity of AT\,2024wpp is also similar to that of a Type Ic-BL SN at this phase.  While it is thus tempting to attribute these features to some sort of ``hidden'' SN component, these features then disappear rather than strengthen, as would be expected if they truly represented optically-thick, radioactively-heated material expanding homologously (i.e., a traditional SN component).  A more likely scenario is that they represent higher-density regions in the wind in which full ionization has not been achieved, or they may originate from the swept-up shell at the edge of the wind.  A definitive explanation will likely require radiative transfer modelling of different scenarios.

\subsection{The origin of late-time narrow spectral features}

All three LFBOTs with sensitive late-time spectroscopy show spectral features exclusively of H and He\footnote{AT\,2018cow also shows plausible oxygen features \citep{Perley+2019}, but they are much weaker than the H and He features.}, making clear that the progenitor is not H-poor (though in the case of AT\,2018cow it may be at least partially H-depleted).  If it is a massive star it must therefore have retained at least some fraction of its envelope.

The actual origin of these spectral features remains unclear.  In all three events they show up only at late times when the luminosity of the transient has declined substantially from peak---quite specifically \emph{not} the phase where strong CSM interaction is expected.  The irregular line profiles are not consistent with being pre-shock CSM from mass loss, and they are too narrow to be SN ejecta lines.  

Compared to previous events, AT\,2024wpp provides two new clues about the nature of this material in this system.  First, there are two distinct components to the emission: one at or near the systemic redshift, and one blueshifted.  This indicates that the material must have a highly asymmetric distribution, with multiple physically separated components moving in different directions; it is not a simple spherical or equatorial outflow as in the toy model proposed for AT\,2018cow by \citet{Margutti+2019}.
Second, the emission features are relatively stable (in velocity and strength) over at least several weeks.  This indicates that the structure itself must also be fairly stable.\footnote{We note that stronger, broader, single-peaked H/He features present in AT\,2018cow and CSS\,161010 at late times were less stable: while they persisted in all late-time spectra of these events, they showed changes in profile and relative intensity during their evolution \citep{Perley+2019,Margutti+2019,Gutierrez+2024}.  A possible conclusion is that there is some variation of the Doppler-broaded components of these features but not the line centres.}

It is not obvious how an explosion simultaneously produces a powerful, high-velocity explosion with a high covering fraction while also accelerating one or more separate components to a high and (largely) constant velocity.   However, one possibility is that one or both emission lines originate from a tidal stream:  in the classical TDE model, half of the material escapes the system and the remainder circularizes and accretes \citep{Rees1988,Guillochon+2016}, naturally leading to two distinct structures.  In this scenario, the blueshifted line could originate from unbound material from a disrupted progenitor with the red line originating from bound material closer to the central engine.   However, it is not clear why both lines would appear at the same time in this scenario, given that the bound material would be located far deeper inside the wind.  The two lines could originate from two different tidal streams moving in different directions as a result of a multiple-partial-disruption scenario---but the simple, smooth light curve evolution of the transient in general does not suggest multiple disruptions have occurred.

Another possibility is that the structure could represent a surviving binary (or tertiary) companion being ablated by the X-rays and wind of the compact object.  Double-peaked (although not fully-separated or net-velocity-offset) intermediate-width line profiles with peak separations of $\approx$ 2000~km~s$^{-1}$ have in fact been seen from the ablated companions of pulsars \citep{Strader+2019}.   Any such companion would have to be located outside the optical photosphere at $>$3$\,\times$10$^{14}$ cm for these lines to be visible as early as 30\,d, so its role in the progenitor system would have to be somewhat incidental.

More observations of 18cow-like objects at late phases, and more robust models, will be needed to distinguish these and other possibilities.

\subsection{Limits on a plateau and disk transition}
\label{sec:plateau}

Simple TDE models predict a correlation between the rise time and the black hole mass, with a scaling law of $t_{\rm rise} \propto M_{\rm BH}^{1/2}$ \citep{vanVelzen+2019,VanVelzen+2020}.  Recently, \cite{Mummery+2024} have additionally proposed that the late-time ``plateau'' luminosity of TDEs in the optical should scale as $L_{\rm plat} \propto M_{\rm BH}^{3/2}$.  While we do not detect an optical plateau in AT\,2024wpp, it is possible to place a deep upper limit on its luminosity.  This, and our rise-time measurement, can then be used to test the consistency between these two scaling relations which, together, imply $L_{\rm plat} \propto  t_{\rm rise}^3$.  The rise time of AT\,2024wpp is at least double that of AT\,2018cow (Figure \ref{fig:latelc}), which would imply that the plateau luminosity should be larger by a factor of 8 ($\approx$ 2.2 mag).  Our late HST limit is about 2.4 mag above the first detection of AT\,2018cow on the plateau phase, so we do not quite reach this depth.   However, it should be noted that the plateau phase is not truly flat; and our observations may be in slight tension with the predicted correlation if this were accounted for.  Absent this correction (which is not included in the scaling relation of \citealt{Mummery+2024}), our constraint on the black hole mass is only mildly constraining:  we set a limit on the black hole mass of $<3\times10^{4} M_\odot$.

\begin{table}
	\centering
	\caption{Host-galaxy emission line fluxes}
	\label{tab:hostlineflux}
	\begin{tabular}{lcc}
		\hline
Species & $\lambda$ & Flux \\
 & (\AA) & ($10^{-17}$ erg cm$^{-2}$ s$^{-1}$) \\
		\hline
H$\alpha$   & 6563 &  61.31 $\pm$ 1.04 \\
H$\beta$    & 4861 &  17.41 $\pm$ 0.73 \\
H$\gamma$   & 4340 &   6.19 $\pm$ 0.86 \\
$[$\OII$]$  & 3727 &  60.31 $\pm$ 1.95 \\
$[$\OIII$]$ & 4959 &   8.26 $\pm$ 0.73 \\
$[$\OIII$]$ & 5007 &  26.44 $\pm$ 0.75 \\
$[$\NII$]$  & 6584 &   2.98 $\pm$ 1.05 \\
$[$\NII$]$  & 6548 &   6.45 $\pm$ 1.05 \\
$[$\SII$]$  & 6717 &  14.84 $\pm$ 2.38 \\
$[$\SII$]$  & 6731 &  15.41 $\pm$ 2.38 \\
      \hline
	\end{tabular}
{\par \begin{flushleft}
Note --- Fluxes have not been corrected for Galactic extinction.
\end{flushleft}}
\end{table}

\section{Summary and Conclusions}
\label{sec:conclusions}

We provide a summary of the essential properties of AT\,2024wpp in the context of the broader LFBOT and transient populations below.

\begin{itemize}
  \item The peak bolometric luminosity is $10^{45}$ erg~s$^{-1}$, reached on a timescale of less than two days after outburst.  This is a factor of 2--3 higher than what has been seen in previous LFBOTs, and an order of magnitude in excess of SN shock breakouts or any SN at (radioactive) peak.  It is similar only to the most energetic known TDEs.
  \item Despite its rapid post-peak fading, AT\,2024wpp remains more luminous than any normal SN for at least the first 30 days.  The integrated radiative energy of the event is about $10^{51}$ erg, comparable to superluminous supernovae. 
  \item AT\,2024wpp is a luminous X-ray source, although the X-ray luminosity is much less than the optical luminosity ($\sim10^{43}$ erg\,s$^{-1}$ during the first week).  The X-rays decline rapidly, but rebrighten between 40--80 days.
  \item The fading of the optical light curve is smooth.  The bolometric luminosity decays as $t^{-3.5}$, and is driven by gradual contraction of the photosphere with little temperature change.  
  \item The early photospheric expansion speed from the time-dependent SED is about 0.2$c$.  This is approximately consistent with the velocity of the fastest ejecta inferred from radio/millimetre modelling, and from subtle broad features in early-time spectra.
  \item The photosphere reached a maximum radius of $\sim$1.5\,$\times$\,$10^{15}$\,cm (100 AU) before the radius began to decrease.  This transition, at $\sim$5 days, was not accompanied by the appearance of any spectral features, much as was the case for previous LFBOTs.  The photosphere remains hot and compact for the rest of the observable evolution.
  \item The early optical spectra are devoid of narrow lines from shock interaction, even for spectra taken on the optical rise near bolometric maximum.  
  \item Ultraviolet spectroscopy at 20 days shows a featureless spectrum approximately consistent with a 21000\,K blackbody with only minor deviations.  P-Cygni or other spectral features from a stellar wind, or from the ejecta itself, are absent.
  \item Lines of H and He (and no other elements) appear in the optical spectrum starting at about 35 days.  
  The observed lines have a complex profile with two components, each of which has a characteristic velocity width of $\delta v \approx 2000$~km~s$^{-1}$ without a narrow peak.  The blueshifted line component has a systemic velocity of $\Delta v = -$6600 km~s$^{-1}$ ($\sim$0.02$c$).  The small $\delta v/\Delta v$ implies the presence of fast-moving blobs or streams of material.
  \item The properties inferred by the first 100 days of radio observations indicate an energetic shock expanding down a steep density profile ($n \propto r^{-3}$). 
  The radio shock undergoes a transition at $\sim$100 days that causes it to fade rapidly, behaviour also seen in previous LFBOTs. 
  \item The host galaxy is a generally unremarkable low-mass galaxy with moderate metallicity, similar to the Large Magellanic Cloud.  The explosion is clearly off-centre (but within the stellar disk), and no star-formation at the location is detectable in the UV, although the observations do not constrain minor star-forming regions.
\end{itemize}

The large radiative energy and long-lived high-luminosity X-ray emission require that the power source have a large energy supply that can be converted to radiation efficiently; it must activate almost immediately after the transient ``turns on'', yet continue to remain active over a long timescale, with the luminosity falling off approximately as $t^{-3.5}$.   
As has been argued in previous work (e.g., \citealt{Perley+2019}), a black hole is far more likely given the very low ejecta masses, though an exotic magnetar formed in an ultra-stripped binary might also be able to explain some features of the system \citep{Li+2024}.

The photometry, spectroscopy, X-ray, and radio observations can be explained self-consistently in a scenario in which the black hole central engine drives a fast ($\sim0.2c$) wind and irradiates it with X-rays.  The X-rays keep the low-density wind and inner CSM almost completely ionized, suppressing any line formation and allowing the transient to maintain a hot photosphere within the expanding wind, which slowly contracts with time as the wind gradually tapers off.  

The ejecta profile, low ejecta mass, association with star-forming galaxies, and dominance of H/He lines in the spectra are all similar to properties previously seen in Type Ibn supernovae.  This gives added credence to the notion that their progenitors might be linked (\citealt{Perley+2022,Metzger+2022}; see also \citealt{Fox+2019}).  
In this scenario, the primary difference between Type Ibn SNe and LFBOTs would be the far higher accretion power in the latter, which drives a more powerful shock and sterilizes the inner envelope, preventing line formation.
However, this conclusion depends on a simplified shock model that may not be fully applicable here.  Future work will examine the radio properties of LFBOTs in more detail.

Most of the observations discussed above can be broadly understood under a scenario in which the core of a massive star collapses to a black hole and accretes its envelope \citep{Kashiyama+2015,Quataert+2019}.  
However, the extremely powerful nature of the central engine 
is difficult to reconcile with the H/He dominated nature of the system in this scenario: it indicates that there must simultaneously exist a fast-rotating BH and a non-stripped star, which is difficult to achieve in single-stellar evolution.  
Additionally, the coherent slow-moving structures inferred from late spectroscopy seem to indicate a much more asymmetric form of mass loss prior to the explosion than would be expected from any single stellar evolution scenario.

A binary scenario may be able to explain these features:  in particular, the massive star -- black hole merger model of \cite{Metzger+2022} naturally explains the H+He-rich nature, recent mass loss, and TDE-like behaviour of the system simultaneously.  It is not obvious how this scenario would produce the late-time ``streams'', but the highly asymmetric mass loss expected in a binary scenario has better potential to explain the late-phase spectroscopy than single-star models.   Alternatively, the IMBH TDE model remains a possibility, although because substantial pre-existing circumstellar material is not expected in this scenario a new interpretation of the radio behaviour would be required.

Our interpretations are broadly similar to those proposed in the literature of this event to date.  \cite{LeBaron+2026} also favour an interpretation in which the optical lines are suppressed due to X-ray ionization of a continuous wind from the engine, and attribute a black hole / massive stellar merger as the most likely progenitor.  They attribute the late-time line profiles to equatorial vs. polar outflows from super-Eddington accretion \citep{Yoshioka+2024}, differing from the interpretation presented here.  The accompanying paper by \cite{Nayana+2025} also favours a binary-merger model, emphasizing similarity of the CSM profile they derive to other LFBOTs.

With the Rubin Observatory's Legacy Survey of Space and Time \citep{Ivezic+2019} scheduled to begin in early 2026, the rate of discovery of 18cow-like events is expected to greatly increase.  The resulting order-of-magnitude improvement in the sample size (from $\sim$10 to hundreds) will be essential to confirm their association with recent star-formation and to map out the extremes of the population.  Similar rare events even more luminous than AT\,2024wpp (and MUSSES2020j) could exist; their discovery would further push the boundaries of existing models.

At the same time, the remarkable properties of AT\,2024wpp demonstrate the clear need to continue to monitor the sky at high cadence for relatively nearby ($z<0.1$), bright objects that can be followed up immediately and extensively by smaller facilities, and tracked out to very late times with larger ones.  The existence of a population of slower, more luminous 18cow-like events also raises the possibility that there might also exist faster but lower-luminosity events being missed at current survey cadences.   The continued operation of the current wide-area facilities and the commissioning of new ones, particularly ones operating at high cadence or in the ultraviolet, 
will be key to resolving the mysteries of these events in the coming years.

\section*{Acknowledgements}

We thank Eliot Quataert, Luc Dessart, Ben Margalit, Ross Ferguson, Aaron Tohuvavohu, Brad Cenko, and Jamie Kennea for useful discussions.   We thank the referee for helpful suggestions that improved the manuscript.  

Based on observations obtained with the Samuel Oschin Telescope 48-inch and the 60-inch Telescope at the Palomar Observatory as part of the Zwicky Transient Facility project. ZTF is supported by the National Science Foundation under Grant No. AST-2034437 and a collaboration including Caltech, IPAC, the Oskar Klein Center at Stockholm University, the University of Maryland, University of California, Berkeley , the University of Wisconsin at Milwaukee, University of Warwick, Ruhr University Bochum, Cornell University, Northwestern University and Drexel University. Operations are conducted by COO, IPAC, and UW.  

The Liverpool Telescope is operated on the island of La Palma by Liverpool John Moores University in the Spanish Observatorio del Roque de los Muchachos of the Instituto de Astrofisica de Canarias with financial support from the UK Science and Technology Facilities Council.

Based on observations made with the Nordic Optical Telescope, owned in collaboration by the University of Turku and Aarhus University, and operated jointly by Aarhus University, the University of Turku and the University of Oslo, representing Denmark, Finland and Norway, the University of Iceland and Stockholm University at the Observatorio del Roque de los Muchachos, La Palma, Spain, of the Instituto de Astrofisica de Canarias. The NOT data were obtained under program ID 68-501.

This work made use of data supplied by the UK \emph{Swift} Science Data Centre at the University of Leicester. We are grateful to Phil Evans, Aaron Tohuvavohu, and Jamie Kennea for advice on the \emph{Swift}/XRT data reduction. 

Some of the data presented herein were obtained at the W.~M. Keck Observatory, which is operated as a scientific partnership among the California Institute of Technology, the University of California, and NASA. The Observatory was made possible by the generous financial support of the W.~M. Keck Foundation. The authors wish to recognize and acknowledge the very significant cultural role and reverence that the summit of Maunakea has always had within the indigenous Hawaiian community. We are most fortunate to have the opportunity to conduct observations from this mountain.

Some observations reported here were obtained at the MMT Observatory, a joint facility of the University of Arizona and the Smithsonian Institution.

The National Radio Astronomy Observatory and Green Bank Observatory are facilities of the U.S. National Science Foundation operated under cooperative agreement by Associated Universities, Inc. This paper makes use of the following ALMA data: ADS/JAO.ALMA\#2023.1.01730.T ALMA is a partnership of ESO (representing its member states), NSF (USA) and NINS (Japan), together with NRC (Canada), NSTC and ASIAA (Taiwan), and KASI (Republic of Korea), in cooperation with the Republic of Chile. The Joint ALMA Observatory is operated by ESO, AUI/NRAO and NAOJ.

Based on observations collected at the European Organisation for Astronomical Research in the Southern Hemisphere under ESO programme 2114.D-5014(E).   We thank John Pritchard and Paula Sanchez Saez, and the entire observatory staff, for their excellent support. 

Based on observations collected at the European Organisation for Astronomical Research in the Southern Hemisphere, Chile, as part of ePESSTO+ (the advanced Public ESO Spectroscopic Survey for Transient Objects Survey – PI: Inserra). ePESSTO+ observations were obtained under ESO program ID 112.25JQ.

This research is based on observations made with the NASA/ESA Hubble Space Telescope obtained from the Space Telescope Science Institute, which is operated by the Association of Universities for Research in Astronomy, Inc., under NASA contract NAS 5–26555. These observations are associated with programs 16714, 17477, and 17889.

Based on observations obtained at the Southern Astrophysical Research (SOAR) telescope, which is a joint project of the Minist\'{e}rio da Ci\^{e}ncia, Tecnologia e Inova\c{c}\~{o}es (MCTI/LNA) do Brasil, the US National Science Foundation’s NOIRLab, the University of North Carolina at Chapel Hill (UNC), and Michigan State University (MSU).

The Pan-STARRS1 Surveys (PS1) and the PS1 public science archive have been made possible through contributions by the Institute for Astronomy, the University of Hawaii, the Pan-STARRS Project Office, the Max-Planck Society and its participating institutes, the Max Planck Institute for Astronomy, Heidelberg and the Max Planck Institute for Extraterrestrial Physics, Garching, The Johns Hopkins University, Durham University, the University of Edinburgh, the Queen's University Belfast, the Harvard-Smithsonian Center for Astrophysics, the Las Cumbres Observatory Global Telescope Network Incorporated, the National Central University of Taiwan, the Space Telescope Science Institute, the National Aeronautics and Space Administration under Grant No. NNX08AR22G issued through the Planetary Science Division of the NASA Science Mission Directorate, the National Science Foundation Grant No. AST–1238877, the University of Maryland, Eotvos Lorand University (ELTE), the Los Alamos National Laboratory, and the Gordon and Betty Moore Foundation.

This research made use of \texttt{ccdproc}, an Astropy package for image reduction \citep{Craig+2025}.

G.S. and A.Y.Q.H. acknowledge support in part from a Sloan Research Fellowship (Award Number FG-2024-21320) from the Alfred P. Sloan Foundation.   C.S. and A.Y.Q.H. acknowledge support in part from National Aeronautics and Space Administration (NASA) grant 80NSSC24K0377, from Hubble Space Telescope (HST) grant HST-GO-17477.006-A, and from a Scialog award from the Research Corporation for Science Advancement (``Early Science with the LSST''). 

P.C. acknowledges support from the Zhejiang Provincial Top-Level Research Support Program.

I.A. and J.C. are supported by the National Science Foundation award AST 2505775, NASA grant 24-ADAP24-0159, Scialog award SA-LSST-2024-102a, and the Discovery Alliance Catalyst Fellowship Mentors award 2025-62192-CM-19

A.A. acknowledges support from the Ministry of Education Yushan Fellow Program (MOE-111-YSFMS-0008-001-P1) and from the National Science and Technology Council, Taiwan (NSTC 114-2112-M-008-021-MY3).

T.-W.C. acknowledges support from the Ministry of Education Yushan Fellow Program (MOE-111-YSFMS-0008-001-P1) and from the National Science and Technology Council, Taiwan (NSTC 114-2112-M-008-021-MY3).

E.R.C.~acknowledges support from the National Aeronautics and Space Administration through the Astrophysics Theory Program, grant 80NSSC24K0897.

G.D. acknowledges support from the European Union’s Horizon Europe research and innovation programme under the Marie Skłodowska-Curie grant agreement No 101199369.

D.F.'s contribution to this material is based upon work supported by the National Science Foundation under Award No. AST-2401779.

A.G.Y.'s research is supported by ISF, IMOS and BSF grants, as well as the André Deloro Institute for Space and Optics Research, the Center for Experimental Physics, a WIS-MIT Sagol grant, the Norman E Alexander Family M Foundation ULTRASAT Data Center Fund, and Yeda-Sela;  AGY is the incumbent of the The Arlyn Imberman Professorial Chair.

L.G. acknowledges financial support from AGAUR, CSIC, MCIN and AEI 10.13039/501100011033 under projects PID2023-151307NB-I00, PIE 20215AT016, and CEX2020-001058-M.

M.G. acknowledges support from an STFC PhD studentship and from the Faculty of Science and Technology at Lancaster University

C.P.G. acknowledges financial support from the Secretary of Universities and Research (Government of Catalonia) and by the Horizon 2020 Research and Innovation Programme of the European Union under the Marie Sk\l{}odowska-Curie and the Beatriu de Pin\'os 2021 BP 00168 programme, from the Spanish Ministerio de Ciencia e Innovaci\'on (MCIN) and the Agencia Estatal de Investigaci\'on (AEI) 10.13039/501100011033 under the PID2023-151307NB-I00 SNNEXT project, from Centro Superior de Investigaciones Cient\'ificas (CSIC) under the PIE project 20215AT016 and the program Unidad de Excelencia Mar\'ia de Maeztu CEX2020-001058-M, and from the Departament de Recerca i Universitats de la Generalitat de Catalunya through the 2021-SGR-01270 grant.

C.L. is supported by DoE award \#\,DE-SC0025599.

Zwicky Transient Facility, W. M. Keck Observatory and MMT Observatory access was supported by Northwestern University and the Center for Interdisciplinary Exploration and Research in Astrophysics (CIERA).

A.M. gratefully acknowledges support from an STFC PhD studentship and the Faculty of Science and Technology at Lancaster University.

T.E.M.B. is funded by Horizon Europe ERC grant no. 101125877.

M.N. is supported by the European Research Council (ERC) under the European Union’s Horizon 2020 research and innovation programme (grant agreement No.~948381).

N.R. is supported by a Northwestern University Presidential Fellowship Award. We gratefully acknowledge the support of the NSF-Simons AI-Institute for the Sky (SkAI) via grants NSF AST-2421845 and Simons Foundation MPS-AI-00010513.

A.S. acknowledges the Warwick Astrophysics PhD prize scholarship made possible thanks to a generous philanthropic donation.

S.J.S. acknowledges funding from STFC Grant ST/Y001605/1, a Royal Society Research Professorship and the Hintze Family Charitable Foundation. 

%%%%%%%%%%%%%%%%%%%%%%%%%%%%%%%%%%%%%%%%%%%%%%%%%%
\section*{Data Availability}

All photometry is provided in the supplementary data files.  Spectroscopy will be made available on WISEREP, and is available on request to the lead author.

%%%%%%%%%%%%%%%%%%%% REFERENCES %%%%%%%%%%%%%%%%%%

\bibliographystyle{mnras}
\bibliography{ref}

\clearpage

$^{\ljmu}$ Astrophysics Research Institute, Liverpool John Moores University, 146 Brownlow Hill, Liverpool L3 5RF, UK \\
$^{\cornell}$ Department of Astronomy, Cornell University, Ithaca, NY 14853, USA \\
$^{\zhejiang}$ Institute for Advanced Study in Physics, Zhejiang University, Hangzhou 310027, China \\
$^{\weizmann}$ Department of Particle Physics and Astrophysics, Weizmann Institute of Science, 234 Herzl St, 7610001 Rehovot, Israel \\
$^{\taui}$ School of Physics and Astronomy, Tel Aviv University, Tel Aviv 6997801, Israel \\
$^{\caltech}$ Division of Physics, Mathematics and Astronomy, California Institute of Technology, Pasadena, CA 91125, USA \\
$^{\oxford}$ Astrophysics sub-Department, Department of Physics, University of Oxford, Keble Road, Oxford, OX1 3RH, UK \\
$^{\syracuse}$ Department of Physics, Syracuse University, Syracuse, NY 13210, USA \\
$^{\esochile}$ {European Southern Observatory, Alonso de C\'ordova 3107, Vitacura, Casilla 19001, Santiago, Chile} \\
$^{\unc}$ Department of Physics and Astronomy, University of North Carolina at Chapel Hill, Chapel Hill, NC 27599-3255, USA \\
$^{\ncu}$ Graduate Institute of Astronomy, National Central University, 300 Jhongda Road, 32001 Jhongli, Taiwan \\
$^{\dirac}$ {DIRAC Institute, Department of Astronomy, University of Washington, 3910 15th Avenue NE, Seattle, WA 98195, USA} \\
$^{\berkeley}$ {Department of Astronomy, University of California, Berkeley, CA 94720, USA} \\
$^{\lbnl}$ Physics Division, Lawrence Berkeley National Laboratory, 1 Cyclotron Road, MS 50B-4206, Berkeley, CA 94720, USA \\
$^{\ifa}$ Institute for Astronomy, University of Hawaii, 2680 Woodlawn Drive, Honolulu HI 96822 \\
$^{\ista}$ Institute of Science and Technology Austria, Am Campus 1, 3400 Klosterneuburg, Austria \\
$^{\finca}$ Finnish Centre for Astronomy with ESO (FINCA), FI-20014 University of Turku, Finland \\
$^{\ipac}$ IPAC, California Institute of Technology, 1200 E. California Blvd, Pasadena, CA 91125, USA \\
$^{\umn}$ School of Physics and Astronomy, University of Minnesota, Minneapolis, MN 55455, USA \\
$^{\sorbonne}$ Sorbonne Universite, F-75014 Paris, France \\
$^{\cnrs}$ Institut d'Astrophysique de Paris (IAP), CNRS  \\
$^{\lancaster}$ Department of Physics, Lancaster University, Lancaster, LA1 4YB, UK \\
%$^{\leicester}$ School of Physics and Astronomy, University of Leicester, University Road, Leicester, LE1 7RH, United Kingdom \\
$^{\coo}$ Caltech Optical Observatories, California Institute of Technology, Pasadena, CA 91125, USA \\
$^{\cfa}$ Center for Astrophysics | Harvard \& Smithsonian, 60 Garden Street, Cambridge, MA 02138, USA \\
$^{\ice}$ Institute of Space Sciences (ICE-CSIC), Campus UAB, Carrer de Can Magrans, s/n, E-08193 Barcelona, Spain \\
$^{\ieec}$ Institut d'Estudis Espacials de Catalunya (IEEC), 08860 Castelldefels (Barcelona), Spain \\
$^{\warsawobs}$ Astronomical Observatory, University of Warsaw, Al. Ujazdowskie 4, 00-478 Warszawa, Poland \\
$^{\cardiff}$ Cardiff Hub for Astrophysics Research and Technology, School of Physics \& Astronomy, Cardiff University, Queens Buildings, The Parade, Cardiff, CF24 3AA, UK \\
$^{\turku}$ Department of Physics and Astronomy, University of Turku, 20014 Turku, Finland \\
$^{\northwestern}$ Department of Physics and Astronomy, Northwestern University, 2145 Sheridan Rd, Evanston, IL 60208, USA \\
$^{\ciera}$ Center for Interdisciplinary Exploration and Research in Astrophysics (CIERA), Northwestern University, 1800 Sherman Ave, Evanston, IL 60201, USA \\
$^{\skai}$ NSF-Simons AI Institute for the Sky (SkAI), 172 E. Chestnut St., Chicago, IL 60611, USA \\
$^{\carnegie}${Observatories of the Carnegie Institution of Washington, 813 Santa Barbara St, Pasadena, CA 91101, USA} \\
$^{\mitkavli}${MIT Kavli Institute for Astrophysics and Space Research, 70 Vassar St, Cambridge, MA 02139} \\
$^{\tcd}$ School of Physics, Trinity College Dublin, The University of Dublin, Dublin 2, Ireland \\
$^{\icen}$ Instituto de Ciencias Exactas y Naturales (ICEN), Universidad Arturo Prat, Chile \\
$^{\queens}$ Astrophysics Research Centre, School of Mathematics and Physics, Queens University Belfast, Belfast BT7 1NN, UK \\
$^{\okc}$ The Oskar Klein Centre, Department of Astronomy, Stockholm University, Albanova University Center, SE 106 91 Stockholm, Sweden \\
$^{\ncnr}$ Astrophysics Division, National Centre for Nuclear Research, Pasteura 7, 02-093 Warsaw, Poland \\
$^{\tarapaca}$ Instituto de Alta Investigaci\'on, Universidad de Tarapac\'a, Casilla 7D, Arica, Chile \\
$^{\ucla}$ Department of Physics \& Astronomy, University of California Los Angeles, PAB 430 Portola Plaza, Los Angeles, CA 90095-1547, USA \\
$^{\warwick}$ Department of Physics, University of Warwick, Gibbet Hill Road, Coventry CV4 7AL, UK \\
$^{\maryland}$ Department of Astronomy, University of Maryland, College Park, MD 20742, USA \\
$^{\jssi}$ Joint Space-Science Institute, University of Maryland, College Park, MD 20742, USA \\
$^{\gsfc}$ Astrophysics Science Division, NASA Goddard Space Flight Center, Mail Code 661, Greenbelt, MD 20771, USA \\
$^{\cddd}$ Center for Data Driven Discovery, California Institute of Technology, Pasadena, CA 91125, USA \\

%%%%%%%%%%%%%%%%%%%%%%%%%%%%%%%%%%%%%%%%%%%%%%%%%%

%%%%%%%%%%%%%%%%% APPENDICES %%%%%%%%%%%%%%%%%%%%%

\appendix

\section{Observational Details}
\label{sec:observationdetails}

\subsection{Palomar 60-inch telescope}

The SEDM on P60 \citep{Blagorodnova+2018} was triggered via the web interface on Fritz \citep{vanderWalt+2019,Coughlin+2023} to obtain spectroscopy of AT\,2024wpp almost immediately after its discovery.  Observations begain at MJD 60579.4667, 2.3 hours after the first alert (and within 1 hour after the second alert caused the candidate to pass the Bright Transient Survey software filter).  The initial trigger was for spectroscopic observations as part of the Bright Transient Survey.  An $r$-band acquisition image of the field using the SEDM Rainbow Camera was obtained before the telescope proceeded to spectroscopy using the central IFU.  Additional Rainbow Camera imaging was obtained on subsequent nights in $u$, $g$, $r$, and $i$ filters.

Imaging observations were automatically processed and image differenced against Pan-STARRS referencing imaging using the P60 pipeline, following the method of \cite{Fremling+2016}.  The flux of the source was determined via PSF photometry.

The IFU observations were processed and the source spectrum extracted using the SEDM data reduction pipeline \citep{Rigault+2019,Kim+2022}.  Unfortunately, the early spectrum was affected by camera artefacts due to a recurring detector malfunction and is not useable. 

\subsection{Liverpool Telescope}

\subsubsection{IO:O}

Observations were taken with the IO:O optical imager on the Liverpool Telescope \citep{Steele+2004} in the Sloan Digital Sky Survey (SDSS) $u$, $g$, $r$, $i$, and $z$ filters.  Reduced images were downloaded from the telescope archive and processed with custom image-subtraction and analysis software. Image stacking and alignment is performed using \textsc{SWarp} \citep{Bertin2010} where required. Image subtraction on $g$, $r$, $i$, and $z$ observations is performed using a pre-explosion reference image in the appropriate filter from Pan-STARRS1.  Photometry in these bands was measured using PSF fitting methodology relative to Pan-STARRS1 or SDSS standards and is based on techniques described in \cite{Fremling+2016}.

For the SDSS $u$-band, the source location was not covered by SDSS and as a result image subtraction could not be performed. Aperture photometry was performed using \texttt{sep} \citep{Barbary+2016}, a python implementation of \texttt{SExtractor} \citep{Bertin+1996}. A 2.5" aperture was used to maximise flux measured from the transient whilst avoiding contamination from the nearby host galaxy. The $u$-band airmass coefficient was measured using two observations of the standard star 114-654 \citep{Landolt+1972} on the night of 2024-09-30. This standard star was then used to measure the brightness of the star BD-17 520, present in our $u$-band images of AT\,2024wpp, which was then used to calibrate subsequent epochs. For our final two epochs BD-17 520 was saturated. The star SDSS J024201.54-165333.8, close to the edge of our frames with photometry measured by SDSS, was visible in the deeper images and used instead.  The magnitudes of these secondary standard stars (and others relevant for calibrating other observations with small numbers of viable stars in the field) are provided in Table \ref{tab:standardstars}.

\begin{table*}
	\centering
	\caption{Secondary calibration stars}
    \label{tab:standardstars}
	\begin{tabular}{llllllll}
		\hline
RA          & dec        & $u$              & $g$              & $r$              & $i$      & $J$ & $H$ \\   
02:42:10.24 & -17:00:23.5 & 12.35 $\pm$ 0.01 &   &  & &  & \\ 
02:42:01.54 & -16:53:33.8 & 19.82 $\pm$ 0.04 &   &  & &  & \\ 
02:42:07.24 & -16:56:36.3 & 19.70 $\pm$ 0.10 & 18.54 $\pm$ 0.02 & 18.11 $\pm$ 0.01 & 17.97 $\pm$ 0.01  & & \\ 
02:42:04.30 & -16:56:38.7 & 20.60 $\pm$ 0.13 & 19.13 $\pm$ 0.01 & 18.60 $\pm$ 0.01 & 18.37 $\pm$ 0.01 & & \\ 
02:42:06.31 & -16:54:35.0 &                  & 19.52 $\pm$ 0.02 & 19.18 $\pm$ 0.01 & 19.08 $\pm$ 0.01 & & \\ 
02:41:59.29 & -16:57:43.4 &                  & 21.38 $\pm$ 0.07 & 20.39 $\pm$ 0.05 & 19.67 $\pm$ 0.02 & & \\ 
02:41:57.82 & -16:57:56.4 &                  & 21.07 $\pm$ 0.05 & 20.62 $\pm$ 0.04 & 20.40 $\pm$ 0.03 & & \\ 
\hline
02:41:53.12 & -16:55:54.7 &                  &                  & 20.29 $\pm$ 0.02 & 19.02 $\pm$ 0.02 & & \\ 
02:42:00.70 & -16:57:32.0 &                  &                  & 20.84 $\pm$ 0.02 & 19.29 $\pm$ 0.02 & & \\ 
02:42:08.16 & -16:57:29.1 &                  &                  & 21.72 $\pm$ 0.02 & 19.82 $\pm$ 0.02 & & \\ 
\hline
02:41:59.39 & -17:01:03.0 & & & &  &                  & 15.85 $\pm$ 0.17 \\
02:41:58.72 & -16:59:52.5 & & & &  & 15.59 $\pm$ 0.06 & 14.84 $\pm$ 0.07 \\ 
        \hline        
      \hline
	\end{tabular}
{\par \begin{flushleft}
Note --- Magnitudes are given only in filters employed for calibration of the transient in LT IO:O imaging ($u$), VLT FORS2 imaging ($u$, $g$, $r$, $i$), and LT LIRIC imaging ($J$, $H$).  A wider selection of stars was used for other filters and instruments.  Values in the top segment of the table are from our LT:IOO calibration ($u$) or PS ($griz$); values in the middle section are from our VLT calibration; values in the bottom section are from our LT:LIRIC calibration.\\
\end{flushleft}}
\end{table*}

\subsubsection{LIRIC}
Near infrared data was taken with LIRIC on the Liverpool Telescope in the FELH1500 filter and Barr J filter. The FELH1500 filter combined with the sCMOS detector result in a pseudo H-band filter \citep{Batty+2022}. 

The observations were dithered using a nine-position large offset pattern. The first set of observations was a 100$\times$30s exposure in the H band taken on MJD 60588.103229. The second set of observations was taken at MJD 60594.109514 in H band (20$\times$30s) and J band (16$\times$30s).  In an attempt to ensure sufficient bright stars were contained within the sequence for photometric calibration, not all dither positions contained the target.

The data was reduced manually, following the usual steps: the raw images were bias subtracted using the available overscan, and sky subtracted using three nearest exposures on each side of an image. They were stacked manually due to the lack of bright sources in most of the exposures; the stack shifts were based on known pixel offsets per each dither and a model of telescope wobble during the duration of the observations. The aperture photometry was performed with \texttt{SExtractor} and calibrated using the 2MASS All-Sky Point Source Catalog \citep{Skrutskie+2003}.

For the first night, we recover a source of magnitude $H = 18.34^{+0.35}_{-0.30}$, for a total exposure of 900s on the source. For the second night, we obtain $3\sigma$ upper limits of $H > 18.67$ mag (180\,s exposure on the source) and $J > 16.10$ mag. J-band images did not have bright objects in the dithered fields around AT\,2024wpp and could not be stacked, so the upper limit is given for a single 30s exposure.

\subsubsection{SPRAT}

We obtained two spectra using the SPectrograph for the Rapid Acquisition of Transients (SPRAT; \citealt{Piascik+2014}) at MJD 60580.182 and MJD 60582.219. The first spectrum was observed using a blue optimised configuration, and the second a red optimised configuration. The spectra were reduced using a custom python code utilising the packages \texttt{numpy} \citep{Harris+2020}, \texttt{scipy} \citep{scipy+2020}, \texttt{astropy} \citep{astropy+2022}, \texttt{matplotlib} \citep{Hunter+2007} and \texttt{lacosmic} \citep{vanDokkum+2001}. Flux calibration was carried out using the standard star Hiltner 102 \citep{Stone+1977}, observed on the same nights. Airmass differences between science and standard star observations were then corrected for using Table 1 from La Palma Technical Note No. 31\footnote{\url{https://www.ing.iac.es/Astronomy/observing/manuals/ps/tech_notes/tn031.pdf}}.

\subsection{SOAR/GHTS}
\label{sec:ghts}

We obtained three epochs of longslit spectroscopy of AT\,2024wpp with the Goodman High Throughput Spectrograph (GHTS; \citealt{Clemens2004}) mounted on the Southern Astrophysical Research (SOAR) telescope on 27 September 2024, 28 September 2024, and 1 October 2024.  The observations consisted of 3 $\times$ 300\,s exposures. All three observations were taken with a grating of 400 lines/mm and a 1.0'' wide slit mask in the M1 spectroscopic setup (hereafter 400M1) with $2 \times 2$ binning using the GHTS Red Camera. The 400M1 spectra cover a wavelength range of 3800 -- 7040 $\AA$.

We reduced the spectra using a custom Python routine \citep{Kaiser+2021} with a slightly
improved technique for the flux calibration (Kaiser et al. in prep.). Observations in the 400M1 configuration are affected by a broad sensitivity dip near 4800 \AA, and the masking of Balmer lines in typical spectrophotometric standards causes the derived sensitivity functions to omit this correction.  We derived a relative sensitivity function correction using observations of three DC white dwarfs from 24 April 2025, which we applied prior to the derivation of the spectrophotometric standard-based sensitivity function (which is now a ``pseudo-sensitivity'' function). The (pseudo-)sensitivity function for flux calibration was created using the reference spectrum from \citet{Moehler+2014} and observations of the spectrophotometric standard LTT 7987 obtained the night of 11 November 2025 in the same setup as the previous observations.

We also obtained imaging observations at the location of AT\,2024wpp on 28 November 2024, 3 December 2024, 15 December 2024, and 22 December 2024 with SOAR/GHTS in imaging mode. We took 2 $\times$ 300\,s exposures each in $g$ and $r$-band during the first two epochs, 2 $\times$ 300\,s exposures in $g$, $r$, and $i$-band on the third epoch, and 3 $\times$ 300\,s exposures in $g$ and $r$-band.  After bias-correction, flat fielding, background subtraction and astrometric correction using \textsc{Astrometry.net} \citep{Lang2010}, the images were stacked with \textsc{SWarp}.  We performed image subtraction on the stacked images using \texttt{SFFT} \citep{Hu2022} with an archival template from the DECam Legacy Survey \citep{Dey+2019}. Aperture photometry was conducted with \textsc{SExtractor} \citep{Bertin+1996} and zeropoints were measured from the Pan-STARRS1 catalogue \citep{Chambers+2016}.

\subsection{P200/DBSP}

AT\,2024wpp was observed using the Double Beam Spectrograph (DBSP, \citealt{Oke+1982}) on the Palomar 200-inch telescope (P200) on four occasions.  The observations were reduced using DBSP-DRP \citep{dbsp_drp} and Pypeit \citep{pypeit}.

\subsection{NOT/ALFOSC}

We obtained spectroscopic observations using the Alhambra Faint Object Spectrograph and Camera (ALFOSC\footnote{\href{http://www.not.iac.es/instruments/alfosc}{{http://www.not.iac.es/instruments/alfosc}}}) mounted on the 2.56~m Nordic Optical Telescope (NOT) located at the Roque de los Muchachos Observatory on La Palma (Spain). 
The spectra were reduced in a standard manner using a custom fork of \texttt{PypeIt} \citep{pypeit}. 

\subsection{Lick/Kast}

We obtained spectra on two consecutive nights at MJD 60590.341 and MJD 60591.358 on the Kast Double Spectrograph mounted on the 3~m Shane telescope. We observed with a slit width of 1.5”. For the red side we used the 300/7500 grating (2.55\,\AA\,pixel$^{-1}$\ dispersion) and for the blue side we used the 600/4310 grism (1.02\,\AA\,pixel$^{-1}$\ dispersion). Spectra were reduced using the UCSC spectral pipeline \citep{Siebert+2019}, with flux calibration carried out using the star Feige 34 observed on the 2nd night.

\subsection{WINTER}

We observed AT 2024wpp in the J- and shortened H-bands with the Wide-field Infrared Transient explorer \citep[WINTER;][]{2020SPIE11447E..9KL, 2024SPIE13096E..3JF}, a camera mounted on the 1-m telescope at Palomar Observatory. Data were reduced using the standard WINTER pipeline built using \texttt{mirar} \citep{mirar}. We performed six epochs of imaging, with detections in the J and Hs bands during the first set of observations beginning at 2024-10-05T08:45:23 UTC. Subsequent epochs resulted in non-detections.

\subsection{MMT/Binospec}

We observed AT\,2024wpp with Binospec \citep{Fabricant+2019} mounted on the 6.5\,m MMT telescope at the Fred Lawrence Whipple Observatory. We used the 270 lines/mm grating and achieved wavelength coverage between 3820-9210~\AA\ at $R\sim1340$. The reduction is performing following standard procedures implemented in \texttt{pypeit} \citep{pypeit}.  The Binospec sensitivity function is known to be unstable, and the resulting spectrum could not be flux calibrated to a precision sufficient for this work, but it is consistent with the lack of any narrow features at this epoch.

\subsection{Pan-STARRS}

Pan-STARRS (PS) comprises two 1.8-m telescope units (Pan-STARRS1 and Pan-STARRS2), both situated at the summit of Haleakala on the Hawaiian island of Maui \citep{Chambers+2016}. Pan-STARRS1 (PS1) has a 1.4 gigapixel camera and a 0.26\arcsec pixel scale, providing a $\sim$7 square degree field-of-view (FOV). Pan-STARRS2 (PS2) possesses a 1.5 gigapixel camera and a slightly larger FOV. Both PS1 and PS2 are equipped with the same filter system ($grizy_{\rm PS}$; see \citealt{Tonry+2012}). Both telescopes also possess a broad $w_{\rm PS}$ filter, which is a composite of $g \! + \! r \! + \! i$. Images were obtained using both telescopes, and these were processed with the image processing pipeline \citep{Magnier+2020a}. All images were astrometrically and photometrically calibrated \citep{Magnier+2020b}, before having reference frames subtracted from the target frames \citep{Waters+2020}. PSF photometry was then performed on these difference images \citep{Magnier+2020c}. 

We note that at the sky location of AT\,2024wpp, the STScI PS1 $3 \pi$ stamp server reference frames are shallower than our late-time stacked target images \citep{Flewelling2020}. Subtraction of a reference template shallower than the science image can lead to substantial systematic errors in measuring reliable photometry for the transient ($\lesssim 0.8$~mag for AT\,2024wpp at late times).
To account for this effect, we revisited the sky position of AT\,2024wpp in September 2025, and acquired deep $grizy_{\rm PS}$ reference imaging. Specifically, our updated reference stacks possess total exposure times,
$t_g = 2400$,
$t_r = 2400$,
$t_i = 800$,
$t_z = 1200$ and
$t_y = 2000$~seconds.
These stacks are sufficiently sensitive to act as reference templates for our late-time science observations,
with a recovery fraction of 50~per~cent for fake stars down to
$m_g \approx 24.1$,
$m_r \approx 23.9$,
$m_i \approx 23.6$,
$m_z \approx 22.8$ and
$m_y \approx 22.0$~mag.
Using these deep stacks as reference templates 
allows us to extract reliable photometry for AT\,2024wpp by accurately subtracting the host galaxy flux, a vital step given the 
contribution of its nearby host.

\subsection{NTT/EFOSC2}

We obtained seven epochs of photometric observations in the Bessel $U$ and Gunn $g$, $r$, $i$ and $z$ filters with the ESO Faint Object Spectrograph and Camera version 2 (EFOSC2; \citealt{EFOSC2}), mounted on the New Technology Telescope (NTT) at La Silla Observatory, Chile. Images were downloaded from the ESO archive and reduced using the EsoReflex pipeline \citep{esoreflex}. Images were stacked and aligned using SWarp. There were significant fringing effects in the $i$ and $z$ bands, which were manually corrected using a custom method: reference images were created of the respective $i$ and $z$ fringing patterns from an epoch with minimal sources and at least nine dithering positions. A small cut-out of a dark section of the fringing pattern in the science and reference image was created and a median pixel value for each cutout was calculated. This was repeated for a bright section of the fringing pattern in each image. The average fringing effect was calculated by subtracting the median dark value from the median bright value in each image. The science frame was then corrected by subtracting the fringe frame scaled by the ratio between the two average values.  

Image subtraction was performed using reference images taken by EFOSC2 for the $g$, $r$, $i$ and $z$ filters. The subtraction was performed in a standard manner: each science image was convolved with the PSF of the reference image and vice versa. The scaling factor was determined by comparing the SExtractor magnitudes of selected stars in the science image to the reference image. The final subtracted image was produced by subtracting the scaled, convolved reference image from the convolved science image. Aperture photometry was performed on the subtracted images. For the $U$ band, aperture photometry was performed on the stacked, corrected images. A selection of bright stars was used to find the zero-points. For the  $g$, $r$, $i$ and $z$ filters, the PanSTARRS 1 (PS1) catalogue magnitudes were compared to the SExtractor magnitudes. For the $U$ band, the reference magnitudes were calculated for two bright stars in an LT frame using an SDSS source in the frame as a standard. The aperture diameter was determined using the estimated seeing for each epoch.

\subsection{VLT/FORS2}
\label{sec:fors2}

We obtained three epochs of spectroscopic observations with the FORS2 instrument \citep{Appenzeller+1998} installed at VLT/UT1 under DDT Programme 2114.D-5014(E), PI Perley. Two exposures were taken each night, 2$\times$1320s for the first epoch and 2$\times$1380s for the other two epochs; all spectra were taken using the 300V grating. The spectra were reduced in a standard manner (they were de-biased, corrected for cosmic rays, flat-fielded, wavelength calibrated using arc-lamp lines and skylines, flux-calibrated, and corrected for telluric absorption) using a custom python code utilising packages \texttt{numpy} \citep{Harris+2020}, \texttt{scipy} \citep{scipy+2020}, \texttt{astropy} \citep{astropy+2022}, \texttt{ccdproc} \citep{Craig+2025}, and \texttt{lacosmic} \citep{vanDokkum+2001}. The spectra taken at the same epoch were reduced individually and then stacked together.

We also obtained three epochs of photometric observations in the $u_{\rm High}$, $g_{\rm High}$, $R_{\rm Special}$ and $I_{\rm Bessel}$ filters (hereafter, $u$, $g$, $r$, and $i$) with FORS2. The images were downloaded from the ESO archive and reduced using the EsoReflex pipeline. They were then stacked and aligned using SWarp. Image subtraction was performed using late-time reference images taken by FORS2 for each filter: $u$ and $g$ references were taken on 2025-07-22 and $r$ and $i$ images were taken on 2025-08-03.  The subtraction was performed in a standard manner: each science image was convolved with the PSF of the reference image and vice versa. The scaling factor was determined by comparing the SExtractor magnitudes of selected stars in the science image to the reference image. The final subtracted image was produced by subtracting the scaled, convolved reference image from the convolved science image. Aperture photometry was performed on the subtracted image. Zero-points for the $g$ and $i$ filters were computed by comparing the PS1 magnitudes (in the equivalent filters) of a selection of stars in the convolved science images to their SExtractor magnitudes.  For the $u$ band, the reference magnitudes were calculated for two bright stars in an LT frame using an SDSS source in the frame as a standard.   For $r$ band, a number of the PS1 calibration stars are saturated in the December, January, and reference imaging: to obtain a more consistent calibration we use the November image to refine the magnitudes of several fainter stars with only marginal detections in PS1 (refined values are given in the middle section of Table \ref{tab:standardstars}), then use these to calibrate the later images.  A similar procedure was used for $i$-band, although it was only needed for measuring the host galaxy magnitude using the reference image.

\subsection{Magellan/IMACS}
We obtained one optical spectrum of AT\,2024wpp on 2024-11-06 using the Inamori Magellan Areal Camera and Spectrograph \citep[IMACS;][]{Dressler2011} mounted on the 6.5-m Magellan-Baade telescope under decent conditions ($\sim$ 0.7\farcs). The observations consist of two 1500-second exposures with a 300 lines/mm grating, resulting in a spectral resolution R $\sim$ 1000. The spectra were reduced with SimSpec \footnote{\url{https://astro.subhashbose.com/software/simspec}}, a semi-automated, modular, long-slit spectroscopic data reduction pipeline based on {\tt PyRAF} \citep{Pyraf}. The data reduction includes basic data processing (bias subtraction, flat fielding), cosmic-ray removal, wavelength calibration (using arc lamp frames taken immediately after the target observation), and relative flux calibration with a spectroscopic standard observed the same night as the science object. 

\subsection{Keck Cosmic Web Imager}
\label{sec:kcwi}

We obtained two epochs of observations using the Keck Cosmic Web Imager (KCWI) Integral Field Unit spectrograph \citep{Morrissey+2018}.  The first epoch was acquired on 2024 Oct 2 using the medium slicer, the BL grating centred at 4400\AA, and the RL grating centred at 7150\AA.  The second epoch was acquired on 2025 Jan 1 using the medium slicer, the BL grating centred at 4600\AA, and the RL grating centred at 7100\AA.
Observations were reduced using the official data reduction pipeline, with minor modifications:  the differential atmospheric refraction correction was based on linear shifting and centred at 5600\,\AA\ for both channels, cosmic ray rejection was performed using a mask generated from multiple 2D frames, sky subtraction was performed on individual frames using median subtraction, and a supplementary telluric correction was performed.  Elliptical apertures were used to extract the transient and the host galaxy; the flux scale for the host was then corrected using profile-fitted photometry from the Legacy Survey DR10.  Additionally, the object spectrum from 2024 Oct 2 was saturated in the central slit, and so was extracted using neighbouring slits and flux-normalized using our photometry.

\subsection{Keck/LRIS}

AT\,2024wpp was observed using LRIS on four occasions (October 6, October 9, November 3, and November 9).   The first three occasions employed the B400 grating and R400 grism, while on the last occasion we employed the B600 grating and R400 grism.  

Data were reduced using LPipe \citep{Perley2019} using standard techniques and spectrophotometric standards observed in the same configuration.  The basic pipeline reduction of the observation from October 9 is affected by low-level ($\sim$5\%) residuals owing from Balmer features in the spectrum of the standard star BD+28\,4211 close to the peak of the LRIS-blue sensitivity curve (the issue is similar to the one affecting the GHTS observations; \S \ref{sec:ghts}).  The sensitivity curve derived from G191-B2B on October 6 using the same LRIS-blue setup is free of this effect.  To remove these residuals, we first take the ratio of the two sensitivity functions in the blue region of the spectrum, then normalize it via a polynomial fit to isolate the higher-frequency features imposed by BD+28\,4211.  We damp the resulting correction curve beyond 5000\AA\ using a logistic function, and then use the final correction curve to correct the sensitivity function for October 9 and update the flux calibration of the observation of AT\,2024wpp.

\subsection{HST/WFC3}
\label{sec:hstphot}

We obtained two epochs in the F814W, F555W, F336W and F225W filters with the Wide Field Camera 3 (WFC3; \citealt{HST_WFC3}) mounted on the \emph{Hubble Space Telescope} (HST). The images downloaded from the MAST archive had already been reduced and stacked using the AstroDrizzle pipeline \citep{AstroDrizzle}.  
The count rate $C$ was calculated using a 5-pixel (0.1925$\arcsec$) radius circular aperture, and the AB magnitude was calculated using the zeropoints and aperture (encircled-energy) corrections from the WFC3 handbook.

%%%%%%%%%%%%%%%%%%%%%%%%%%%%%%%%%%%%%%%%%%%%%%%%%%

\bsp	% typesetting comment
\label{lastpage}
\end{document}